\documentclass[a4paper,fleqn,usenatbib]{mnras}

\usepackage{newtxtext,newtxmath}

\usepackage[T1]{fontenc}
\usepackage{ae,aecompl}

\usepackage{graphicx}	

\usepackage{epstopdf}
\usepackage{subfigure}
\usepackage{relsize}
\usepackage{commath}
\usepackage{algorithm}
\usepackage[noend]{algpseudocode}
\usepackage{multirow}
\usepackage{bigdelim}
\usepackage{bm}

\usepackage{tikz}
\usetikzlibrary{shapes.geometric, arrows}

\tikzstyle{startstop} = [rectangle, rounded corners, minimum width=3cm, minimum height=1cm,text centered, draw=black, fill=red!30, text width = 7cm]
\tikzstyle{io} = [trapezium, trapezium left angle=70, trapezium right angle=110, minimum width=3cm, minimum height=1cm, text centered, draw=black, fill=blue!30, text width=7cm]
\tikzstyle{decision} = [diamond, minimum width=3cm, minimum height=1cm, text centered, draw=black, fill=green!30]
\tikzstyle{arrow} = [thick,->,>=stealth]
\tikzstyle{process} = [rectangle, minimum width=3cm, minimum height=1cm, text centered, text width=7cm, draw=black, fill=orange!30]

\floatname{algorithm}{Procedure}

\title[Estimator for IGM EOS]{A Novel Estimator for the Equation of State of the IGM by Ly$\alpha$ Forest Tomography}

\author[]{
Hendrik M\"uller,$^{1,2}$\thanks{hmueller@mpifr-bonn.mpg.de}
Christoph Behrens,$^{1}$
David J.E. Marsh,$^{1}$
\\
$^{1}$Institut f\"ur Astrophysik, Universit\"at G\"ottingen, Friedrich-Hund Platz 1, D-37077 G\"ottingen, Germany\\
$^{2}$Present Address: Max-Planck-Institut f\"ur Radioastronomie at Bonn, Auf dem H\"ugel 69, D-53121 Bonn (Endenich), Germany
}

\date{Accepted XXX. Received YYY; in original form ZZZ}

\pubyear{2020}

\begin{document}
\label{firstpage}
\pagerange{\pageref{firstpage}--\pageref{lastpage}}
\maketitle

\begin{abstract}
We present a novel procedure to estimate the Equation of State of the intergalactic medium in the quasi-linear regime of structure formation based on Ly$\alpha$ forest tomography and apply it to 21 high quality quasar spectra from the UVES\_SQUAD survey at redshift $z=2.5$. Our estimation is based on a full tomographic inversion of the line of sight. We invert the data with two different inversion algorithms, the iterative Gauss-Newton method and the regularized probability conservation approach, which depend on different priors and compare the inversion results in flux space and in density space. In this way our method combines fitting of absorption profiles in flux space with an analysis of the recovered density distributions featuring prior knowledge of the matter distribution. Our estimates are more precise than existing estimates, in particular on small redshift bins. In particular, we model the temperature-density relation with a power law and observe for the temperature at mean density $T_0 = 13400^{+1700}_{-1300}\,\mathrm{K}$ and for the slope of the power-law (polytropic index) $\gamma = 1.42 \pm 0.11$ for the power-law parameters describing the temperature-density relation. Moreover, we measure an photoionization rate $\Gamma_{-12} = 1.1^{+0.16}_{-0.17}$. An implementation of the inversion techniques used will be made publicly available.
\end{abstract}

\begin{keywords}
(cosmology:) large scale structure of Universe-- quasars: absorption lines --(methods): data analysis
\end{keywords}



\section{Introduction}
In the cosmological standard model small initial density perturbation evolved to the large scale structure of matter in the Universe that is visible today \citep{Mukhanov2005, Planck2016, Planck2020}. This large scale structure consists of dense objects such as galaxies and galaxy clusters, and the intergalactic space which is diffusely populated with baryonic matter, the so called intergalactic medium (IGM). The physics of the IGM has been targeted by many studies in the past. Among many applications, the clustering properties of matter in the IGM can be used to study dark matter \citep{Seljak2006, Hui2017, Rogers2020}. Moreover, the thermal history of the IGM provides information on the reionization epoch, when the first galaxies and stars formed at high redshift, and other related cosmic heating processes. \citep{McQuinn2010, Compostella2013, Daloisio2015, McQuinn2016, Liu2020}

A powerful probe of the IGM is the Ly$\alpha$ forest. The Ly$\alpha$ forest consists of densely packed, narrow absorption lines at higher frequencies (bluewards) of the Ly$\alpha$ emission line in the spectra of distant galaxies. It appears due to intervening hydrogen along the line of sight, and thus having differently redshifted absorption lines \citep{Bahcall1965, Gunn1965, Bi1997}. At every point along the line of sight the IGM absorbs a fraction of the light with an optical depth which is proportional to the density of neutral hydrogen. Since every point along the line of sight corresponds to a specific Hubble redshift, the normalized flux in the Ly$\alpha$ forest is a tracer of the neutral hydrogen density fluctuations in the IGM. \citep{Bi1992, Hui1997}

The profiles of the absorption lines contain information regarding thermal properties of the IGM, peculiar velocities and the density profile of the underlying absorbers. There is, however, a particular difficulty in studying these absorbtion features: The width of the absorption lines is a composition of an intrinsic width of the underlying density profile, and thermal broadening of the line, which is difficult to separate \citep[compare][]{Hui1997, Garzilli2020}. Moreover, peculiar velocities affect the profile of the absorption lines. The Equation of State (EOS) is a relation between the overdensity and the temperature in the IGM, hence describing the thermal part of the line broadening. Thus, it is crucial to determine the EOS of the IGM first, above all the IGM temperature, when deducing the clustering properties of the underlying neutral hydrogen density field from absorption line profiles. Furthermore, constraining the EOS of the IGM is of interest on its own to understand the thermal evolution of the IGM and related heating processes.

There are many attempts to estimate the temperature of the IGM: Among others this includes \citet{Schaye1999, Schaye2000, McDonald2001, Theuns2002, Viel2006, Bolton2008, Viel2009, Lidz2010, Becker2011, Calura2012, Rudie2012, Garzilli2012, Boera2014, Bolton2014, Hiss2018, Rorai2018, Boera2019, Walther2019,  Telikova2019, Garzilli2020, Gaikwad2020, Gaikwad2020b}. In these studies the thermal Doppler broadening and the intrinsic width of the filaments in the IGM are distinguished by Voigt-profile fitting, curvature methods, wavelet studies or studies of the Ly$\alpha$ forest power spectrum. The exact values recovered by the different groups vary, but coincide at temperatures at mean density $T_0 \sim 5000-20000\,\text{K}$ at redshift $z\sim 2-3$. However, note that there remain some notable discrepancies between the different methods. In this work we present a novel estimate by a full inversion technique of the Ly$\alpha$ forest spectra observed in UVES\_SQUAD \citep{Murphy2019} survey.

Fitting tools are only a part of the tomographic use of the Ly$\alpha$ forest. In the past, powerful inversion schemes for the reconstruction of the neutral hydrogen density along a single line of sight at high (echelle) spectral resolution were proposed \citep{Nusser1999, Pichon2001, Gallerani2011, Mueller2020} and shown to yield accurate reconstruction results \citep{Mueller2020}. Moreover, also the neural network method currently presented in \citet{Huang2020} could be used to carry out such reconstructions, but has not been utilized in this way yet. \citet{Rollinde2001} demonstrated that inversion schemes are capable of constraining the EOS of the IGM. The quality of the reconstruction with the scheme proposed in \cite{Pichon2001} depends strongly on the a-priori chosen EOS. Moreover, the schemes proposed in \cite{Gallerani2011} and \cite{Mueller2020} depend on weaker prior assumptions, in this case being independent from any choice for the EOS. In this paper, we compare the inversion results of these approaches while varying the prior assumptions on the EOS for the inversion with the method provided by \cite{Pichon2001}. Our estimates are based on the best match to observational data and the agreement between these different inversion algorithms (which depend on different priors). In this way we additionally perform model fitting in the density domain (rather than on the observed flux). This provides a powerful pipeline for analysis of Ly$\alpha$ forest data regarding the thermal history of the IGM.

The plan for the rest of the paper is as follows: We present the analytic model of the Ly$\alpha$ forest in Sec. \ref{sec: analytic_model}, we present our inversion and estimation techniques in Sec. \ref{sec: method}. The flowchart in Fig. \ref{fig: estimation_procedure} summarizes our procedure. We test it on synthetic data in Sec. \ref{sec: synthetic}. We apply our novel method to observational data from the UVES\_SQUAD survey in Sec. \ref{sec: application}. Our final marginal distributions for $\gamma$ and $T_0$ are shown in Fig. \ref{fig: marginal}. Additionally we simulate synthetic data with the same spectral properties as the observational ones and exactly mimic the estimation done on observational data in Sec. \ref{sec: mock} to perform a consistency test and to verify our estimation procedure. Finally we discuss our results in Sec. \ref{sec: discussion} and finish with our conclusions.

During the rest of this paper we use the \citet{Planck2016} cosmology. The chosen cosmology mainly affects the relation between the neutral hydrogen density and the optical depth by scaling the conversion between comoving distance and redshift bins. However, the error introduced by the cosmological model i.e. the uncertainty in the evolution of the Hubble constant up to redshift $z=2.5$ measured by \citet{Planck2016}, is negligible compared to the accumulation of statistical and synthetic errors during the inversion procedure.

We will make all our implementations (in particular the used inversion schemes) publicly available as part of the \small{REGLYMAN} toolbox \footnote{Available at \url{https://github.com/hmuellergoe/reglyman}} \citep{reglyman}. This toolbox makes strong use of the \small{REGPY} \footnote{\url{https://github.com/regpy/regpy}} library \citep{regpy} and the \small{NBODYKIT} \footnote{\url{https://nbodykit.readthedocs.io}} library \citep{Hand2018}.

\section{Analytic Model} \label{sec: analytic_model}
The observable in the Ly$\alpha$ forest is the normalized flux $F$, i.e. the ratio between the observed flux and the flux that would be observed at full transmission. The optical depth $\tau$ is defined as the negative logarithm of the normalized flux, i.e. it is:
\begin{align}
    F = \exp \left( -\tau \right) \label{eq: normalized_flux}
\end{align}
The optical depth in the Ly$\alpha$ forest is given by \citep{Bahcall1965, Gunn1965, Hui1997}:
\begin{align}  \nonumber
\tau(z_0)&=\sigma_0 c \int_\mathrm{LOS} dx(z) \frac{n_\mathrm{H I}(x,z)}{1+z} \\
& \times \frac{1}{\sqrt{\pi} b_\mathrm{T}(x, z)} \exp \left[ -\left( \frac{ v_\mathrm{H}(z_0)-v_\mathrm{H}(z)-v_\mathrm{pec} (x, z)}{ b_\mathrm{T}(x, z)} \right)^2 \right], \label{eq: tau}
\end{align}
where $\sigma_0$ is the Ly$\alpha$ cross section, $c$ the speed of light, $n_\mathrm{HI}$ the number density of neutral hydrogen, $z$ the Hubble redshift , $v_\mathrm{H}$ the differential Hubble velocity and $v_\mathrm{pec}$ peculiar velocities (redshift space distortions). For Eq. \eqref{eq: tau} the Voigt profile is approximated to a Gaussian, which is valid for low column density regions. $b_\mathrm{T}$ is the thermal broadening parameter which satisfies \citep{Hui1997}:
\begin{align}
b_\mathrm{T}(x, z)=\sqrt{\frac{2 k_\mathrm{B} T(x, z)}{m_\mathrm{p}}}, \label{eq: broadening}
\end{align}
where $k_\mathrm{B}$ denotes the Boltzmann-constant, $T$ the actual temperature and $m_\mathrm{p}$ the proton mass. We assume a power law EOS as it is widely assumed in Ly$\alpha$ forest studies \citep{Hui1997, Sanderbeck2016}:
\begin{align}
    T(x, z) = T_0(z) \Delta^{\gamma-1}, \label{eq: eos}
\end{align}
where $T_0$ denotes the temperature at mean density and $\Delta=\rho_\mathrm{b}/\langle \rho_\mathrm{b} \rangle$ is the baryonic density perturbation. $T_0$ defines the scale of the temperature, and thus by Eq. \eqref{eq: broadening} the scale of thermal broadening. The neutral hydrogen density $n_\mathrm{HI}$ and the baryonic density perturbation $\Delta$ are related by \citep{Hui1997, Nusser1999}:
\begin{align}
    n_\mathrm{HI}(x, z) = \hat{n}_\mathrm{HI} (z) \Delta^\alpha(x, z), \label{eq: neutral_fraction}
\end{align}
where $\alpha=2.7-0.7 \gamma$ and $\hat{n}_\mathrm{HI}$ denotes the neutral hydrogen density at mean density. For Eq. \eqref{eq: neutral_fraction} we assumed that hydrogen and helium is highly ionized and the photoheating rate dominates ionizing equilibrium. See Appendix \ref{app: neutral_hydrogen_density} for a detailed derivation of this property. At redshift $z=2.5$ it is found to be (see Appendix \ref{app: neutral_hydrogen_density}):
\begin{align}
\hat{n}_\mathrm{HI}(z=2.5) = \frac{1.871 \cdot 10^{-14}\,\mathrm{m}^{-3}\mathrm{s}^{-1}} {T_0(z=2.5)^{0.7} \Gamma(z=2.5)}, \label{eq: mean_dens} 
\end{align}
where $\Gamma$ denotes the photo-ionization rate and $T_0$ is expressed in units of Kelvin. Often the shifted photoionization rate $\Gamma_{-12} = \Gamma / (10^{-12}\,\mathrm{s}^{-1})$ is used instead of $\Gamma$ in studies of the IGM.

\section{Method} \label{sec: method}

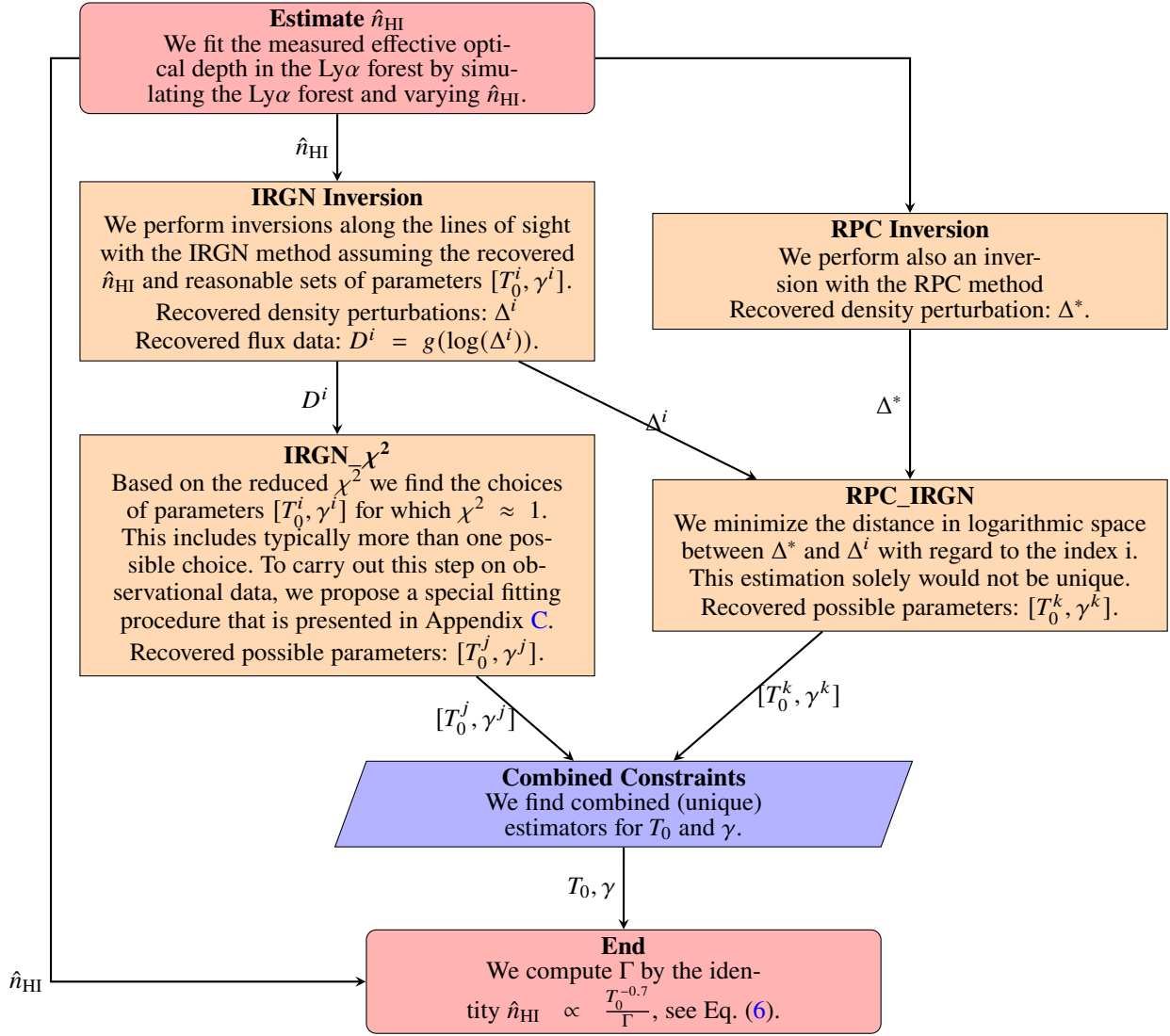
\begin{figure*}
\begin{tikzpicture}[node distance=3cm]
\node (nHI) [startstop] {\large{\textbf{Estimate} $\hat{n}_\mathrm{HI}$}\\ We fit the measured effective optical depth in the Ly$\alpha$ forest by simulating the Ly$\alpha$ forest and varying $\hat{n}_\mathrm{HI}$.};

\node (irgn) [process, below of = nHI] {\large{\textbf{IRGN Inversion}} \\ We perform inversions along the lines of sight with the IRGN method assuming the recovered $\hat{n}_\mathrm{HI}$ and reasonable sets of parameters $[T_0^i, \gamma^i]$.\\ Recovered density perturbations: $\Delta^i$\\ Recovered flux data: $D^i=g(\log(\Delta^i))$.};

\node (rpc) [process, right of = irgn, xshift = 5cm] {\large{\textbf{RPC Inversion}} \\ We perform also an inversion with the RPC method \\ Recovered density perturbation: $\Delta^*$.};

\node (irgn_chi) [process, below of = irgn, yshift=-1cm] {\large{\textbf{IRGN\_}$\bm{\chi^2}$} \\ Based on the reduced $\chi^2$ we find the choices of parameters $[T_0^i, \gamma^i]$ for which $\chi^2 \approx 1$. This includes typically more than one possible choice. To carry out this step on observational data, we propose a special fitting procedure that is presented in Appendix \ref{app: fitting_procedure}.\\Recovered possible parameters: $[T_0^j, \gamma^j]$.};

\node (rpc_irgn) [process, right of = irgn_chi, xshift = 5cm] {\large{\textbf{RPC\_IRGN}} \\ We minimize the distance in logarithmic space between $\Delta^*$ and $\Delta^i$ with regard to the index i. This estimation solely would not be unique. \\ Recovered possible parameters: $[T_0^k, \gamma^k]$.};

\node (combined) [io, below of = irgn_chi, yshift=-0.5cm, xshift=4cm] {\large{\textbf{Combined Constraints}} \\ We find combined (unique) estimators for $T_0$ and $\gamma$.};

\node (end) [startstop, below of = combined, yshift=0.5cm] {\large{\textbf{End}} \\ We compute $\Gamma$ by the identity $\hat{n}_\mathrm{HI} \propto \frac{T_0^{-0.7}}{\Gamma}$, see Eq. \eqref{eq: mean_dens}.};

\draw [arrow] (nHI) -- node [anchor=east] {\large{$\hat{n}_\mathrm{HI}$}} (irgn);

\draw [arrow] (nHI) -| node [anchor=south] {} (rpc);

\draw [arrow] (irgn) -- node [anchor=east] {\large{$D^i$}} (irgn_chi);

\draw [arrow] (irgn) -- node [anchor=west] {\large{$\Delta^i$}} (rpc_irgn);

\draw [arrow] (rpc) -- node [anchor=east] {\large{$\Delta^*$}} (rpc_irgn);

\draw [arrow] (irgn_chi) -- node [anchor=east] {\large{$[T_0^j, \gamma^j]$}} (combined);

\draw [arrow] (rpc_irgn) -- node [anchor=west] {\large{$[T_0^k, \gamma^k]$}} (combined);

\draw [arrow] (nHI) -- +(-4, 0) |- node [anchor=east] {\large{$\hat{n}_\mathrm{HI}$}} (end);

\draw [arrow] (combined) -- node [anchor=east] {\large{$T_0, \gamma$}} (end);
\end{tikzpicture}
\caption{Flowchart of estimation method. The method will be discussed throughout Sec. \ref{sec: method}.}
\label{fig: estimation_procedure}
\end{figure*}

It is a common approach in the analysis of the temperature of the IGM to measure the thermal line-width $b_\mathrm{T}$ in the Ly$\alpha$ forest and to deduce $T_0$ from that. However, the absorption feature is also broadened by the intrinsic width of the underlying density field $\Delta$. \cite{Garzilli2020} for example modeled the baryonic density perturbation $\Delta$ as sum of Gaussian peaks with varying center and standard deviation $b_\lambda$ (neglecting peculiar velocities). According to \cite{Garzilli2020} the width of the absorption lines $b$ in the Ly$\alpha$ forest is:
\begin{align}
    b^2=b_\lambda^2+b_\mathrm{T}^2 \label{eq: total_broadening}.
\end{align}
Our new method is motivated by Eq. \eqref{eq: total_broadening}. It is a difficult task to estimate the thermal broadening from the total broadening of the line. However, for Ly$\alpha$ forest tomography, i.e. the problem of recovering the density field in the Ly$\alpha$ forest at high spectral resolution, this problem is tackled by powerful inversion algorithms. It is not possible to study the thermal broadening parameter without constraining the underlying matter density field. Thus, both problems (i.e. inversion of the Ly$\alpha$ forest and constraining the EOS of the IGM) should be solved simultaneously. Such an approach was used by \cite{Rollinde2001}. They used the iterative Gauss-Newton method (IRGN) proposed by \cite{Pichon2001} to invert the Ly$\alpha$ forest. The IRGN algorithm tries to recover the observed Ly$\alpha$ flux by modelling the absorption explicitly by Eq. \eqref{eq: tau}. The basic idea now is that the quality of the inversion with the IRGN method depends strongly on which set of parameters $\hat{n}_\mathrm{HI}$, $T_0$ and $\gamma$ has been assumed prior to the inversion procedure, i.e. whether the true temperature-density relation has been used to model the Ly$\alpha$ forest. \citet{Rollinde2001} chose the set of parameters which match the observations best based on a reduced $\chi^2$ criterion for the recovered flux (i.e. the flux calculated from the recovered density). This method of estimating the temperature-density relation by comparing the recovered flux (where the density has been recovered by the IRGN inversion algorithm) and the observed flux will be called IRGN\_$\chi^2$ throughout this paper. 

Meanwhile, there are inversion schemes available that depend on less strong priors \citep{Gallerani2011, Mueller2020}, namely the regularized probability conservation (RPC) method. This method recovers the neutral hydrogen density in the Ly$\alpha$ forest by a statistical approach which is not affected by any choice for $\gamma$ and $T_0$. It has been demonstrated that the RPC inversion provides a suitable estimator for the true density \citep{Mueller2020}. Thus, it is also possible to compare the recovered density with the IRGN method for different models of the EOS with a density that approximates the true density reasonably well. Choosing the set of parameters that model the recovered density profile best will be called the RPC\_IRGN method throughout this paper. Compared to the IRGN\_$\chi^2$ the comparison between observational data and recovered data is not performed in flux space, but in density space.

The plan for the rest of the Section is as follows: We will discuss every step of the estimation (every cell in Fig. \ref{fig: estimation_procedure}) in the following subsections. Our final estimation procedure is shown as a flowchart in Fig. \ref{fig: estimation_procedure}.

The two inversion schemes IRGN and RPC were intensively studied and tested throughout the former publication \citet{Mueller2020}. We summarize some of our main findings in Appendix \ref{app: inversion_schemes} and will only provide short discussions of the inversion schemes in the following subsections.

\subsection{Estimate $\hat{n}_\mathrm{HI}$}
In this paper we examine the three independent parameters $[\Gamma, T_0,\gamma]$. However, it is beneficial for this analysis to estimate the set of parameters $[\hat{n}_\mathrm{HI}(T_0, \Gamma), T_0, \gamma]$. We estimate in a first step $\hat{n}_\mathrm{HI}$ (upper red box in Fig. \ref{fig: estimation_procedure}). In a second step (blue box in Fig. \ref{fig: estimation_procedure}) we estimate $T_0$ and $\gamma$ while keeping $\hat{n}_\mathrm{HI}$ constant, i.e. in this second step $T_0$ only affects the thermal broadening, see Eq. \eqref{eq: eos}. Finally we compute $\Gamma$ from our estimates for $\hat{n}_\mathrm{HI}$ and $T_0$ by Eq. \eqref{eq: mean_dens} in order to find estimates for the physical set of parameters $[\Gamma, T_0, \gamma]$.

The estimation of $\hat{n}_\mathrm{HI}$ is a long performed task in the study of the Ly$\alpha$ forest. $\hat{n}_\mathrm{HI}$ appears as multiplicative factor for the optical depth and thus determines the scale of the observed optical depth. This becomes most obvious within the fluctuating Gunn-Peterson approximation. The fluctuating Gunn-Peterson approximation is a large scale approximation of Eq. \eqref{eq: tau}, i.e. the Voigt-profile is approximated by a Dirac-delta function:
\begin{align}
    \tau \approx \frac{\sigma_0 c \hat{n}_\mathrm{HI}}{H} \Delta^\alpha, \label{eq: gunn_peterson}
\end{align}
where $H$ denotes the Hubble constant. One could estimate $\hat{n}_\mathrm{HI}$ by Eq. \eqref{eq: gunn_peterson} and by the observed effective optical depth:
\begin{align}
\tau_\mathrm{eff}=-\log ( \langle F \rangle),
\end{align}
where $\langle F \rangle$ denotes the mean normalized flux. The effective optical depth was measured by e.g. \cite{Kirkman2005, Faucher2008, Becker2013} and used for estimation of the neutral hydrogen density at mean density perturbation. It is clear that clustering in the IGM and peculiar velocities (which are not reflected in the fluctuating Gunn-Peterson approximation) alter the observed effective optical depth. We therefore use the complete forward model in Eq. \eqref{eq: tau} for the estimation of $\hat{n}_\mathrm{HI}$. In particular we choose some reasonable prior choice for $T_0$ and $\gamma$, we simulate the baryonic density perturbation in a simulation box and compute the optical depth by Eq. \eqref{eq: tau} while varying $\hat{n}_\mathrm{HI}$. We take the value that fixes the observed $\tau_\mathrm{eff}$ best. Our simulation is outlined in Sec. \ref{sec: synthetic_data}. While a prior choice for $T_0$ and $\gamma$ is necessary, to compute the optical depth from our simulation, these choices have a small effect on the estimated $\hat{n}_\mathrm{HI}$ which will be accounted for in Sec. \ref{sec: mean_density}. 

As a second step we estimate $T_0$ and $\gamma$ while keeping $\hat{n}_\mathrm{HI}$ constant, i.e. we fit the high spectral resolution profile of the absorption lines in the Ly$\alpha$ forest. There are several methods available to perform this task, above all explicit Voigt-profile fitting. Here we sketch a new method for performing such estimates. We propose two alternative fitting procedures: We compare the IRGN inversion results to the observed data in flux space (we call this approach IRGN\_$\chi^2$) and we compare the IRGN and RPC inversion in density space (we call this method RPC\_IRGN). Both are needed to find statistically significant estimates for the EOS of the IGM.

\subsection{IRGN Inversion} \label{sec: irgn}
We outline the approach of \cite{Rollinde2001} within this section. This method is based on an inversion algorithm for recovering the logarithm of the overdensity $\bm{\mathit{A}}=\log (\bm{\mathit{\Delta}})$ (where the bold symbols denote vectors, i.e. the density perturbation and its logarithm discretized on the observed redshift bins). \cite{Rollinde2001} used a larger vector of observables which also contained more parameters, but we are only interested in the logarithmic density perturbation here. Let us define the forward operator $g$ as the operator which maps $\bm{\mathit{A}}$ to the normalized flux $\bm{\mathit{D}}$, i.e. $g(\bm{\mathit{A}})=\bm{\mathit{D}}$. Hence, g is specified by Eq. \eqref{eq: normalized_flux}, Eq. \eqref{eq: tau}, Eq. \eqref{eq: broadening} and Eq. \eqref{eq: eos}. \cite{Pichon2001} demonstrated that given particular data $\bm{\mathit{D}}$, the posterior of the vector of parameters $\bm{\mathit{A}}$ is (assuming a Gaussian likelihood and a Gaussian prior):
\begin{align} \nonumber
    &p(\bm{\mathit{A}}|\bm{\mathit{D}}) \propto \exp \left( -\frac{1}{2} [\bm{\mathit{D}}-g(\bm{\mathit{A}})]^T \bm{\mathsf{C}}_d^{-1} [\bm{\mathit{D}}-g(\bm{\mathit{A}}) ] \right.\\
    &\hspace{3cm}\left.-\frac{1}{2} [\bm{\mathit{A}}-\bm{\mathit{A}}_0]^T \bm{\mathsf{C}}_0^{-1} [\bm{\mathit{A}}-\bm{\mathit{A}}_0] \right), \label{eq: posteriori}
\end{align}
where $\bm{\mathsf{C}}_0$ and $\bm{\mathsf{C}}_d$ denote the covariance matrix of the initial guess $\bm{\mathit{A}}_0$ and noise. Eq. \eqref{eq: posteriori} is only valid if $\bm{\mathit{A}}$ is Gaussian distributed. However, this is a reasonable assumption since the density perturbation $\bm{\mathit{\Delta}}$ can be approximated by a lognormal distribution \citep{Coles1991, Choudhury2001, Choudhury2005, Gallerani2006}, see also Appendix \ref{app: just_sim}. We introduce a linearization $g(\bm{\mathit{A}}) \approx g(\bm{\mathit{A}}_0)+\bm{\mathsf{G}} \left( \bm{\mathit{A}}-\bm{\mathit{A}}_0 \right)$. Here $\bm{\mathsf{G}}$ denotes the matrix of functional derivatives of $g$. Then the estimator $\langle \bm{\mathit{A}} \rangle$ which maximizes the posterior of the linearized problem satisfies:
\begin{align} \nonumber
\langle \bm{\mathit{A}} \rangle = \bm{\mathit{A}}_0 +&\bm{\mathsf{C}}_0
\bm{\mathsf{G}}^T (\bm{\mathsf{C}}_d + \bm{\mathsf{G}} \bm{\mathsf{C}}_0 \bm{\mathsf{G}}^T)^{-1} \\
& \cdot [\bm{\mathit{D}} +\bm{\mathsf{G}} (\langle \bm{\mathit{A}} \rangle -\bm{\mathit{A}}_0) -g(\langle \bm{\mathit{A}} \rangle)], \label{eq: pichon}
\end{align}
The implicit Eq. \eqref{eq: pichon} is solved with a fixed point iteration, i.e. we apply the right hand side of Eq. \eqref{eq: pichon} iteratively until convergence is achieved, i.e. until the residuum becomes noise-like. For more details on the derivation and application of Eq. \eqref{eq: pichon} we refer to \cite{Pichon2001} and \cite{Mueller2020}. According to our nomenclature in \cite{Mueller2020} we call this method "iterative Gauss-Newton method" (IRGN).

The forward operator $g$ (and thus also the derivative $\bm{\mathsf{G}}$) appearing in Eq. \eqref{eq: pichon} depends strongly on the choice of the constants $T_0$ and $\gamma$ in Eq. \eqref{eq: eos}. Hence $\gamma$ and $T_0$ affect the recovered logarithmic density perturbation $\langle \bm{\mathit{A}} \rangle$ and the recovered normalized flux $g(\langle \bm{\mathit{A}} \rangle)$.

We invert the Ly$\alpha$ forest data at a high spectral resolution with the IRGN method by assuming $[\gamma^i, T_0^i]$ on a uniform grid of values which cover the whole range of reasonable parameters, e.g. $T_0 \in [5000, 30000]\,\text{K}$ and $\gamma \in [1.2, 2]$. The selection of the exact values tested in this sample plays only a minor role as long the whole range is sufficiently covered. We store the recovered density perturbations $\Delta^i$ and the recovered flux data $D^i$ for later usage.

\subsection{RPC Inversion} \label{sec: rpc_inversion}
We invert the same spectra with the RPC method and recover the density profile $\Delta^*$.

In this section we outline the recent development in Ly$\alpha$ forest tomography presented in \cite{Mueller2020} based on the approach by \cite{Gallerani2011} and sketch a novel way of using this approach for constraining the EOS of the IGM. If thermal broadening would be absent, Eq. \eqref{eq: tau} and Eq. \eqref{eq: neutral_fraction} would describe a one-to-one correspondence between the baryonic density perturbation and the observed optical depth. In the presence of thermal broadening this one-to-one correspondence is not satisfied since $\Delta$ also appears in the standard deviation of the Gaussian. Nevertheless, \cite{Gallerani2011} proposed to also assume this one-to-one relation in this case. Let us denote the probability density function of the observed flux by $P_\mathrm{F}$ (which can be measured from the sample of observed lines of sight) and the probability density function of $\Delta$ by $P_\Delta$ (which has to be assumed prior to the inversion procedure). Let $F_\mathrm{max}$ be the maximal flux that can be distinguised from full emission. Then this flux can be identified by an overdensity $\Delta_b$ corresponding to the bright limit:
\begin{align}
    \int_{F_\mathrm{max}}^1 P_\mathrm{F} dF = \int_0^{\Delta_\mathrm{b}} P_\Delta d\Delta.
\end{align}
According to \cite{Gallerani2011} the flux $F_*$ in every bin is now identified with a density perturbation $\Delta_*$ in that bin by the equation:
\begin{align}
    \int_{F^*}^{F_\mathrm{max}} P_\mathrm{F} dF = \int_{\Delta_\mathrm{b}}^{\Delta^*} P_\Delta d\Delta. \label{eq: gallerani}
\end{align}
Similar equations are derived for the minimal flux that can be distinguished from full absorption \citep{Gallerani2011}. We found that for the high quality data that are used throughout this paper $F_\mathrm{max} = 0.99$ is a good choice. The assumption that the overdensity is lognormal distributed is a good model for $P_\Delta$ which was successfully used by \cite{Gallerani2011} and the consecutive work \cite{Kitaura2012}. The lognormal model is used in a large number of publications as a model for the density perturbations in the quasi-linear regime. We discuss the reliability of the lognormal model more detailed in Appendix \ref{app: just_sim}. We demonstrated in \cite{Mueller2020} that the reconstruction results are improved by reformulating Eq. \eqref{eq: gallerani} as an optimization problem and adding a penalty term, i.e. by solving the problem:
\begin{align} \nonumber
    &\Delta^* \in \text{argmin}_\Delta \\
    &\;\;\;\;\;\;\;\; \left\{ \Psi(\Delta) =  \frac{1}{2}\norm{\int_{F^{*}}^{F_\mathrm{max}} P_{F} dF-\Phi (\Delta)}_{L^2}^2 +\frac{\alpha}{2} \norm{\frac{\partial \Delta}{\partial z}}_{L^2}^2 \right\}   \label{eq: optim}
\end{align}
instead of Eq. \eqref{eq: gallerani}. Here, $\Phi$ denotes the operator $\Phi: \Delta^* \mapsto \int_{\Delta_\mathrm{b}}^{\Delta^*} P_\Delta d\Delta$ where $\Delta^*$ is the vector of density perturbations and $\Phi(\Delta^*)$ is evaluated pointwise. Eq. \eqref{eq: optim} is solved by a gradient descent algorithm with small numerical computation time. This method is called the "regularized probability conservation approach" (RPC). For more details on the implementation and the mathematical details of the RPC method we refer to \cite{Mueller2020}. In short, the first term of $\Psi(\Delta)$ is the reformulation of Eq. \eqref{eq: gallerani} as an optimization problem, i.e. only minimizing this term would be equivalent to solving Eq. \eqref{eq: gallerani}. The second term is the norm of the gradient of the overdensity multiplied with a regularization parameter $\alpha$. This penalty term introduces suppression of small scale fluctuations of the recovered density profile due to the propagation of noise. $\alpha$ has been tuned manually on synthetic data to return proper inversion results. By default, $\alpha$ should be chosen such that the first term and the second term in Eq. \eqref{eq: optim} are of similar order. One particular advantage of the RPC method is that is does not depend explicitly on a chosen set of parameters $[\Gamma, T_0, \gamma]$, the only crucial assumption is the probability density function $P_\Delta$ which could be identified from hydrodynamic simulations \citep[c.f. the discussion in][]{Gallerani2011, Kitaura2012}.

For later usage we store the recovered density perturbation $\Delta^*$.

\subsection{IRGN\_$\chi^2$} \label{sec: irgn_chi2_theory}
At next level, see Fig. \ref{fig: estimation_procedure}, we have to examine the inversion results of the former steps. In this subsection we discuss the IRGN\_$\chi^2$ method which uses the inversion results from the IRGN inversion method. We first compute the recovered flux $\bm{\mathit{D}}^i = g(\langle \log( \bm{\mathit{\Delta}}^i )  \rangle^i)$ for every set of parameters from our inversion results $\Delta^i$ with the IRGN method. \cite{Rollinde2001} proposed to examine the recovered flux based on the reduced $\chi^2$:
\begin{align}
    \chi^2=\frac{1}{N_\mathrm{pix}} \sum_{j=1}^{N_\mathrm{pix}} \frac{1}{\sigma_j^2} \left( \bm{\mathit{D}}^i_j-\bm{\mathit{D}}^\mathrm{obs}_j \right)^2,
\end{align}
where $N_\mathrm{pix}$ denotes the number of pixels in each line of sight, $\sigma_j$ the standard deviation of noise in pixel $j$ and $\bm{\mathit{D}}^\mathrm{obs}$ the observed, noisy data. It is well known that $\chi^2>1$ means that the structure in the flux data is badly resolved, while $\chi^2<1$ characterizes overfitting of noise. According to \cite{Rollinde2001} the exact values $[\gamma, T_0]$ are identified with $\chi^2 \approx 1$. If we overestimate thermal broadening (e.g. by overestimating the temperature $T_0$), this promotes smoothing in data space. Therefore, the small scale structures in the normalized flux are inaccurately estimated, such that $\chi^2$ becomes bigger than one. Similarly underestimating thermal broadening leads to $\chi^2<1$.

When applying our estimation to real observational spectra additional sources of uncertainties occur which could bias the estimated parameters. In fact, a high $\chi^2$ could be the consequence of poorly fitting only a part of the spectrum or underestimating the error (in particular for large overdensities). Moreover, the iterative inversion procedure could have stopped too early. All these uncertainties shift the computed reduced $\chi^2$ towards larger (smaller) values. We fight these errors by masking out the wavelengths at which only bad fits to the observed data are available, i.e. we remove every wavelength where the distance between the recovered flux and the observed flux exceeds roughly five times the noise-level. It is more likely that the inversion process failed for these wavelengths (e.g. optimization method did not converge) rather than an absorption feature has been detected. Moreover, instead of just taking the set of parameters for which $\chi^2$ drops to one, we fit the curve $\chi^2(T_0)$ for every $\gamma^i$. Our exact fitting procedure is outlined in Appendix \ref{app: fitting_procedure}. In a nutshell, the reduced $\chi^2$ for smaller temperatures than the true one typically converges to a constant lower limit. The reduced $\chi^2$ for temperatures larger than the true temperatures decreases linearly. The intersection between this linear fit and the lower limit for very small temperatures approximates the exact temperature.

\subsection{RPC\_IRGN}
We discuss in this subsection the RPC\_IRGN method which uses the inversion results with the IRGN method $\Delta^i$ and with the RPC method $\Delta^*$. We pick those values $[\gamma^i, T_0^i]$ for which the correspondence between $\Delta^i$ and $\Delta^*$ is maximal. This approach is driven by the assertion that $\Delta^*$ mimics reasonably well the true density profile which has been demonstrated in \cite{Mueller2020}. We examine the correspondence between $\Delta^i$ and $\Delta_*$ based on the $L^2$ distance of the logarithms of the density perturbations, i.e. we minimize:
\begin{align}
    d_\mathrm{log} = \sqrt{ \sum_j \left( \log(\Delta^i_j)-\log(\Delta^*_j) \right)^2},
\end{align}
where $j$ runs over all bins in the spectra. Taking the logarithm is needed to compute a meaningful distance. The reconstruction methods perform badly at large overdensities and small underdensities due to line saturation. The standard $L^2$-distance would be dominated by the reconstruction of large overdensities and the information regarding similarity would be lost. Therefore, we only use the pixels with moderate overdensities to perform the comparison between the reconstruction with the RPC method and with the IRGN method.

\subsection{Combined Constraints} \label{sec: combined_constraints}
As demonstrated by \cite{Pichon2001} and \cite{Rollinde2001} the IRGN\_$\chi^2$ approach suffers from a degeneracy between $T_0$ and $\gamma$. $\chi^2=1$ defines a line in the 2D parameter space of $T_0$ and $\gamma$. As will be shown in Sec. \ref{sec: synthetic} the RPC\_IRGN approach is to first order only sensitive to $\gamma$. Consequently, combining both estimates can remove the degeneracy between $\gamma$ and $T_0$. The set of parameters $\gamma$ and $T_0$ which matches the constraints from both methods (IRGN\_$\chi^2$ and RPC\_IRGN) is a unique estimator for the Equation of State parameters (blue box in Fig. \ref{fig: estimation_procedure}). We will discuss this in more details in Sec. \ref{sec: synthetic}.

Now we are left with estimates for $\gamma$, $T_0$ (blue box in Fig. \ref{fig: estimation_procedure}) and $\hat{n}_\mathrm{HI}$ from the start (upper red box in Fig. \ref{fig: estimation_procedure}). These measurements are recombined by Eq. \eqref{eq: mean_dens} to compute an estimate for the photoionization rate, $\Gamma_{-12}$.

\subsection{Limitations}
The inversion schemes (IRGN, RPC) try to decompose the thermal profile from the profile of the underlying density field. Hence, they are only applicable for echelle-resolution spectra. More precisely, this decomposition only carries strong information about the thermal history of the IGM for high resolution reconstructions (when thermal broadening and instrumental broadening are of similar order). We examined the resilience of these algorithms against observational noise in \cite{Mueller2020}. Although both algorithms show robustness against noise, for the purpose of estimating $T_0$ and $\gamma$ only high quality data (high signal to noise ratio) should be used. For larger noise contributions the algorithms typically start fitting small scale noise instead of properly fitting the thermal profile. We test the impact of noise on the estimation in Sec. \ref{sec: noisy_spectra}.

\section{Numerical Tests on Synthetic Data} \label{sec: synthetic}

\subsection{Synthetic Data} \label{sec: synthetic_data}
We create synthetic data with the semi-analytic approach described by \cite{Choudhury2001} and \cite{Gallerani2006} which is based on the lognormal model for the density \citep{Coles1991}. The lognormal model is well motivated and used in a large number of publications. We discuss several aspects of the lognormal model in Appendix \ref{app: just_sim}. We demonstrate that the lognormal model can statistically reproduce the structures of real spectra. Moreover, we demonstrate in Appendix \ref{app: just_sim} that the lognormal model is a reasonable analytic description of the true density distribution, but cannot reproduce the skewness in the expected probability density distributions. The lognormal model, however, is inadequate for describing highly non-linear overdensities, or overdensities at very small scale \citep{Choudhury2005}, although the latter might hold for dark matter primarily and lesser for the ordinary matter distribution \citep{Gallerani2006}. As both regimes are not accessible to Ly$\alpha$ forest inversion anyhow (due to line saturation and line broadening) the lognormal model is a simple and sufficient analytic description of the overdensity ditribution.

An implementation of our simulation is available within the \small{REGLYMAN} \citep{reglyman} toolbox which makes strong use of the \small{NBODYKIT} toolbox too. In a nutshell, we create in a first step a Gaussian random field (i.e. the linear overdensity field) with a covariance which is specified by the matter autocorrelation function. More precisely, we start with a Gaussian white noise model and multiply it with the matter power spectrum. We take the inverse Fourier transform of this quantity which is a Gaussian random field with the desired autocorrelation. We project in a second step the linear density field to the quasilinear regime by taking the exponential of the linear field and normalizing. Since the linear density perturbation was Gaussian distributed, the quasilinear density perturbation is now lognormal distributed, i.e. it is the exponential of a Gaussian random field. We compute the peculiar velocities by the Zel'dovich approximation \citep{Zeldovich1970, White2014} from the linear density field: Firstly we calculate the first order Lagrangian displacement field $\bf{\Psi}$ from the linear density field $\delta_\mathrm{L}$ by:
\begin{align}
    \mathcal{F}\bf{\Psi}(\bf{k}) = -i\frac{k_z}{|\bf{k}|^2} \mathcal{F}\delta_\mathrm{L}(\bf{k}),
\end{align}
where $\mathcal{F}$ denotes the Fourier Transform. We then estimate peculiar velocities by:
\begin{align}
    v_\mathrm{pec} = a H f \Psi,
\end{align}
where $a$ is the cosmic scaling factor, $H$ the Hubble constant and $f$ the linear growth rate. Lastly, we compute the Ly$\alpha$ forest from the density perturbation according to Eq. \eqref{eq: tau}.

Our simulation is very fast and allows for the computation of small scales (binning in the forest) and large scales (box width) in parallel due to its semi-analytic nature. The large transverse box size allows us to draw a large sample of lines of sight from one box with large transverse separation length, such that their small scale behaviour in longitudinal direction, i.e. along the lines of sight could be approximated to be independent. Moreover, the speed of the simulation enables us to test varying physical parameters (such as Hubble constant, matter power spectrum or Jeans length) easily.

Among the chosen cosmology we additionally have to specify a Jeans-length for this procedure that describes biasing between dark matter and ordinary matter. The baryonic matter power spectrum $P_\mathrm{B}$ is related to the dark matter power spectrum $P_\mathrm{DM}$ by the relation \citep{Fang1993}:
\begin{align}
    P_\mathrm{B} (k, z) = \frac{P_\mathrm{DM}}{(1+x_b^2k^2)^2},
\end{align}
where $x_b$ denotes the Jeans length \citep[compare the prescriptions in][]{Choudhury2001, Gallerani2006}. We discuss the effect of the Jeans scale on the prior density distribution along with the lognormal model in Appendix \ref{app: just_sim}. In general, the amount of structure depression will also depend on the thermal history of the IGM. Hence, the uncertainties in $x_\mathrm{b}$ need to be considered when applying our estimation to observational data. We adopt the value that was used by \citet{Zaroubi2006} and which is compatible to the findings in \citet{Choudhury2001}. The Jeans length affects the prior density distribution of the overdensity field $P_\Delta$ for the RPC algorithm, but it turns out to play only a minor role for the results with the IRGN algorithm.

We compute independent lines of sight at redshift $z=2.5$ and spectral resolution $R = \frac{\lambda}{\Delta \lambda} = 100000$. To mimic instrumental artifacts we add noise according to the noise model:
\begin{align}
    \sigma_F^2 = \frac{F^2}{SNR^2}+\sigma_0^2,
\end{align}
with signal-to-noise ratio $SNR$ and a small noise contribution $\sigma_0$ which dominates at very small fluxes $F$. Finally we rebin our spectra to a spectral resolution $R=50000$, which is compatible to high resolution spectra taken with the UVES \citep{Dekker2000} or the HIRES \citep{Vogt1994} spectrometers.

\subsection{Estimation Methods} \label{sec: estimation}
For our synthetic data we assume a temperature of $T_0 = 10^4\,\text{K}$, $\gamma = 1.4$ and $\hat{n} = 3.5 \cdot 10^{-5}\,\mathrm{m}^{-3}$. We compute 100 nearly independent (i.e. with a transverse separation length of at least $20\,h^{-1}\,\mathrm{Mpc}$ between each pair of lines) lines of sight of $10\,h^{-1}\,\mathrm{Mpc}$ comoving length at redshift $z=2.5$. Moreover we use a high signal to noise level $SNR=50$ and a small noise contribution $\sigma_0 = 0.01$ which was found in previous investigations \citet{Mueller2020} to return proper inversion results. Moreover, box size was found to have only a minor effect on the synthetic spectra as long as the separation length between neighboring lines of sight remains at the order of at least several $h^{-1}\mathrm{Mpc}$. For this subsection we ignore peculiar velocities in the computation of the lines of sights. We will quantify in Sec. \ref{sec: pec_vel} how much this affects our estimation. 

We first recover $\hat{n}_\mathrm{HI}$ from our set of synthetic data. We simulate the neutral hydrogen density fluctuation in a second box with the same properties, but a different random seed. We compute the effective optical depth while shifting the parameter $\hat{n}_\mathrm{HI}$. Unsurprisingly we recover the artificial neutral hydrogen density parameter as we compare the effective optical depth from our synthetic data with the effective optical depths computed from the same sort of simulation just with varying random seed. For the simulation we had to make prior choices for $T_0$ and $\gamma$. While these parameters mainly have an affect on the width of absorption lines and the smoothness of the flux profile, they also affect the observed mean optical depth. We will therefore add the uncertainty in the prior choice of $\gamma$ and $T_0$ to the error budget when applying our estimation to observational data (Sec. \ref{sec: mean_density}).

\begin{figure}
    \centering
    \includegraphics[width=0.5\textwidth]{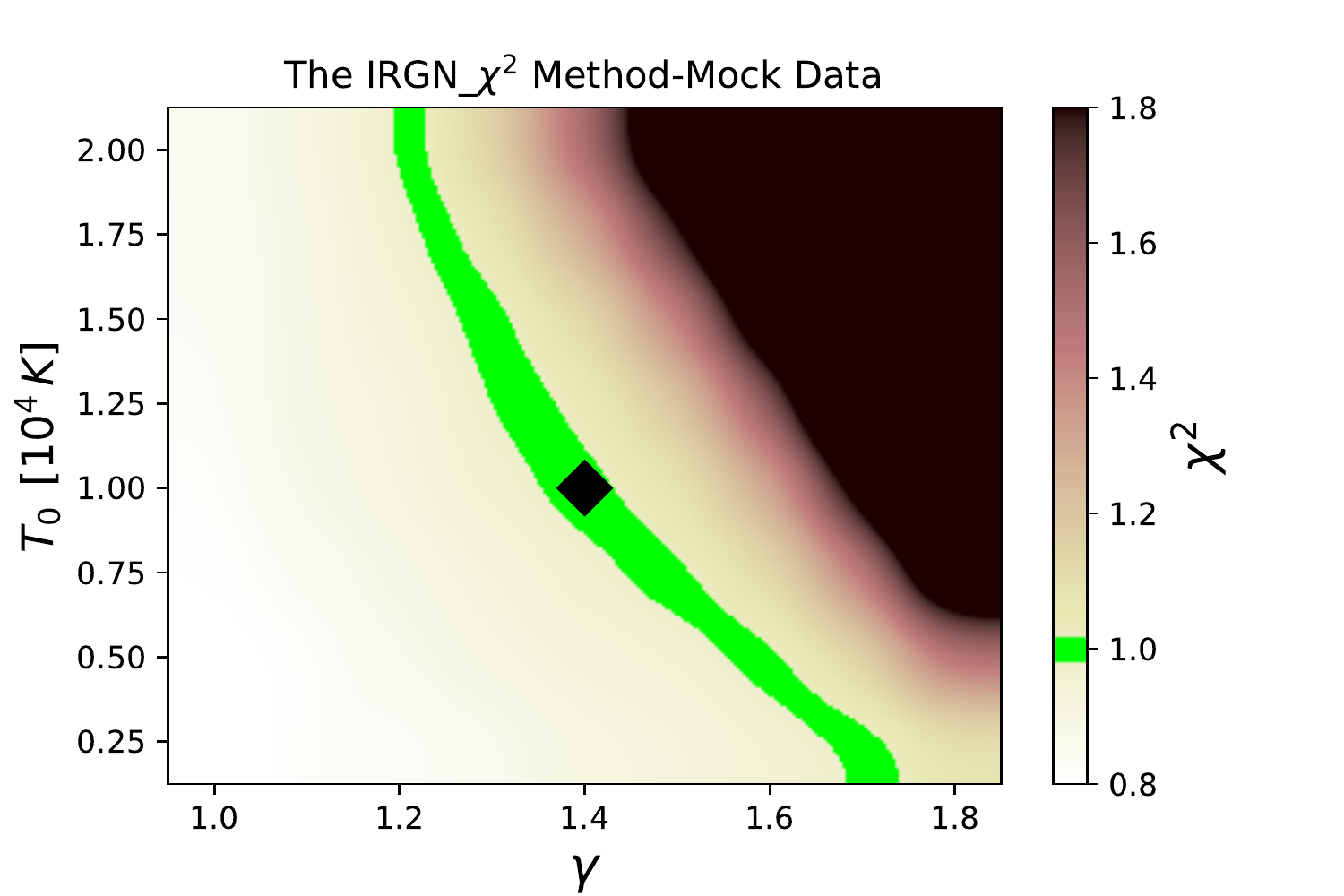}
    \caption{Reduced $\chi^2$ as function of $\gamma$ and $T_0$. The exact set of parameters (which was used for the creation of synthetic data) is indicated with the black diamond in the center. The $\chi^2 \approx 1$, i.e. $\chi^2 \in [0.98, 1.02]$, line is green overplotted.}
    \label{fig: irgn_eos}
\end{figure}

\begin{figure}
    \centering
    \includegraphics[width=0.5\textwidth]{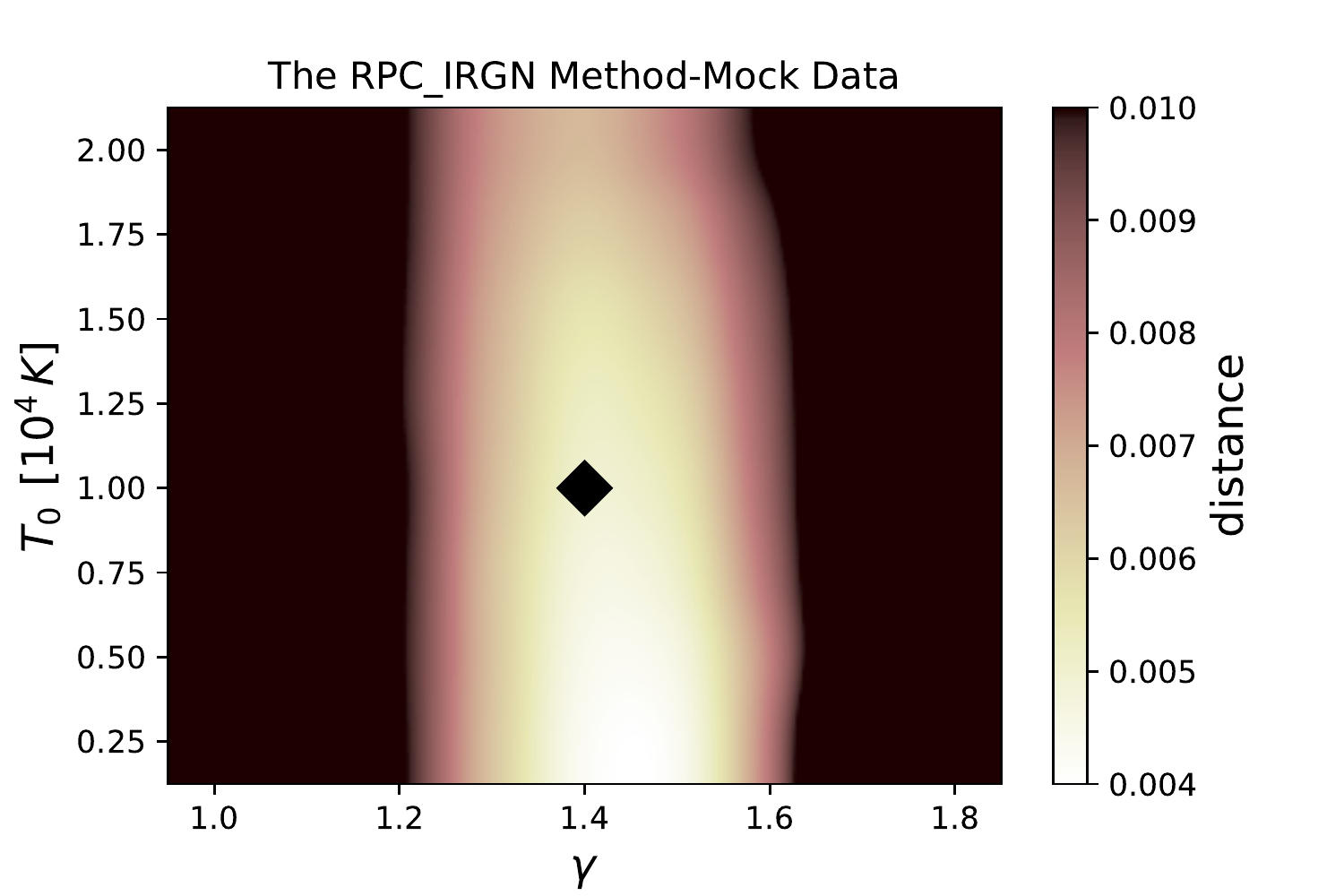}
    \caption{The logarithmic distance between the density profile recovered with the IRGN method (assuming different parameters $\gamma$ and $T_0$) and the density profile recovered with the RPC method as function of $\gamma$ and $T_0$ for the whole sample of lines of sight. The true parameters (which was used for the creation of synthetic data) are indicated with the black diamond in the center of the graphic.}
    \label{fig: rpc_eos}
\end{figure}

\begin{figure}
    \centering
    \includegraphics[width=0.5\textwidth]{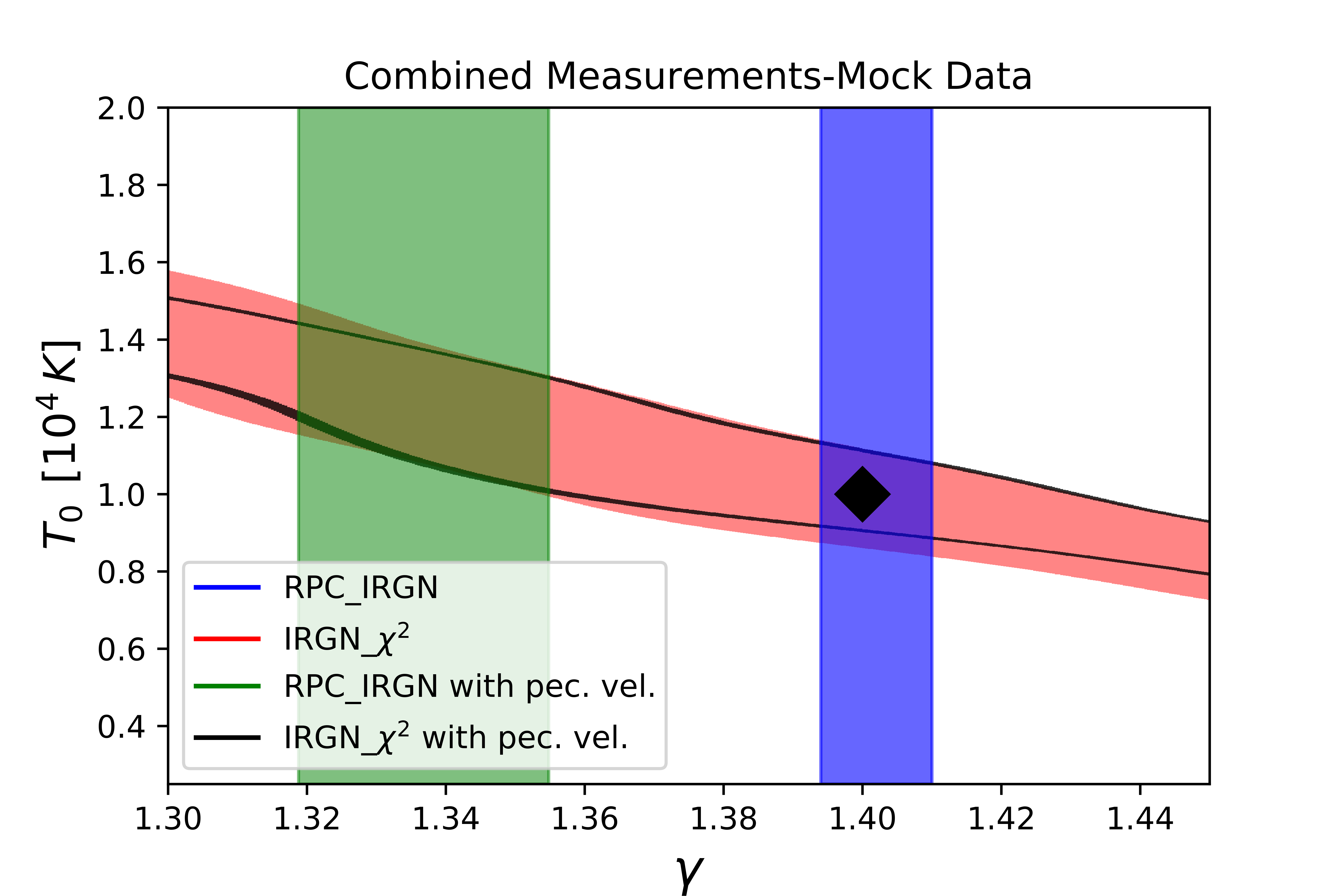}
    \caption{Joint estimate with the RPC\_IRGN (blue shaded) and the IRGN\_$\chi^2$ (red shaded) methods on a set of 100 synthetic lines of sight. The exact set of parameters for $\gamma$ and $T_0$ (black diamond) lies well inside the joint confidence interval. The black lines and the green shaded area show the results if additionally peculiar velocities are considered. The black line indicate the edges of the $\chi^2 \approx 1$ line and the green shaded region the RPC\_IRGN results.}
    \label{fig: eos}
\end{figure}

We apply our inversion steps to these synthetic data. For this study we assume that $\hat{n}_\mathrm{HI}$ has been estimated exactly, i.e. we use the value that was used for the creation of synthetic data. We perform the inversion with the IRGN method on a grid of values $\gamma^i \in [1.0, 1.1, 1.2, 1.3, 1.4, 1.5, 1.6, 1.7, 1.8]$ and $T_0^j \in [2500, 5000, 7500, 10000, 12500, 15000, 17500, 20000]\,\text{K}$. We compute the joint reduced $\chi^2$ and the joint absolute distance for all 100 lines of sight in our sample of synthetic lines of sight. Our joint results for the full sample are shown in Fig. \ref{fig: irgn_eos} and Fig. \ref{fig: rpc_eos}. We use a Gaussian interpolation on our grid of parameters in Fig. \ref{fig: irgn_eos} and Fig. \ref{fig: rpc_eos} (and will use this interpolation also on Fig. \ref{fig: eos} later). This interpolation is purely for plotting purposes and will not be needed for the analysis of observed data.

Similar to the results proposed in \cite{Rollinde2001} $\chi^2 \approx 1$ defines a line in the two dimensional parameter space (green line in Fig. \ref{fig: irgn_eos}). In the most likely region $1.3 \leq \gamma \leq 1.6$ this line is well approximated by a linear relation between $\gamma$ and $T_0$. The slope of this line is characteristic for the chosen redshift $z=2.5$. Future works should extend this study to other redshifts. If a global relation between $\gamma$ and $T_0$ could be established, the degrees of freedom in estimating the thermal history of the IGM would be reduced by one parameter.

In Fig. \ref{fig: rpc_eos} we show the absolute distance between the inversion with the RPC and the IRGN method. This distance only depends very weakly on $T_0$. This is reasonable. We substituted the strong dependency of $\tau$ on $T_0$ in $\hat{n}_\mathrm{HI}$, such that $T_0$ only appears as free parameter in the standard deviation of the Gaussian kernel in Eq. \eqref{eq: tau}. Hence, $T_0$ only plays a minor role for the overall scale of the estimated density field fluctuations amplitudes. On the other side $\gamma$ also affects the standard deviation of the Gaussian kernel (see Eq. \eqref{eq: eos}), but it appears outside of the kernel function too (see Eq. \eqref{eq: neutral_fraction}). All in all, our special choice for the set of parameters allows us to find estimates for the slope parameter $\gamma$ nearly independently from $T_0$, at least for temperatures up to $20000\,\mathrm{K}$ which are tested in Fig. \ref{fig: rpc_eos} and which are considered to be relevant for the Ly$\alpha$ forest at redshift $z=2.5$. Lastly it should be mentioned that the minimum in the distance visible in Fig. \ref{fig: rpc_eos} is not a sharp minimum. Hence, when applying to real observational data we expect a scatter in the estimation of $\gamma$ from different lines of sight.

The IRGN\_$\chi^2$ and the RPC\_IRGN method are complementary. Fig. \ref{fig: eos} demonstrates that our method is working on synthetic data as the true values for $T_0$ and $\gamma$ (black diamond) lies well inside the estimated regions. For the RPC\_IRGN method we fitted the minimum for every line of sight and took the mean of these estimates. Note that for this analysis of synthetic data we optimized the stopping rules and regularization parameters for the inversion procedure with the RPC method and IRGN procedure based on the known synthetic solution.

\subsection{Fitting Procedure} \label{sec: synthetic_fitting}
\begin{figure}
    \centering
    \includegraphics[width=0.5\textwidth]{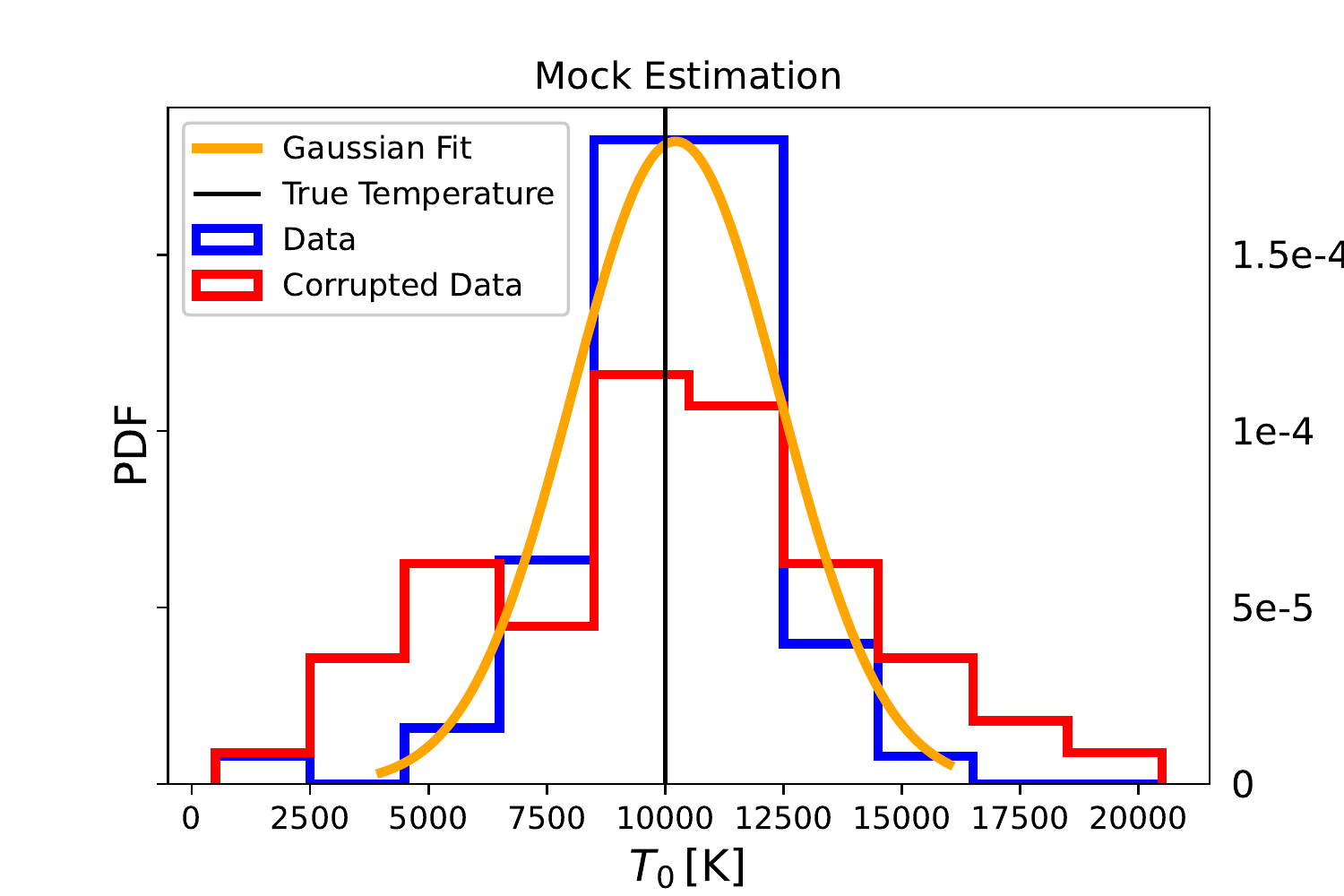}
    \caption{Histogram of recovered temperatures with our fitting procedure from synthetic data. The synthetic data were created with a temperature of $T_0 = 10000\,\text{K}$. The yellow curve is a Gaussian fit to the histogram: We find a mean $\mu = 10220\,\mathrm{K}$ and a standard deviation $\sigma = 2190\,\mathrm{K}$. The red histogram corresponds to results of our fitting procedure from corrupted data, i.e. the recovered fluxes were manually distorted in a small window. The recovery remains robust, although error bars increase.}
    \label{fig: temperature_histogram}
\end{figure}

We now verify the fitting procedure outlined in Appendix \ref{app: fitting_procedure} on our set of synthetic data. For this we fit every line of sight in our sample of 100 synthetic lines of sight individually. We assume the correct $\gamma = 1.4$ at redshift $z=2.5$ (i.e. the parameter that was used for the creation of synthetic data) and vary the temperature at mean density $T_0$. For every line of sight we perform the fitting procedure sketched in Fig. \ref{fig: fitting_procedure}. We show in Fig. \ref{fig: temperature_histogram} a histogram of all our estimates from our synthetic lines of sight. The distribution is well described by a Gaussian centered around $\approx 10000\,\text{K}$. In fact we compute for the average $\langle T_0 \rangle = (10200 \pm 300)\,\text{K}$. This indicates that when averaging over a suitable set of lines of sights our fitting routine applied to the IRGN\_$\chi^2$ step returns a proper estimator for the real temperature once $\gamma = 1.4$ stays fixed. 

We proposed in Sec. \ref{sec: irgn_chi2_theory} that our fitting procedure might be able to handle also bad fits and corrupted data to a reasonable level, see also Appendix \ref{app: fitting_procedure}. We now verify this assertion on our numerical data set. We show the results of the fitting procedure to corrupted data in Fig. \ref{fig: temperature_histogram} by the red histogram. We corrupted the recovered flux in a ten pixel wide window in every spectrum by a constant value of $0.2$. The distribution of estimated temperatures remains approximately Gaussian, but with a larger standard deviation compared to the estimation from the non-corrupted data set. While we try to mask these regions of failed reconstructions out in the application to real data, the estimation suggests that the fitting procedure might be able to handle remaining bad pixel fits to a reasonable level.

\subsection{Peculiar Velocities} \label{sec: pec_vel}
As described by Eq. \eqref{eq: tau} the redshift space coordinate of an absorption feature has two independent components, the Hubble redshift $v_\mathrm{H}$ and peculiar velocities $v_\mathrm{pec}$. It is difficult to estimate the contribution of peculiar velocities from the flux along single lines of sights. Thus, for the inversions we have to ignore peculiar velocities. We quantify in this section how much this assumption is affecting our estimation. For this we take a reduced sample of 25 lines of sights out of our simulation, calculate the peculiar velocities from the density field in the box and model the Ly$\alpha$ forest with respect to these peculiar velocities. Then we perform the inversions with the RPC method and with the IRGN method while still assuming for the inversion algorithms that no redshift space distortions are present. 

Peculiar velocities affect the spectrum in two different ways: They shift the absorption features in redshift space and they narrow/broaden the absorption lines. The latter effect is often ignored when fitting the absorption profile for the purpose of identifying the thermal history of the IGM. However, peculiar velocities are crucial for doing an inversion of the density field \citep{Nusser1999}. The lack of information regarding peculiar velocities inserts an additional bias in the estimation of the EOS from tomographic inversion. The impact of peculiar velocities on the reconstruction results with the IRGN and with the RPC inversion results is presented in more detail in Appendix \ref{app: inversion_schemes}. In particular, peculiar velocities bias the estimation of $\gamma$ by the RPC\_IRGN method by adding an additional discrepancy between the RPC and IRGN reconstructions.

This additional discrepancy is mainly caused by the difference between the neutral hydrogen density probability density distribution in redshift space and in real space, see Appendix \ref{app: inversion_schemes} and Appendix \ref{app: just_sim}. We investigated whether this discrepancy between RPC and IRGN reconstructions could be partly lifted by amending the prior lognormal model that we needed to assume for the RPC inversion. As we do not want to fit a completely new model, we only slightly amend the true mean of the lognormal model, i.e. the one which was used to create synthetic data. In fact, we found the strongest coincidence between RPC and IRGN reconstruction results for reconstructions with the RPC method which assume a slightly smaller mean value in the lognormal model than the true one.

We perform an analysis assuming no peculiar velocities on synthetic data that in fact include peculiar velocities. In this case the RPC\_IRGN algorithm recovers equation of state $\gamma=1.34 \pm 0.02$, compared to the true value $\gamma=1.4$. We observe a larger statistical variation for the mean of the estimates for $\gamma$ in our sample. This is explained by the reduced sample size. To protect us from underestimating the resulting bias, we include a bias, $\delta_\gamma=0.08$, as an additional source of error (added in quadrature) in our analysis of real data with this method.

On the other hand, peculiar velocities play only a minor role for the IRGN\_$\chi^2$ method. We show in Fig. \ref{fig: eos} with black lines the edges of the $\chi_2 \approx 1$ (e.g. $\chi^2 \in [0.98, 1.02]$) region when finding estimates from synthetic data which had been created with redshift space distortions. To obtain these estimates we cut of all the pixels in which the recovered flux is clearly distorted from the observed normalized flux. This mainly occurs at the boundaries of the studied wavelength interval. These fluxes are clearly not well recovered as the convolution in Eq. \eqref{eq: tau} could not be computed from values outside of the density interval and the integration is not complete at the boundaries. When peculiar velocities are ignored in the creation of synthetic data, then this would only affect a very small number of pixels due to the relatively small thermal broadening parameter. However, when peculiar velocities are considered other parts of the spectrum could be shifted towards the boundary. Hence, we do not consider the flux in pixels which are less than roughly $100\,\mathrm{km/s}$ away from the boundary of the wavelength interval. The line profiles in Fig. \ref{fig: eos} coincide very well, such that the bias introduced by not including peculiar velocities in the IRGN\_$\chi^2$ method is very small and will be ignored in the following. In particular, even in this case that peculiar velocities are ignored the profile shown by the black lines suggests a linear relation between $\gamma$ and $T_0$ in the most likely region of values for $\gamma$. 

There is a straightforward explanation for this finding that the IRGN\_$\chi^2$ estimation is unaffected by peculiar velocities. We do an error in the inversion when not including peculiar velocities in the inversion, i.e. when not respecting the broadening and narrowing of absorption lines by an additional velocity distribution. However, we make the same error when forwardly modeling the optical depth in the Ly$\alpha$ forest from the inverted density field and both effects cancel out when taken together. 

\subsection{Noisy Spectra} \label{sec: noisy_spectra}
\begin{figure*}
\centering
IRGN\_$\chi^2$-Mock Data \\
\subfigure[$\mathrm{SNR}=25$]{
\includegraphics[width=0.3\textwidth]{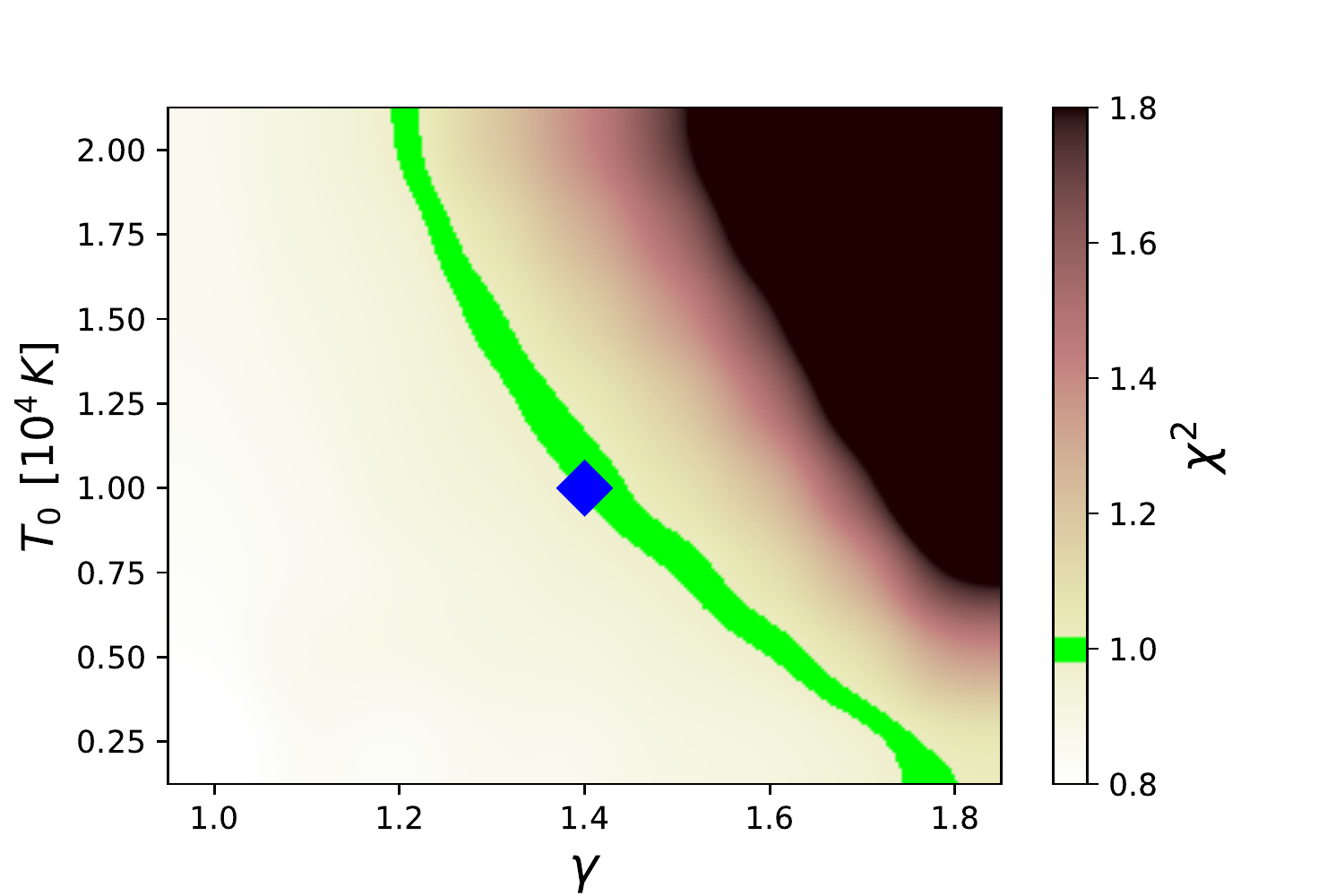}}
\subfigure[$\mathrm{SNR}=10$]{
\includegraphics[width=0.3\textwidth]{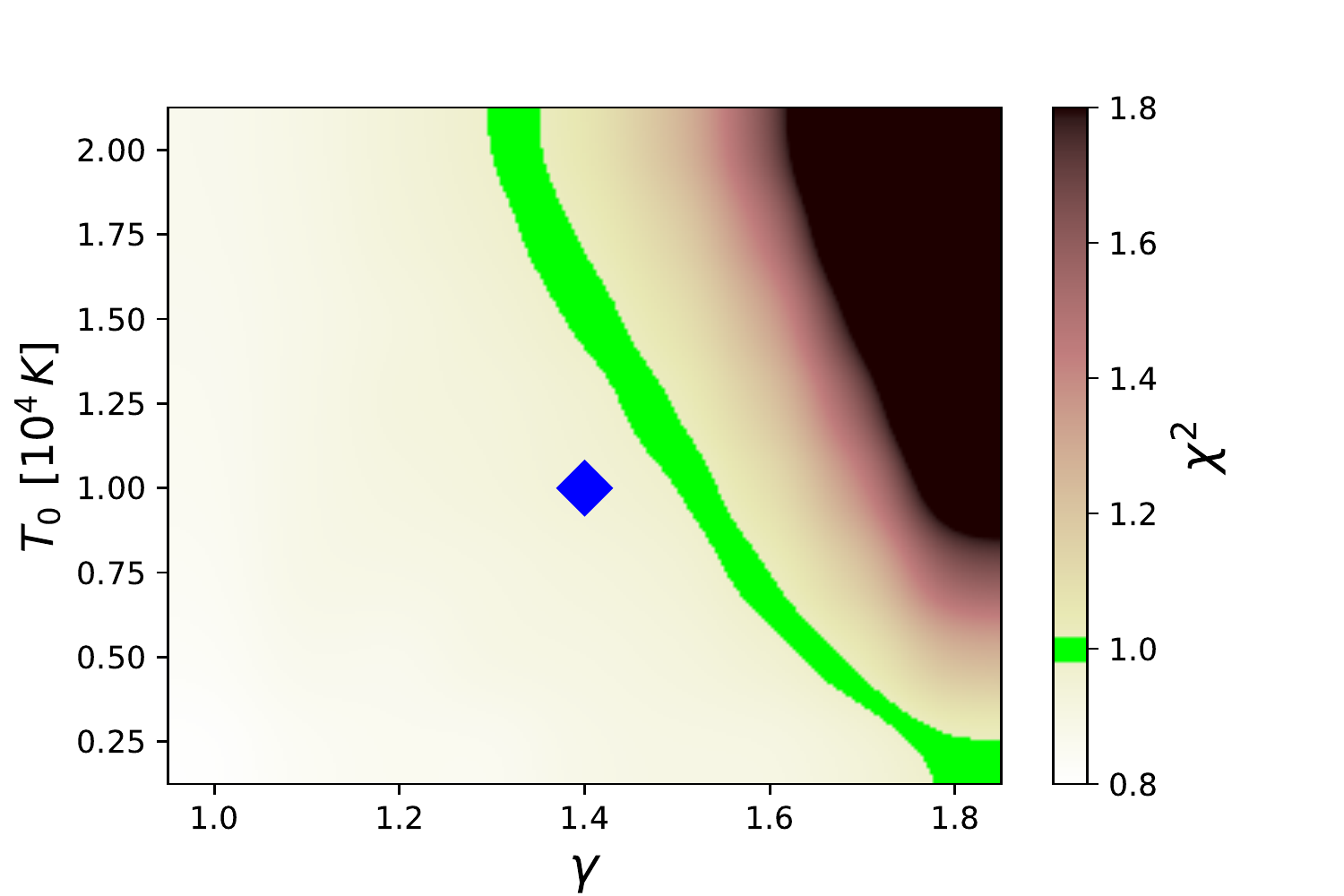}}
\subfigure[$\mathrm{SNR}=2$]{
\includegraphics[width=0.3\textwidth]{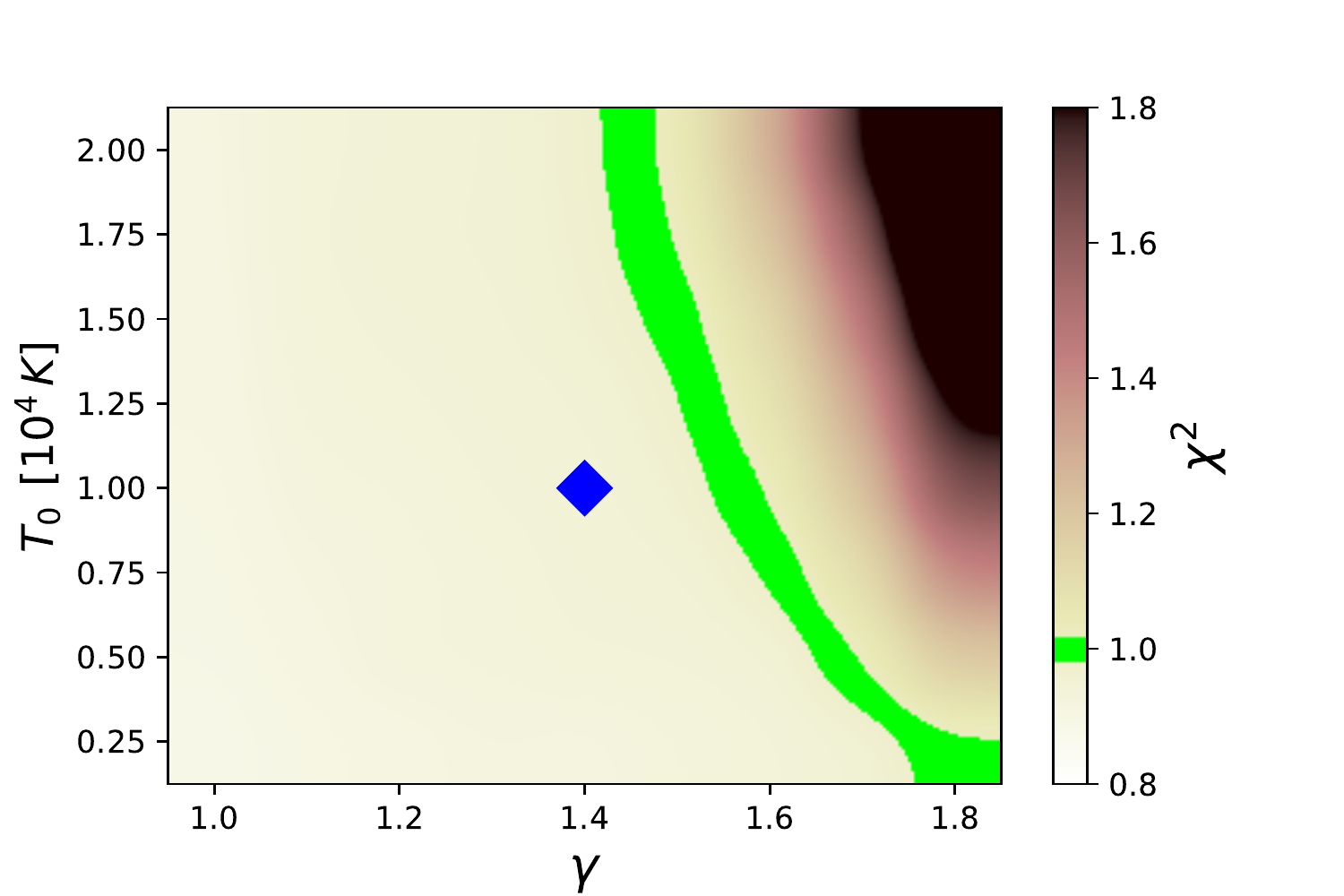}}
\caption{Same as in Fig. \ref{fig: irgn_eos}, but for varying noise levels on the synthetic spectra. The IRGN\_$\chi^2$ becomes inadequate at large noise levels.}
\label{fig: noise}
\end{figure*}

In the former subsections we examined our estimation procedure at high signal to noise ratios. In this Section we discuss the estimation from noisy spectra. We show in Fig. \ref{fig: noise} the fitting results with the IRGN\_$\chi^2$ method for smaller SNR. Based on Fig. \ref{fig: noise} we only suggest the IRGN\_$\chi^2$ method for high quality data SNR>10 as we now explain.

The IRGN method performs a minimization of the two terms in the exponential of Eq. \eqref{eq: posteriori}. The first term (the so called data-fidelity term) measures the proximity of the observed data and the recovered data (i.e. computes the reduced $\chi^2$). The second term (the so called penalty term) measures the proximity of the recovered solution to a prior guess. The scale of the penalty term is set by $\bm{\mathsf{C}}_0$, the prior auto correlation. In an ideal reconstruction the auto-correlation function of the recovered density matches the prior correlation, and the reduced $\chi^2$ is exactly one. If thermal broadening is overestimated, then the recovered flux will be more blurred than the true one. The reduced $\chi^2$ will be greater than one. If thermal broadening is underestimated, the recovered flux will fit the additional pixel to pixel noise contribution. $\chi^2$ drops below 1. In this sense, the penalty term evaluates over- or underestimation of thermal broadening. However, if there is a significant noise contribution, then the penalty term introduces effective smoothing to regularize the recovered density against noise rather than evaluating the over- or underestimation of thermal broadening.

In fact, as shown in Fig. \ref{fig: noise}, smaller SNR push the reduced $\chi^2$ of the exact set of parameters to smaller values. Consequently the $\chi^2 \approx 1$ lines would bias the estimation of $\gamma$ and $T_0$ towards larger values. Furthermore, the error of the estimation increases with decreasing SNR which is indicated by the larger width of the green overplotted lines ($\chi^2 \in [0.98, 1.02]$). However, for all SNR the behavior of $\chi^2$ as a function of $\gamma$ and $T_0$ is similar. Thus, it might be possible to also include lower quality data in the analysis and fight the introduced bias in a post-processing step. This idea is left for future refinements of our method.

The RPC\_IRGN method is robust against moderate noise contributions. This is a consequence of the robustness of the IRGN and the RPC inversion algorithms against noise, for more details see Appendix \ref{app: inversion_schemes} and \cite{Mueller2020}. The portions of the spectra that are highly affected by noise during inversion, large overdensities and small underdensities (in logarithmc scale), are not considered for the comparison of the RPC method and the IRGN method.

\section{Application To Observational Data} \label{sec: application}
We apply our estimation procedure to observational data in this section. Our final results are shown in Fig. \ref{fig: joint_estimate} and Fig. \ref{fig: marginal}. 

\subsection{Neutral Hydrogen Density At Mean Density} \label{sec: mean_density}
\begin{figure}
    \centering
    \includegraphics[width=0.5\textwidth]{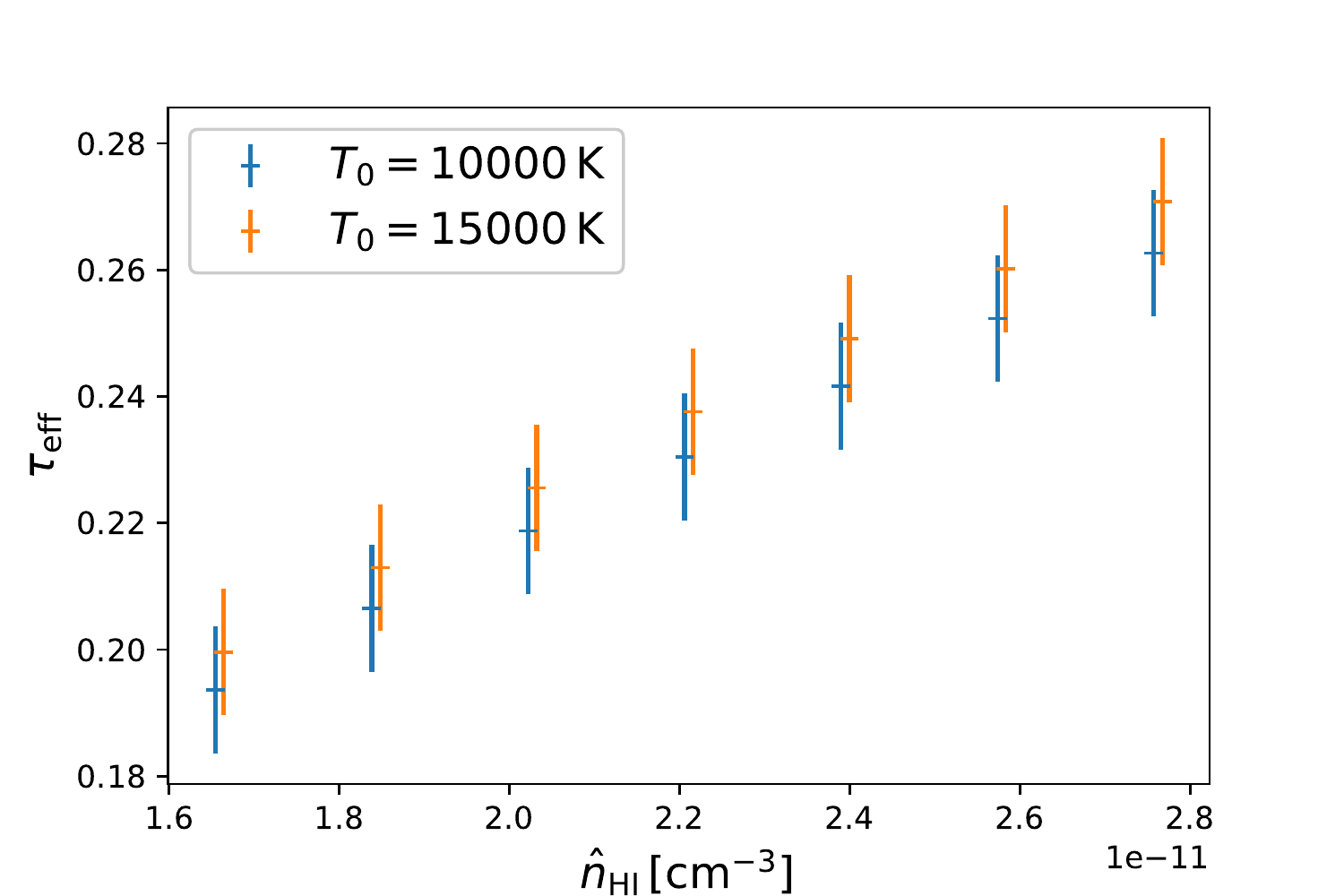}
    \caption{Optical depths from our simulation with varying $\hat{n}_\mathrm{HI}$ for artificial temperature $T_0 = 10000\,\mathrm{K}$ (blue) and for temperature $T_0 = 15000\,\mathrm{K}$ (orange). The errorbars are the statistical errors from varying the simulation seeds in 100 lines of sight of $50\,h^{-1}\,\mathrm{Mpc}$.}
    \label{fig: forward_model}
\end{figure}

We apply the mean optical depth model from \cite{Becker2013}:
\begin{align}
    \tau_\mathrm{eff} \approx 0.751 \cdot \left( \frac{1+z}{4.5} \right)^{2.9}-0.132. \label{eq: tau_eff_becker}
\end{align}
This function was proven to adequately fit the observed effective optical depths. It is also compatible with alternative models such as the fits presented in \cite{Kirkman2005} and \cite{Faucher2008}. In particular for redshift $z=2.5$ we get: $\tau_\mathrm{eff}(z=2.5) \approx 0.23$. We vary the parameter $\hat{n}_\mathrm{HI}$, create each time 100 lines of sight of $50\,h^{-1}\,\mathrm{Mpc}$ length with our simulation outlined in Sec. \ref{sec: synthetic_data} (with peculiar velocities) and compute the effective optical depth. We show our fitting of the optical depth in Fig. \ref{fig: forward_model}. We get $\hat{n}_\mathrm{HI} = \left(22 \pm 3 \right) \cdot 10^{-12}\,\mathrm{cm}^{-3}$. The error originates from the uncertainty in $\tau_\mathrm{eff}$ from \citet{Becker2013}, the variance of estimated $\hat{n}_\mathrm{HI}$ between different random seeds in our simulation box, the prior uncertainty in $T_0$ and $\gamma$ (see Fig. \ref{fig: forward_model}), and the variation in the selection of suitable Jeans scale drawn from the result for $z=2.48$ in \citet{Zaroubi2006}. We summarize all systematic errors for the estimation of $\hat{n}_\mathrm{HI}$ in Tab. \ref{tab: error_mean_dens}. The systematic errors were computed by redoing the same analysis, but varying the unfixed parameters. \citet{Becker2013} measured $\tau_\mathrm{eff}$ roughly with a precision of $\Delta \tau = 0.007$ at redshift $z=2.5$ which translates in an uncertainty of $\sigma_1 \approx 1.3 \cdot 10^{-12}\,\mathrm{cm}^{-3}$. The error for the pressure smoothing scale is calculated from the measurement error in \citet{Zaroubi2006}. We chose $\gamma = 1.4$ and $T_0 = 10000\,\mathrm{K}$ as prior for the simulation and proceed with uncertainties of $\Delta_\gamma = 0.2$ and $\Delta_T = 5000\,\mathrm{K}$ which approximates the bias between current measurements of the thermal history of the IGM (see Sec. \ref{sec: comparison}).

\begin{table}
    \centering
    \begin{tabular}{c|c}
        Error for $\tau_\mathrm{eff}$ from Eq. \eqref{eq: tau_eff_becker} & $\sigma_1 = 1.3 \cdot 10^{-12}\,\mathrm{cm}^{-3}$ \\
        Variation between different seeds & $\sigma_2 = 1.8 \cdot 10^{-12}\,\mathrm{cm}^{-3}$\\
        Unknown Jeans length & $\sigma_3 = 0.7 \cdot 10^{-12}\,\mathrm{cm}^{-3}$ \\
        Unknown $\gamma$ & $\sigma_4 = 0.9 \cdot 10^{-12}\,\mathrm{cm}^{-3}$ \\
        Unknown $T_0$ & $\sigma_5 = 1.3 \cdot 10^{-12}\,\mathrm{cm}^{-3}$ \\
        Total Error & $\sigma = \sqrt{\sum_i \sigma_i^2} \approx 2.81 \cdot 10^{-12}\,\mathrm{cm}^{-3}$
    \end{tabular}
    \caption{Error budget for the estimation of $\hat{n}_\mathrm{HI}$.}
    \label{tab: error_mean_dens}
\end{table}

\subsection{Data Selection} \label{sec: data}
We apply our estimation method to a subsample of the UVES\_SQUAD sample \citep{Murphy2019}. This is a survey of 467 quasar spectra in the redshift range $z=0-5$ observed with the high resolution UVES instrument \citep{Dekker2000}. The spectra are fully reduced and continuum fitted by \citet{Murphy2019}. The spectra are reproducible from raw data. The reduction pipeline only made use of publicly available software including UVES\_POPLER \citep{Murphy2016a}. For more details on the reduction process we refer to \citet{Murphy2019}. The complete survey ranges from quasar spectra with continuum to noise ratio (CNR) of $\mathrm{CNR} = 4$ to $\mathrm{CNR} = 342$ at $2.5\,\text{km/s}$ pixels at wavelength $\lambda = 5500\,\mathring{A}$.

\begin{table}
    \centering
    \begin{tabular}{c|c|c|c}
       QSO  & Redshift & CNR & Dispersion [km/s]\\
J000448-415728	&	2.76	&	113  & 2\\
J005758-264314	&	3.655	&	44 & 2.5\\
J010311+131617	&	2.705	&	66 & 2.5\\
J014214+002324	&	3.37	&	39 & 2.5\\
J015327-431137	&	2.74	&	119 & 2.5\\
J033108-252443	&	2.685	&	64 & 2.5\\
J033244-445557	&	2.6	&	34 & 2.5\\
J034943-381030	&	3.205	&	62 & 1.3\\
J040718-441013	&	3	&	57 & 1.3\\
J042214-384452	&	3.11	&	77 & 2.5\\
J045214-164016	&	2.6	&	40 & 2.5\\
J045523-421617	&	2.66	&	90 & 1.3\\
J064326-504112	&	3.09	&	33 & 2.25\\
J091127+055054	&	2.798	&	36 & 2.5\\
J091613+070224	&	2.786	&	72 & 2.5\\
J094253-110426	&	3.054	&	115 & 2.5\\
J101155+294141  &    2.64   &  36 & 2.5\\
J103909-231326	&	3.13	&	39 & 2.5\\
J111350-153333	&	3.37	&	55 & 2.5\\
J114254+265457	&	2.625	&	99 & 2.5\\
J132029-052335	&	3.7	&	46 & 2.5\\
J134258-135559	&	3.19	&	35 & 2.5\\
J151352+085555	&	2.901	&	40 & 2.5\\
J162116-004250	&	3.703	&	52 & 2.5\\
J193957-100241	&	3.787	&	38 & 2.5\\
J214159-441325	&	3.17	&	42 & 2.5\\
J223408+000001	&	3.025	&	38 & 2.5\\
J224708-601545	&	3.005	&	66 & 2\\
J233446-090812	&	3.317	&	36 & 2.5\\
J235034-432559  &   2.885   &   131& 1.5\\
J235129-142756	&	2.94	&	45 & 2.5
    \end{tabular}
    \caption{Sample of high quality QSO spectra from the UVES\_SQUAD survey \citep{Murphy2019}. The redshift, CNR and the dispersion are taken from \citet{Murphy2019}. The CNR is reported at wavelength of $4500\mathring{A}$. The Dispersion is given in $\mathrm{km/s}$.}
    \label{tab: selected_quasars}
\end{table}

In this publication we investigate $z \approx 2.5$. Moreover, we only use spectra with a sufficiently small noise contribution. In fact we only used spectra with $\mathrm{CNR}>30$ at wavelength $\lambda = 4500\,\mathring{A}$. Moreover, we only use spectra for which the absorption at redshift $z=2.5$ has a rest-frame wavelength greater than $1023\mathring{A}$, to avoid contaminations with absorption from the Ly$\beta$ forest. Our sample of QSO spectra is summarized in Tab. \ref{tab: selected_quasars}. For redshift $2.5$ we only use the wavelength range $\lambda \in [4235, 4275]\,\mathring{A}$, such that uncertainties in redshift are negligible. We divide these spectra in four sets of $10\,\mathring{A}$ in length. These cuts are applied automatically. Thus, at the edges some absorption features might be cut. However, this will not affect the estimation since the boundaries of each set are excluded from IRGN inversion results (the convolution at the boundary of each set is not complete, such that the optical depth could not be fitted).

\begin{figure}
    \centering
    \includegraphics[width=0.5 \textwidth]{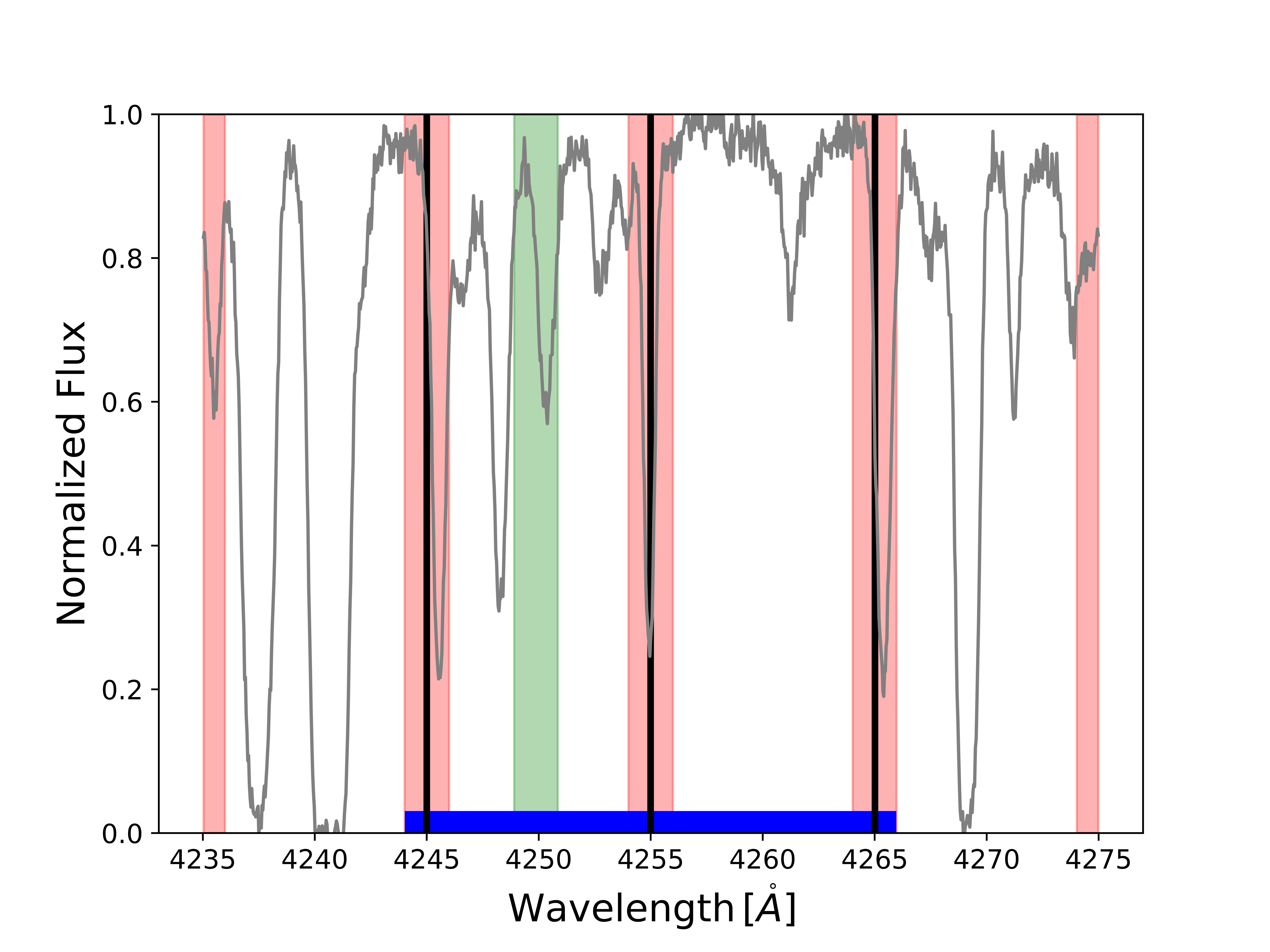}
    \caption{The data preparation process. We show the normalized flux of the QSO J091613-070224 (grey). We verify the continuum fit visually, search for damped Ly$\alpha$ systems in the whole spectrum and mask metal lines (green, here FeIII). We cut the spectrum in four sets (black vertical lines) and select only these sets which show significant absorption sstructure for inversion with the IRGN method (minmal flux smaller than $0.2$). The wavelengths excluded from the sample by this selection process are indicated with the blue bottom line. Lastly we mask the IRGN recovered densities at the boundaries of each set (red shaded) and use only the inner parts for comparison to observed flux (IRGN\_$\chi^2$) and RPC inversion results (IRGN\_RPC).}
    \label{fig: data_prep}
\end{figure}

The spectra were continuum fitted manually by \cite{Murphy2019} with B-splines. We verified by visual inspection that these continuum fits are reasonable by searching for 'unabsorbed' pixels in the wavelength range of interest. We masked the strongest metal transitions between Ly$\alpha$ and Ly$\beta$ rest frame wavelength (O\,VI, C\,II, N\,II, Fe\,III, Fe\,II, Si\,II, N\,I, Si\,III). Damped Lyman $\alpha$ systems are identified with the list provided by \cite{Murphy2019} and were found in the spectra of eight of the quasars in our sample. Again we masked corresponding metal lines in the spectra manually to attain at our final data sample. A walkthrough of our data preperation and estimation process is shown in Fig. \ref{fig: data_prep}.

According to the prescription in \citet{Rollinde2001} we only use the sets for inversion in which the minimal normalized flux drops below $0.2$ for the inversion with the IRGN method, i.e. we only use the sets for inversion if they contain a significant absorption feature. Our final sample consists of $62$ sets. For the RPC method we use all sets to do the reconstruction from a complete sample of the flux probability density distribution. Lastly we evaluate all the reconstructions and reject every set for which the inversion failed, i.e. we were not able to reasonably fit the observed flux.

We found three sets (including the one shown in Fig. \ref{fig: fitting_demonstration} in our sample that do not satisfy the original selection criterion anymore after finally cutting the edges of the spectra from the reconstruction. However, these spectra have single-standing absorption profiles. Thus, they were successfully fitted by the IRGN method providing temperature estimates in great agreement with our overall estimates.

\subsection{RPC\_IRGN}
\begin{figure}
    \centering
    \includegraphics[width=0.5 \textwidth]{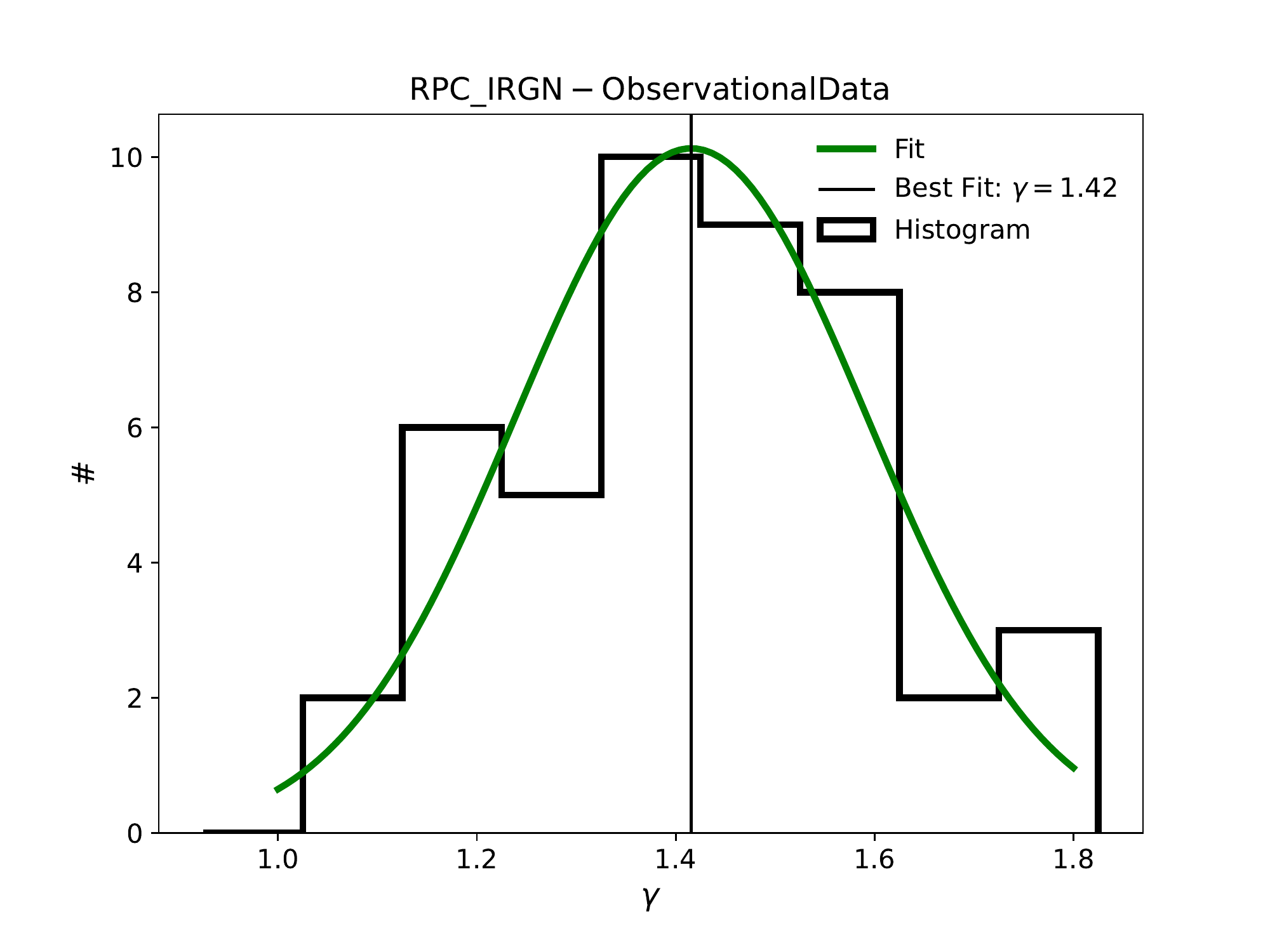}
    \caption{Histogram of estimated values for $\gamma$ from observational data. The green line is a Gaussian fit to the histogram with mean $\mu = 1.44$ and standard deviation $\sigma=0.18$.}
    \label{fig: beta_hist}
\end{figure}

As explained in Sec. \ref{sec: estimation} the distance of the logarithms of the recovered density with the IRGN method and the RPC method is nearly independent of $T_0$. To save computation time it is therefore reasonable to assume a reasonable value for the temperature $T_0$ and only vary $\gamma$ to perform the RPC\_IRGN step. We perform the inversion assuming $T_0 = 10000\,\text{K}$ and vary $\gamma$ on a uniform grid $\gamma \in [1.25, 1.3, 1.35,...1.95, 2.0]$. We performed the inversion for every set in our sample and we store the parameter $\gamma$ which minimizes the distance to the density recovered with the RPC method for every set individually. An examplary estimation for the QSO J033244-445557 is shown in the lower right panel of Subfigure A of Fig. \ref{fig: fitting_demonstration} (red framed). Moreover, we show the observed density perturbations in Subfig. C which demonstrates the overall agreement between the two reconstructions. The minimum in the logarithmic distance is clearly visible. A histogram of all our estimates is shown in Fig. \ref{fig: beta_hist}. The width of the distribution matches well the width of the minimum in Fig. \ref{fig: rpc_eos} and is thus interpreted as the width of the measurement error. 

The systematic error originates from the propagation of the error in $\hat{n}_\mathrm{HI}$ and the fact that the lognormal model is not an exact representation of the true density field. We demonstrate in Appendix \ref{app: just_sim} that the observed density field varies from a lognormal distribution in the low density tail (it is more skewed), but this effect is explained (partly) by the effect of peculiar velocities on the spectrum (see Appendix \ref{app: just_sim}). Thus, we estimate the error related to the use of the lognormal model by the error introduced by peculiar velocities, i.e. the bias observed in Sec. \ref{sec: pec_vel}.

We modeled the logarithm of the true density field as a multivariate Gaussian distributed random field with a spatial correlation function (specified by the matter power spectrum), which was shown to mimic the observed density distribution sufficiently well (recall our discussion in Sec. \ref{sec: synthetic_data}). However, the lognormal prior $P_\Delta$ for the RPC algorithm is a one dimensional probability density distribution approximating the histogram of all measured densities (regardless of their spatial correlation, i.e. whether two pixels are close to each other). In fact, the lognormal prior $P_\Delta$ for the RPC algorithm is described by only one parameter: The mean of the logarithm of the density perturbation $\mu$. The standard deviation of the logarithmic density perturbation is fixed by the assumption that the mean of the non-linear overdensity vanishes (i.e. the density perturbation is normalized). We estimate $\mu$ by the density probability density distribution measured in our simulation box, see Sec. \ref{sec: synthetic_data} for more details on the simulation of the density field. However, we estimated a greater match between RPC and IRGN inversions in Sec. \ref{sec: pec_vel} for slightly smaller means and take this amended prior for reconstruction. We compare the probability density distribution of observational data, the fitted analytic prior and the updated final prior that was used for the reconstruction in Appendix \ref{app: just_sim}, in particular in Fig. \ref{fig: dens_hist}. We estimated the systematic error for $\mu$ by comparing the density distributions calculated in different boxes with our simulation with varying Jeans scale (see Sec. \ref{sec: synthetic_data} and Fig. \ref{fig: jeans} in Appendix \ref{app: just_sim}). The error is estimated by the mean variation from our boxes.

As a consistency check we compared in Appendix \ref{app: just_sim} our estimate for $\mu$ with the mean logarithmic density perturbation of the simulated density distribution in the first subbox of the third run of \small{ILLUSTRIS} \citep{Nelson2015} and found that both estimates are consistent up to roughly one standard deviation of significance.

We repeat our RPC\_IRGN analysis with sets of parameters distorted by these errors and estimate the systematic error from these samples. Our error estimates are summarized in Tab. \ref{tab: error_rpc_irgn}. The statistical error is computed as the statistical variance of the mean of the sample of estimates for each set (portion of a line of sight). To account for probable non-Gaussianities we also add the variation between mean and median to the error. We did the inversion on a uniform grid of values for $\gamma$ with a resolution (half bin size) of $\Delta_\gamma = 0.025$ which adds an additional source of error. Overall we get for the mean value the estimate: $\langle \gamma \rangle = 1.42 \pm 0.11$. 

\begin{table}
    \centering
    \begin{tabular}{c|c}
        Error from the uncertainty of $\hat{n}_\mathrm{HI}$ & $\sigma_1 = 0.046$ \\
        Bias by peculiar velocities from Sec. \ref{sec: pec_vel} & $\sigma_2 = 0.08$ \\
        Uncertainty in mean of prior lognormal field & $\sigma_3 = 0.045$\\
        Statistical Variance of Mean & $\sigma_4 = 0.02$ \\
        Mean-Median Variation & $\sigma_5 = 0.015$ \\
       Grid Resolution & $\sigma_6 = 0.025$ \\
        Total Error & $\sigma = \sqrt{\sum_i \sigma_i^2} = 0.109$
    \end{tabular}
    \caption{Error budget for the estimation of $\gamma$ with the RPC\_IRGN method.}
    \label{tab: error_rpc_irgn}
\end{table}

\subsection{IRGN\_$\chi^2$}
\begin{figure*}
    \centering
    \subfigure[\Large{Estimation}]{
    \includegraphics[width=\textwidth]{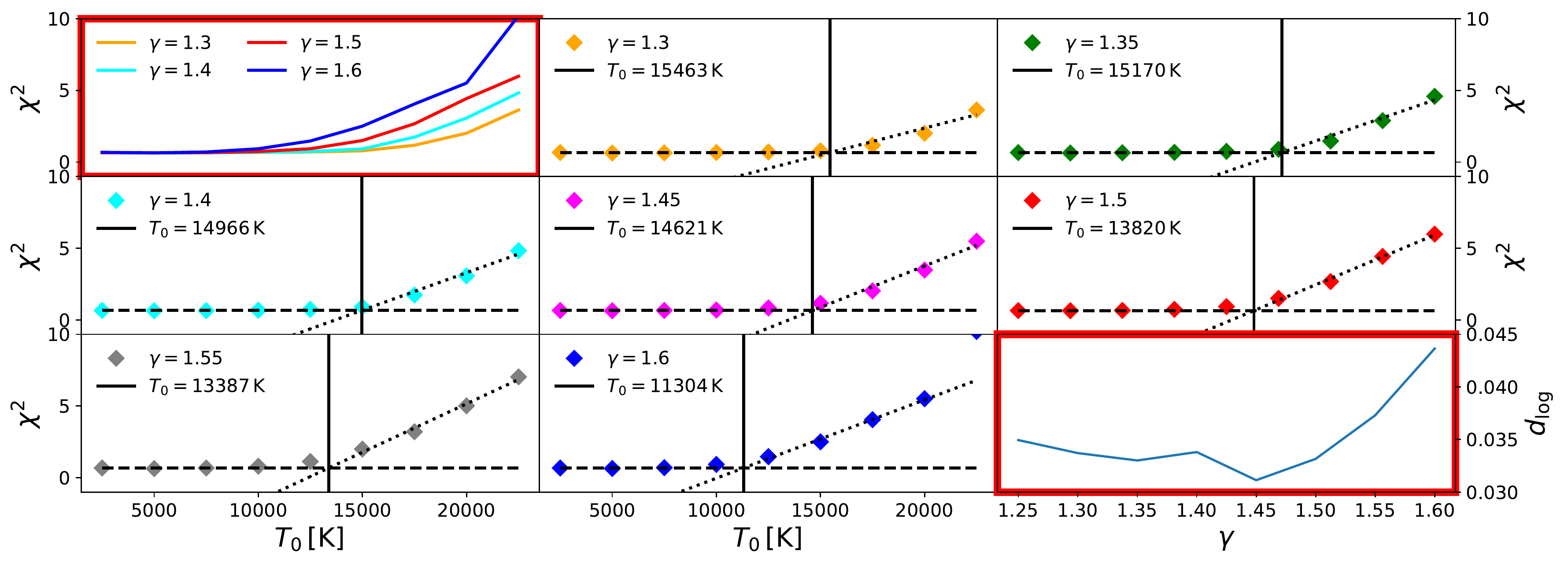}}\\
    \subfigure[\Large{Flux}]{
    \includegraphics[width=0.4\textwidth]{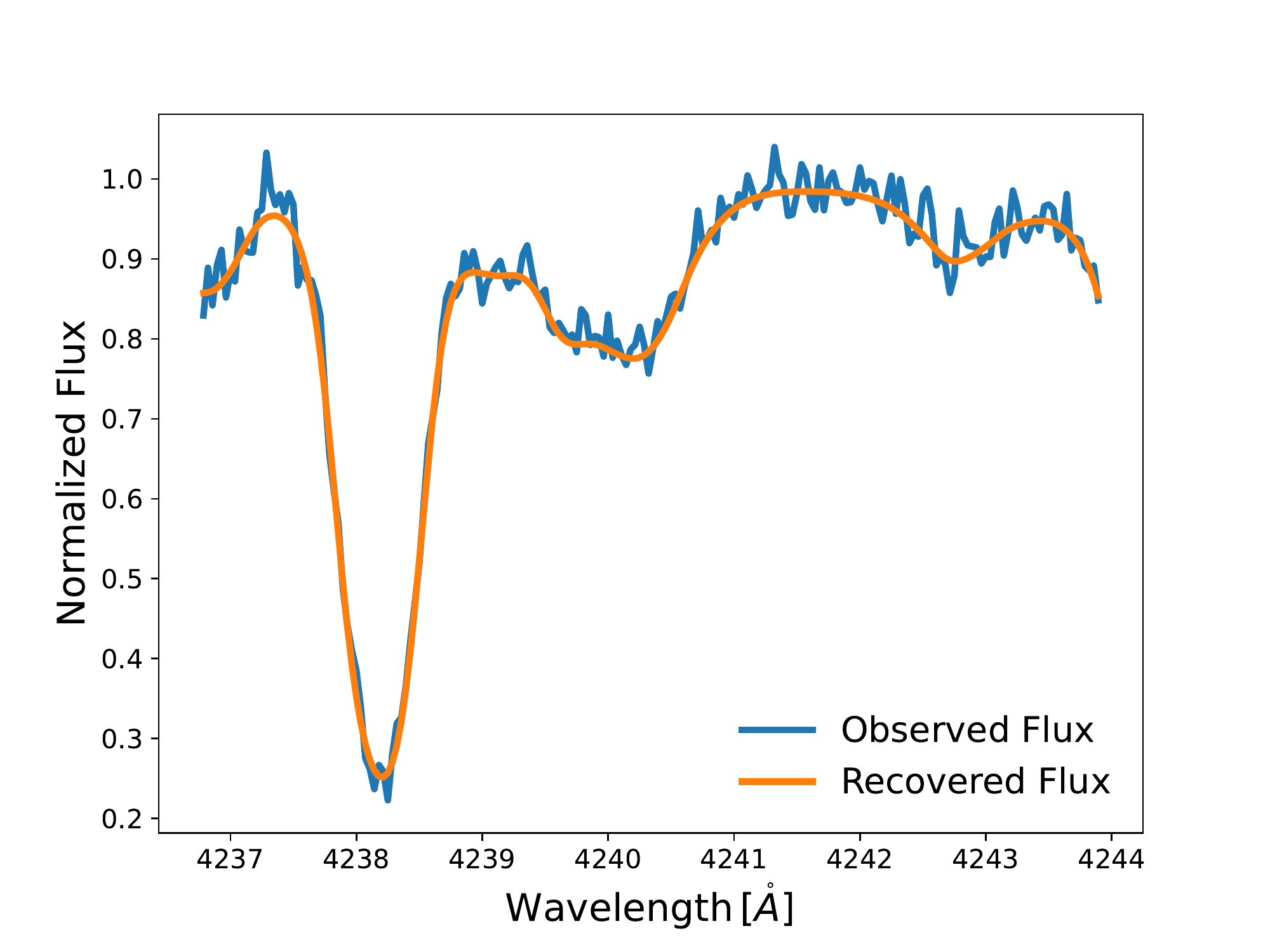}}
    \subfigure[\Large{Density}]{
    \includegraphics[width=0.4\textwidth]{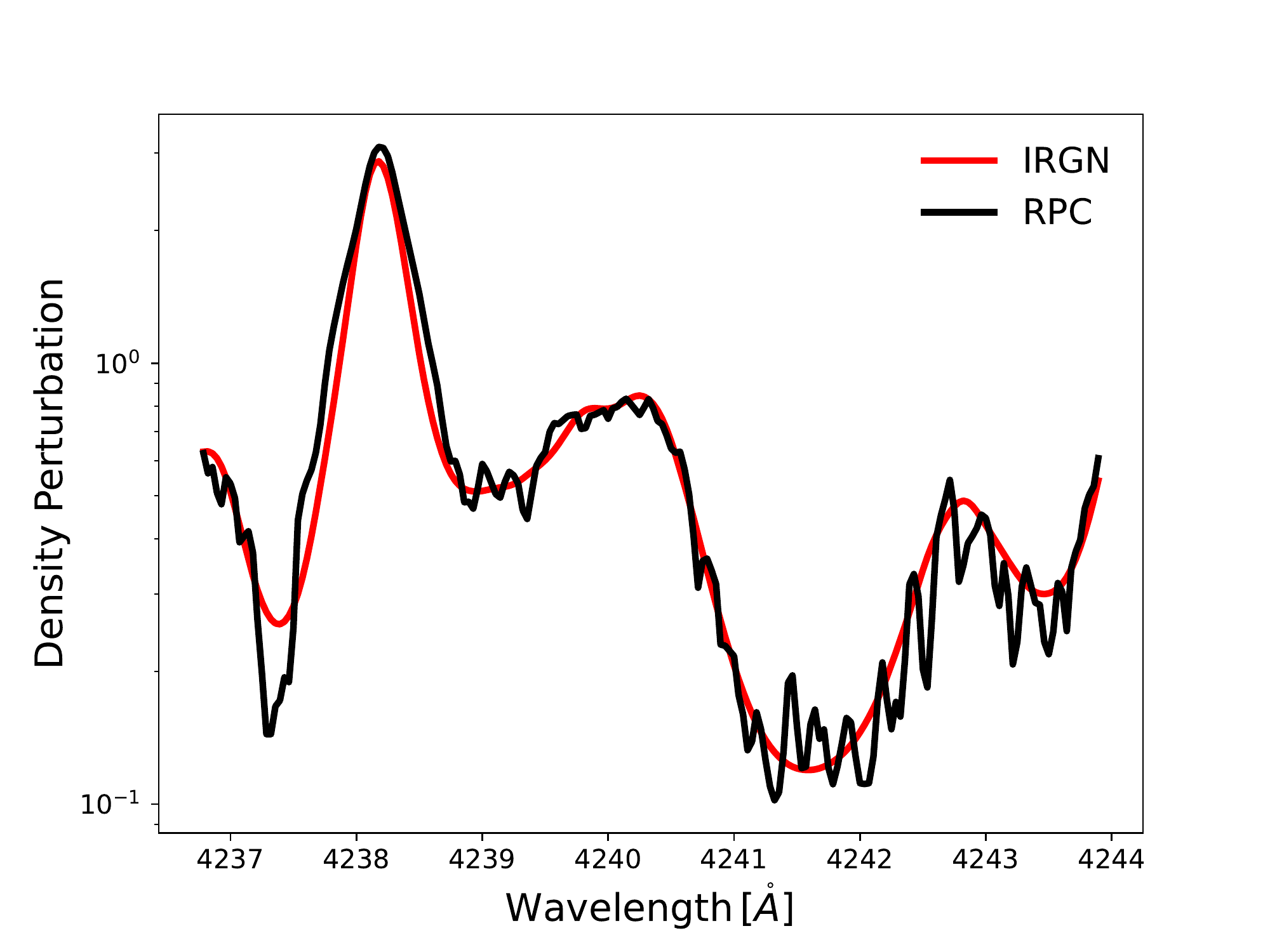}}
    \caption{Exemplary application of our estimation procedure to one set of observational spectra: to the quasar J033244-445557 for wavelength $\lambda \in [4235, 4245]\,\mathring{A}$.\protect\linebreak \textit{Subfigure A}: Summary of the results of our estimation procedure. The red framed panels show the results with the fitting results with the IRGN\_$\chi^2$ method (upper left panel) and with the RPC\_IRGN method (lower right panel). All other panels show $\chi^2$ as computed with the IRGN\_$\chi^2$ method as function of $T_0$ for seven different values of $\gamma$. Upper left panel (red framed): $\chi^2$ as function of $T_0$ for different values of $\gamma$. For clarity we show only four out of seven curves. middle panels: curves for one specific $\gamma$ (diamond) with fitted borderline (dashed line), fitted linear increase (dotted line) and fitted temperature $T_0$ (solid black line). lower right panel: The logarithmic distance calculated with the RPC\_IRGN method for different values of $\gamma$. For this LOS we favor $\gamma = 1.45$.\protect\linebreak \textit{Subfigure B}: The observed, noisy flux and the best reconstruction with the IRGN reconstruction method. \protect\linebreak \textit{Subfigure C}: The recovered density with the IRGN method and with the RPC algorithm for $\gamma = 1.45$, i.e. where the reconstructions reach their minimum logarithmic distance.}
    \label{fig: fitting_demonstration}
\end{figure*}

Based on our estimate for $\gamma$ we perform the IRGN\_$\chi^2$ step for the values $\gamma \in [1.3, 1.35, 1.4, 1.45, 1.5, 1.55, 1.6]$ and $T_0$ on a uniform grid $T_0 \in [2500, 5000, ...,  27500, 30000]\,\text{K}$. We apply our fitting procedure for every $\gamma$ and every line of sight individually. We present in Fig. \ref{fig: fitting_demonstration} an example analysis with our fitting procedure applied to one observation set: the quasar J033244-445557 for wavelength $\lambda \in [4235, 4245]\mathring{A}$. We show the results of our estimation procedure and the best fitting results with the IRGN and RPC algorithm. As a first consistency check we can see that the recovered flux describes the observed flux well (Subfigure B) and that the reconstructions with RPC method and with IRGN algorithm match reasonably well for moderate overdensities, see subfigure C. This indicates that the inversion procedures were successful in fitting the observational data. The two asymptotic regions of a constant $\chi^2$ at smaller temperatures than the estimated $T_0$ and of a linear decay at larger temperatures are clearly visible. We fit both with linear functions (dashed and dotted black lines in the middle panels). The intersection point (solid, vertical black lines) fits well the temperatures at which $\chi^2 = 1$ is expected. Moreover, as expected from our analysis on synthetic data in Sec. \ref{sec: synthetic}, the estimated temperature decreases with increasing $\gamma$. This becomes clearly visible when plotting the $\chi^2 (T_0)$ curves in one figure, see the upper left panel (red framed) in Fig. \ref{fig: fitting_demonstration}. The lines from $\gamma = 1.3$ to $\gamma = 1.6$ are ordered from lowest to uppermost. We find the corresponding estimate for the temperature for fixed $\gamma$ by averaging the estimates from all studied lines of sight weighted by the extent of wavelength coverage in each set (i.e how many pixels are unmasked and used for the inversion).

Our seven estimates for the temperature $T_0$ are plotted in Fig. \ref{fig: joint_estimate}. The errorbars were computed by the standard deviation of the mean of the temperature estimates for each individual line of sight. We show in Appendix \ref{app: meas_hist} histograms of our measurements at every $\gamma$ which justify the use of the average as estimator. We found that variations in $\hat{n}_\mathrm{HI}$ and the strength of the penalty term introduce only minor errors in the temperature estimations with the IRGN\_$\chi^2$ method and can be ignored compared to the statistical error. Based on our proposition in Sec. \ref{sec: estimation} we fit the $\chi^2 = 1$ line with a linear fit $T_0 = m \cdot \gamma +b$. This linear fit is also plotted in Fig. \ref{fig: joint_estimate}. However, the line seems to be over-correlated indicating that the errorbars for the temperatures may be overestimated. The magenta shaded area indicates the $1-\sigma$ predictive distribution of our linear fit parameters. In particular we find: $m=-1.634$ and $b=3.668$, where $m$ and $b$ are reported in units of $10^4\,\mathrm{K}$. The estimated covariance of the linear fit parameters is printed in Tab. \ref{tab: linear_fit},

\begin{table}
    \centering
    \begin{tabular}{c|cc}
    \hline
        & $m$ & $b$ \\ \hline
        $m$ & $\Sigma_\mathrm{mm} = 0.255$ & $\Sigma_\mathrm{bm} = -0.375$ \\
        $b$ & $\Sigma_\mathrm{mb} = -0.375$ &  $\Sigma_\mathrm{bb} = 0.555$
    \end{tabular}
    \caption{Covariance matrix of the linear fit of Fig. \ref{fig: joint_estimate} with the model $T_0 = m \cdot \gamma +b$. The temperatures are reported in units of $10^4\,\mathrm{K}$.}
    \label{tab: linear_fit}
\end{table}

\subsection{Joint Estimation}
\begin{figure}
    \centering
    \includegraphics[width=0.5 \textwidth]{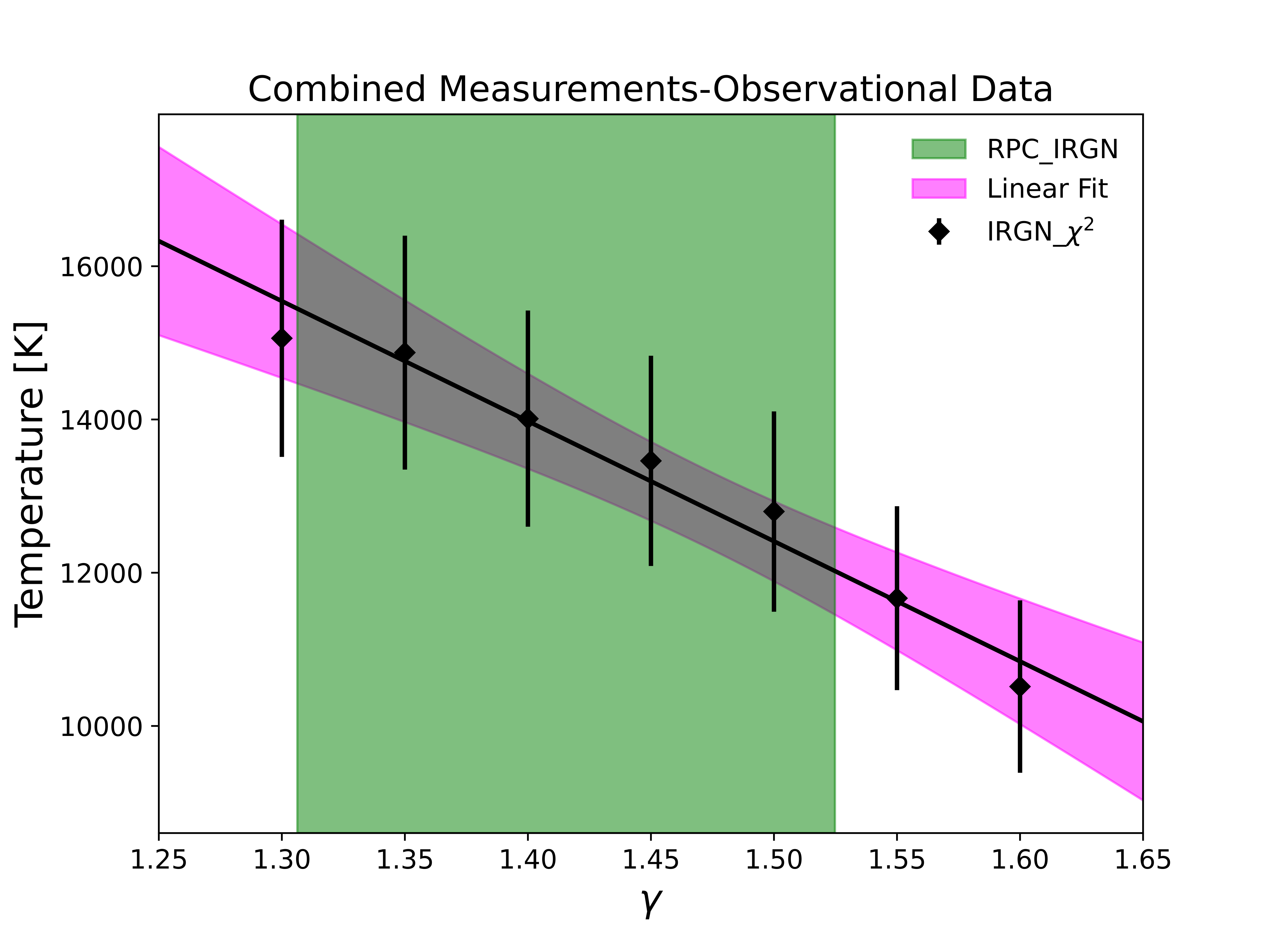}
    \caption{Joint estimation with the RPC\_IRGN method (green shaded, $1-\sigma$ confidence interval) and the estimates with the IRGN\_$\chi^2$ method (black diamonds). The seven data points are fitted with a linear fit (black solid line) with corresponding predictive 1-$\sigma$ error (magenta shaded).}
    \label{fig: joint_estimate}
\end{figure}

\begin{figure}
    \centering
    \includegraphics[width=0.5 \textwidth]{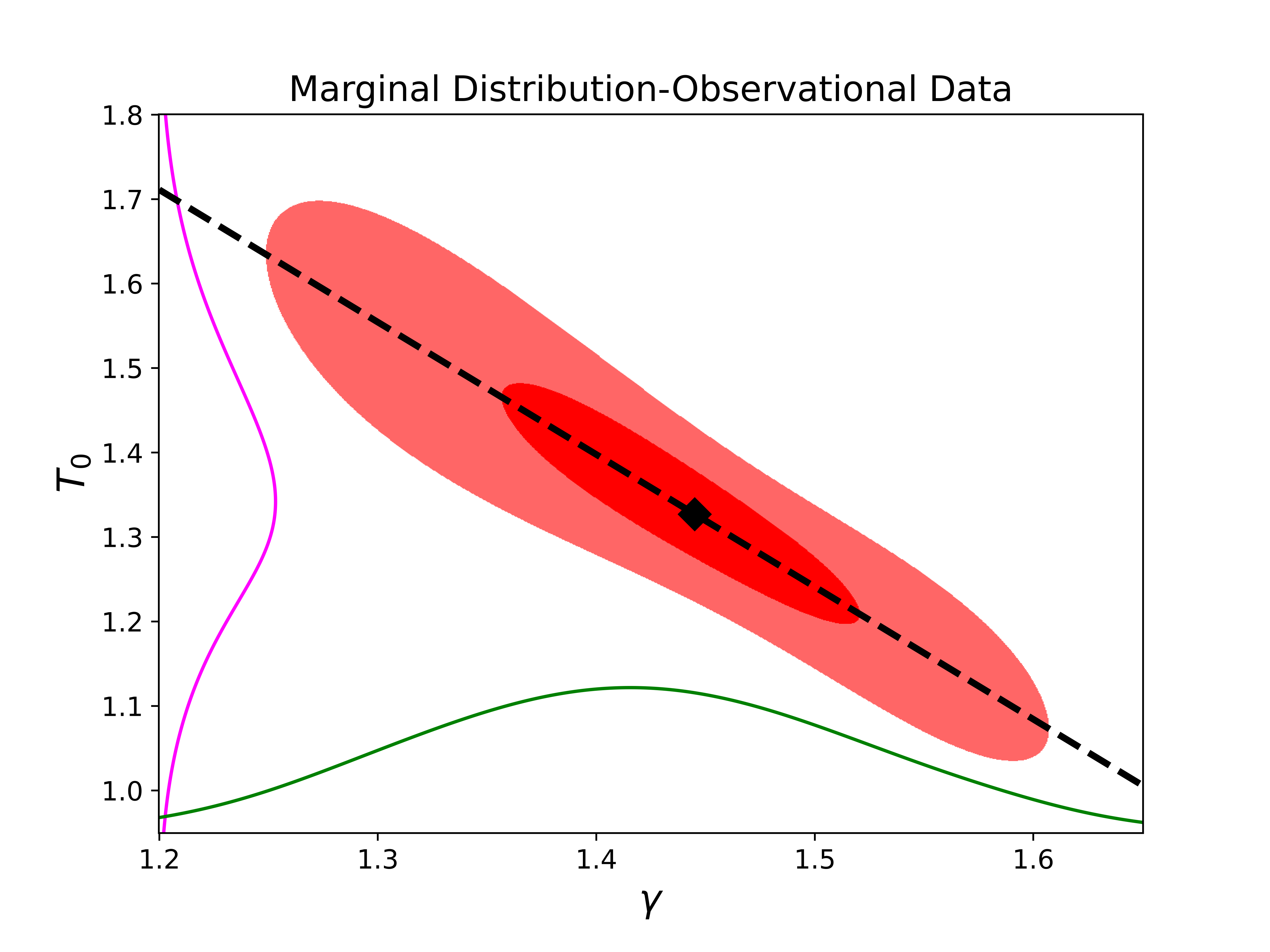}
    \caption{Resulting probability density function $p(\gamma, T_0)$. The black diamond indicates the maximum probability. The $68\%$ confidence level is plotted in dark red, the $95\%$ confidence level in light red. On the axes we plot the arbitrarily normalized marginal distribution for $T_0$ (magenta) and $\gamma$ (green). The black dotted line indicates the linear fit from Fig. \ref{fig: joint_estimate}. The results were obtained by combining the estimates with the RPC\_IRGN method and with the IRGN\_$\chi^2$ method applied to a subsample UVES\_SQUAD survey \citep{Murphy2019}.}
    \label{fig: marginal}
\end{figure}

The estimates with the RPC\_IRGN method (green shaded in Fig. \ref{fig: joint_estimate}) and the IRGN\_$\chi^2$ method (magenta shaded in Fig. \ref{fig: joint_estimate}) can be combined to find joint estimates for $\gamma$ and $T_0$. The temperature estimates (black diamonds) and the corresponding errors (black errorbars) for every of the seven choices of $\gamma$ shown in Fig. \ref{fig: joint_estimate} are the mean and the standard deviation of the observed conditional probability distributions of the temperature given a certain $\gamma$. On a finer grid, we draw this conditional probability for any value of $\gamma$ from our linear fit: The mean is given by the fit value at that specific $\gamma$ and the standard deviation is given by the error in the linear fit predictive distribution, i.e. the vertical width of the magenta shaded region in Fig. \ref{fig: joint_estimate}. The joint probability density distribution is computed numerically by multiplying this conditional distribution with the marginal probability distribution for $\gamma$ estimated with the RPC\_IRGN method (green). The joint probability density function $p(\gamma, T_0)$ is plotted in Fig. \ref{fig: marginal}. We also sketch the marginal distributions for $T_0$ and $\gamma$ in Fig. \ref{fig: marginal}. These two plots present our final estimation results. The marginal distribution $p(T)$ differs slightly from a Gaussian as it is not fully symmetric around the maximum of the distribution. We estimate $T_0$ by the maximum of the distribution and estimate the error towards larger and towards smaller temperatures by the temperature at which the distribution drops below $\exp{(-0.5)} \approx 0.6$, i.e. by fitting a Gaussian to each flank of the marginal distribution. We find for the marginalized $T_0$ distribution the estimate $T_0 = 13400^{+1700}_{-1300}\,\text{K}$.

As pointed out in Sec. \ref{sec: method} we now use this estimate for the temperature at mean density $T_0$ and our estimate for the mean neutral hydrogen density to constrain the photoionization rate $\Gamma$ by Eq. \eqref{eq: mean_dens}. We get $\Gamma_{-12} = 1.1^{+0.16}_{-0.17}$.

\section{Mock Estimation} \label{sec: mock}
We demonstrated in Sec. \ref{sec: synthetic} that our estimation procedure works on high signal-to-noise synthetic spectra. However, for the analysis in Sec. \ref{sec: synthetic} we assumed no systematic uncertainties. In this Section we create mock data with exactly the same parameters ($\gamma=1.42$, $T_0=13400\,\mathrm{K}$, $\Gamma_{-12}=1.1$) and same spectral properties ($SNR, R$) as they were estimated in Sec. \ref{sec: application}. In this simulation we create synthetic spectra which are distorted by peculiar velocities. We exactly mimic the steps that we did in Sec. \ref{sec: application} to perform a consistency check of our estimation and to verify our analysis, in particular the slope of the linear fit, on mock data similar to the observational data set. 

\begin{figure}
    \centering
    \includegraphics[width=0.5 \textwidth]{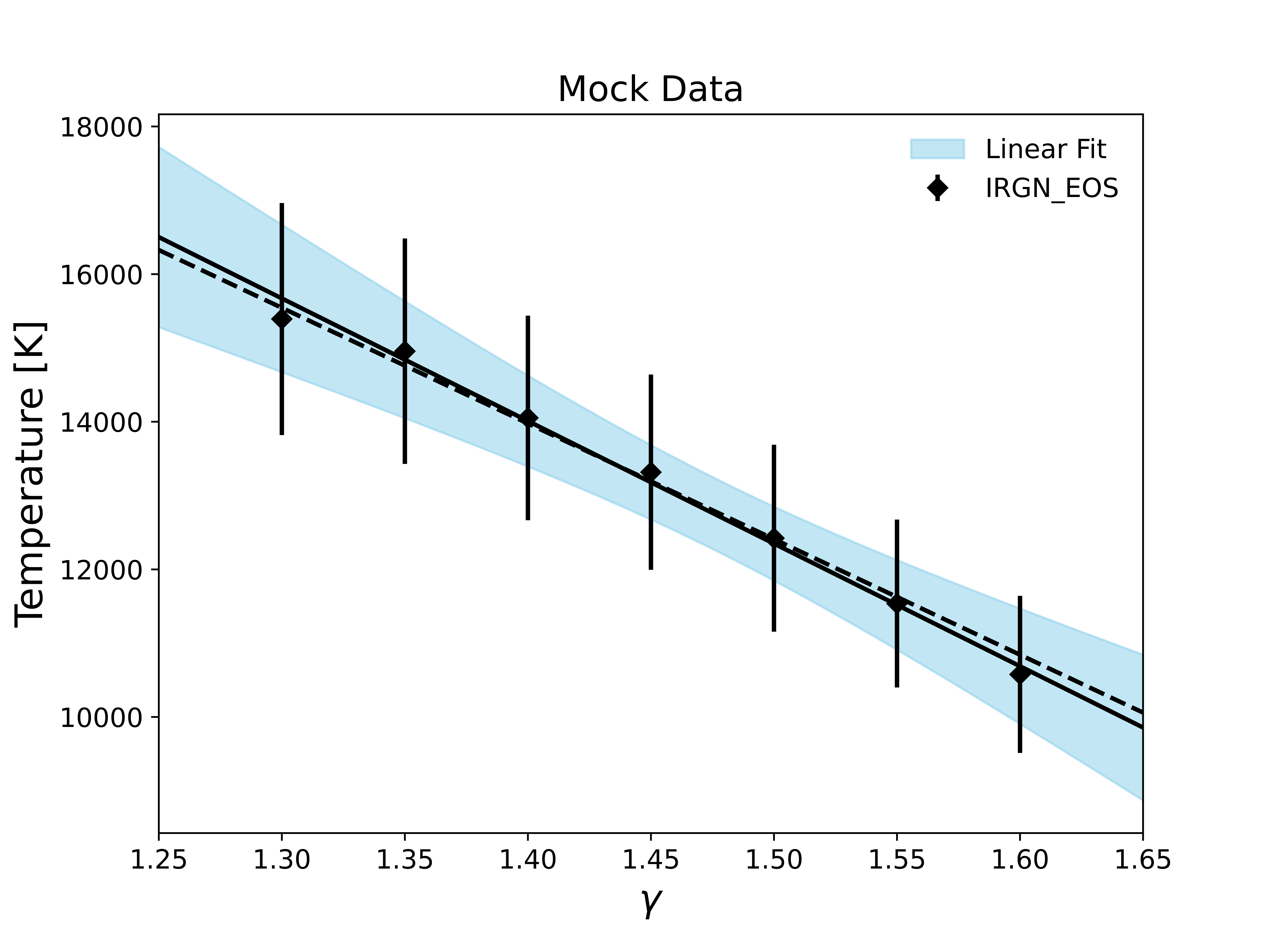}
    \caption{Same as in Fig. \ref{fig: joint_estimate}, but for an estimation from synthetic mock data. The dotted black line indicates the linear fit from Fig. \ref{fig: joint_estimate}, the solid black line shows the linear fit to the plotted data points on mock data. Both linear fits coincide indicating consistency of the estimation procedure.}
    \label{fig: mock}
\end{figure}

The fitting results with the IRGN\_$\chi^2$ algorithm are shown in Fig. \ref{fig: mock} in a similar way as in Fig. \ref{fig: joint_estimate}.We also add the same systematic uncertainties as in the application of real data. The linear relation that we expected from synthetic data and that we observed in the observational data set is again visible. By comparing Fig. \ref{fig: mock} and Fig. \ref{fig: joint_estimate} it becomes obvious that the estimates of the temperature for single values of $\gamma$ is more precise for our mock data sets (the error bars are significantly smaller). The dashed line in Fig. \ref{fig: mock} shows the linear fit from Fig. \ref{fig: joint_estimate}. The dashed line lies inside the predictive distribution (blue shaded) from the linear fit from our mock data. This shows the consistency of our linear fit. The linear fit to observational data and the linear fit to our mock data  coincide. In particular we estimate $T_0 = 13600^{+600}_{-500}\,\text{K}$ from our mock data for the artificial $\gamma=1.42$.

All in all we underestimate the true $\gamma=1.42$. We estimate $\gamma = 1.33$ from our data set distorted with peculiar velocities. However, this underestimation matches the bias $\delta_\gamma$ obtained in Sec. \ref{sec: pec_vel} and which was added to the systematic errors in the estimation of $\gamma$. Hence, the exact value still lies inside the error interval for this mock estimation. We repeat our mock estimation with the same spectra, but not distorted with peculiar velocities. We observe $\gamma=1.44$. This confirms again the accuracy of our method.

The key feature in the precision of our study is the fact that we were able to establish a linear relation between $\gamma$ and $T_0$ with small errorbars. Moreover, we are able to model this relation by simulations (Fig. \ref{fig: mock}). This linear relation might be a general property of the Ly$\alpha$ forest, i.e. the slope and the offset of the linear function might be independent of $T_0$ and $\gamma$ and varying only mildly with redshift. If that would be the case, then the number of degrees of freedom in the estimation of the thermal history of the IGM would be reduced by one. Future studies should examine this point further and extend it to other redshifts. Probably the estimates by the IRGN\_$\chi^2$ method could be combined with classical estimates by Voigt-profile fitting to yield stronger constraints by inserting a strong correlation between the estimated values for $\gamma$ and $T_0$.

\section{Discussion} \label{sec: discussion}
\subsection{Comparison With Recent Estimates} \label{sec: comparison}
\begin{figure}
    \centering
    \includegraphics[width=0.5 \textwidth]{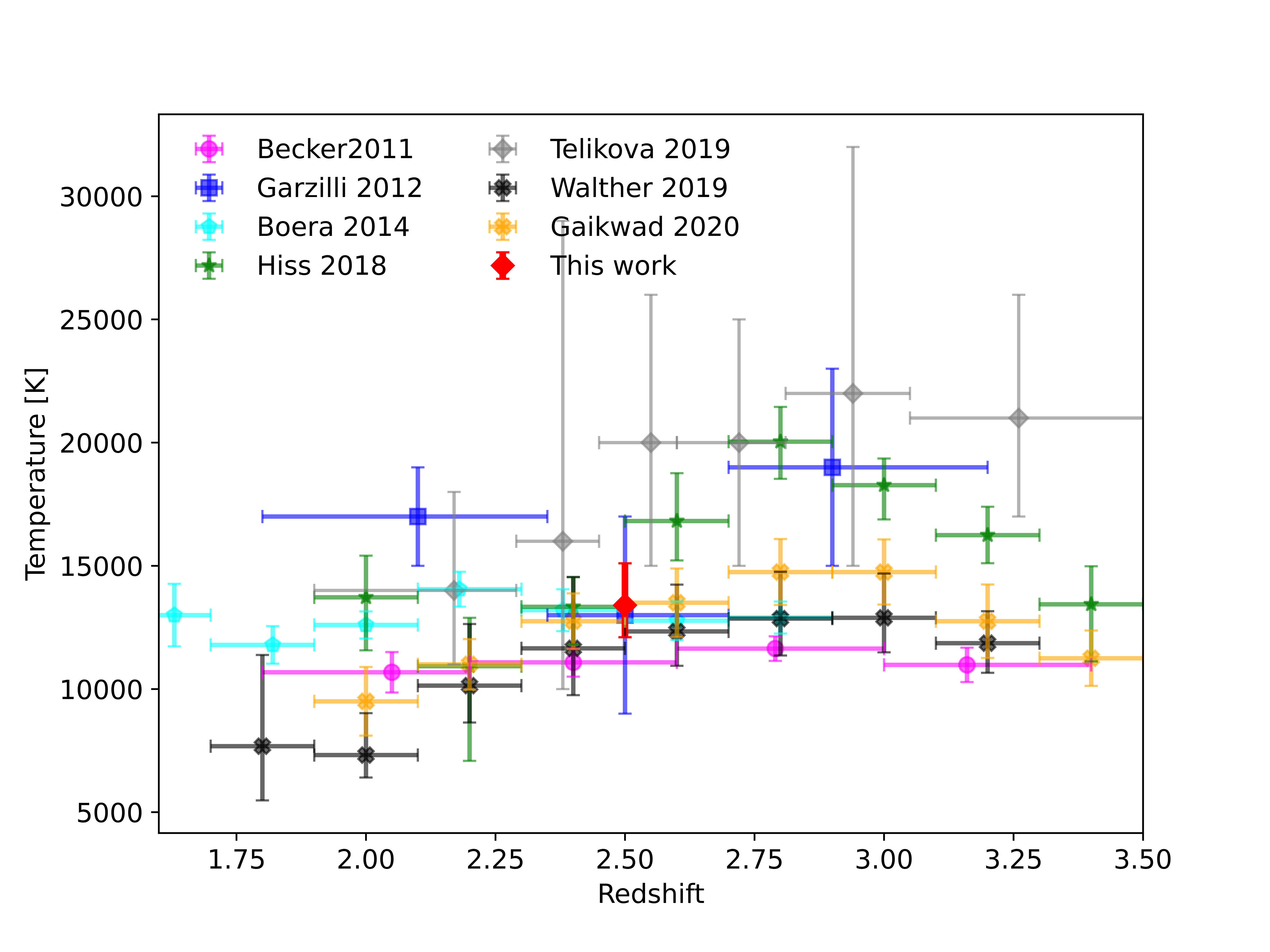}
    \caption{Comparison of recent estimates for $T_0$. We show our estimate with a red diamond and plot the $1-\sigma$ errorbounds of our estimate. Other data points (with $1-\sigma$ errorbounds) are taken from \citet{Becker2011} (magenta circles, a fiducial $\gamma \sim 1.5$ assumed), \citet{Garzilli2012} (blue boxes), \citet{Boera2014} (cyan points, fiducial $\gamma \sim 1.5$ assumed), \citet{Hiss2018} (green stars), \citet{Telikova2019} (grey diamonds), \citet{Walther2019} (black boxes, strong prior results) and \citet{Gaikwad2020b} (orange boxes).}
    \label{fig: T0_results}
\end{figure}

\begin{figure}
    \centering
    \includegraphics[width=0.5 \textwidth]{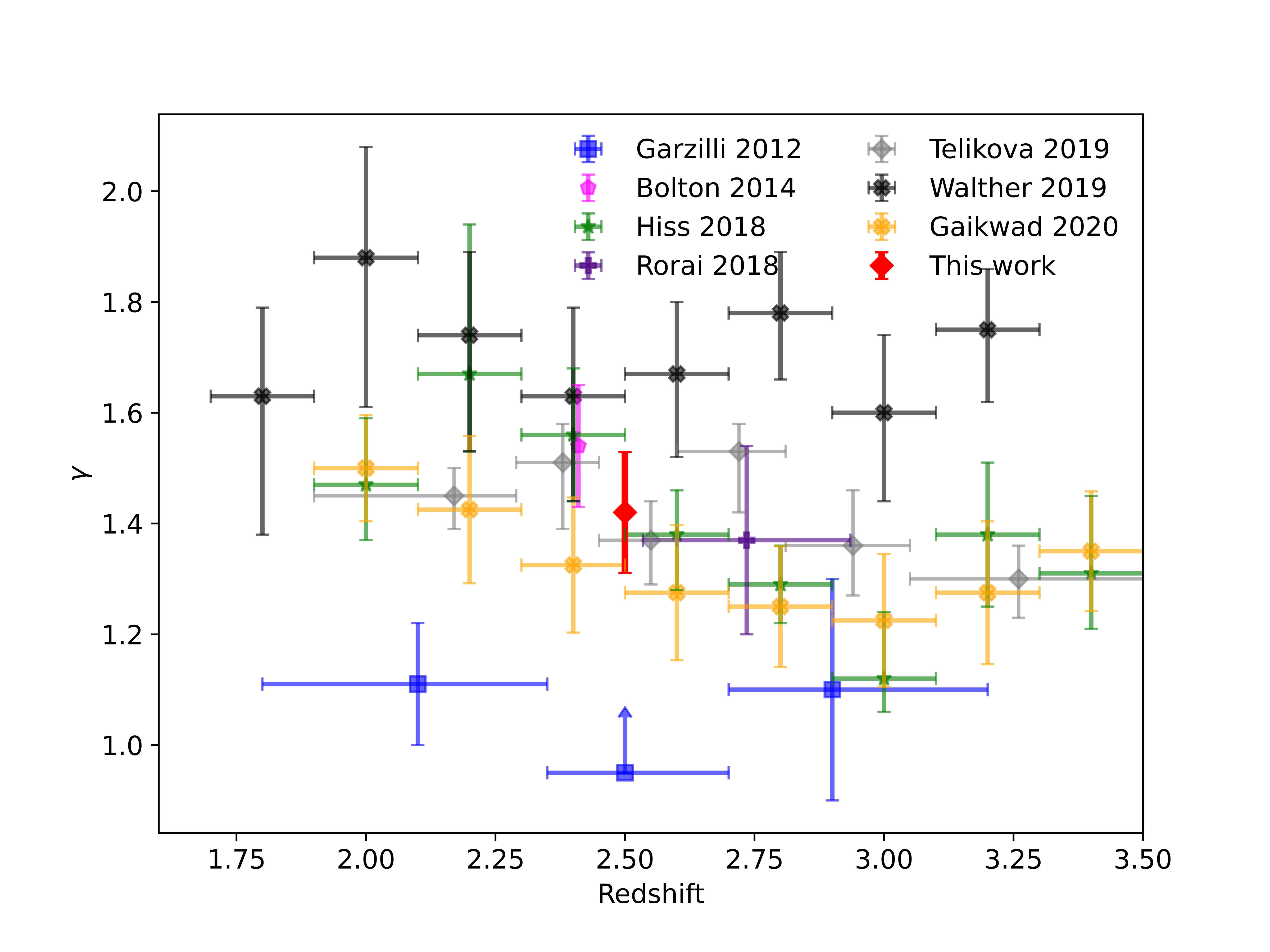}
    \caption{Comparison of recent estimates for $\gamma$. Additionally to the sources in Fig. \ref{fig: T0_results} we plot the results by \citet{Bolton2014} (magenta hexagon) and \citet{Rorai2018} (indigo plus).}
    \label{fig: gamma_results}
\end{figure}

\begin{figure}
    \centering
    \includegraphics[width=0.5 \textwidth]{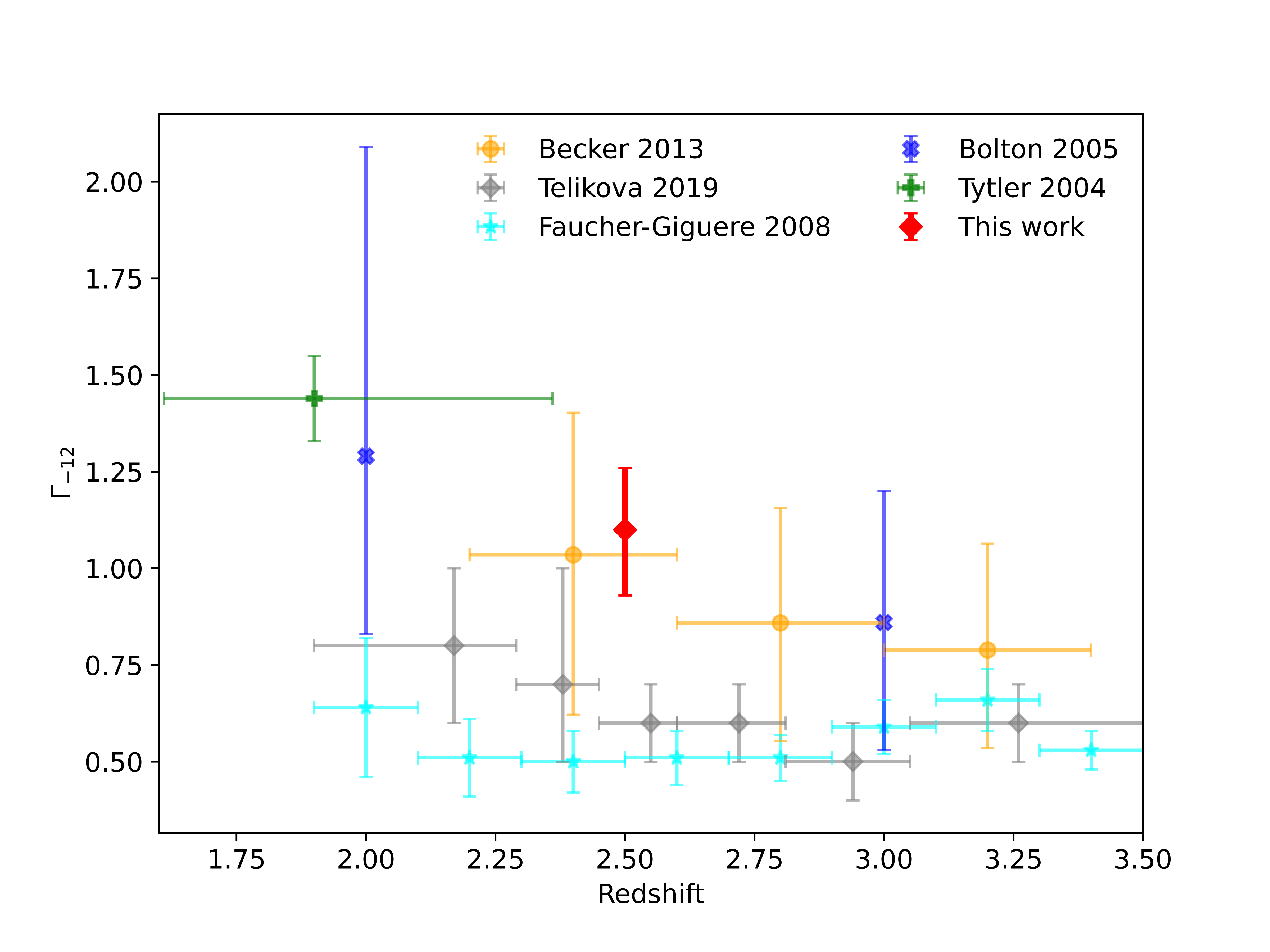}
    \caption{Comparison of recent estimates of the photoionization rate $\Gamma_{-12}$. Our estimate is plotted with a red diamond. The other data points are taken from \citet{Becker2013} (orange circles), \citet{Telikova2019} (grey diamonds), \citet{Faucher2008} (cyan stars), \citet{Bolton2005} (blue crosses) and \citet{Tytler2004} (green diamond).}
    \label{fig: phot_results}
\end{figure}

We compare our estimates with recent estimates for the temperature at mean density in Fig. \ref{fig: T0_results}, with recent estimates of the slope parameter $\gamma$ in Fig. \ref{fig: gamma_results} and with estimates of the photoionization rate in Fig. \ref{fig: phot_results}. 

We predict a slightly larger temperature than in the current study by \citet{Walther2019}, but our estimate for the temperature coincides very well with the measurements by \citet{Garzilli2012}, \citet{Boera2014}, \citet{Hiss2018} and \citet{Gaikwad2020b}. Our results are comparably precise to existing estimates by \citet{Hiss2018} and \citet{Walther2019}. However, in compatible high precision studies such as \citet{Becker2011}, \citet{Boera2014}, \citet{Hiss2018}, \citet{Walther2019} and \citet{Gaikwad2020b} the precision in the temperature estimate comes at the cost of a large uncertainty in redshift. This is caused by accounting for redshift bins of width $\Delta z \sim 0.1-0.2$ to create an artificial sample of absorption lines. However, recent estimates suggest that $T_0$ might vary over the width of this bin (e.g. see the bin-to-bin variations in the estimates represented in Fig. \ref{fig: T0_results}). Our estimate is free of such uncertainties, because our redshift bin is 10 times narrower (this is discussed further below). 

Moreover, our estimate for the slope parameter $\gamma$ of the temperature-density relation matches well the observations by \cite{Hiss2018} and \citet{Telikova2019} and is compatible with the lower limit by \citet{Garzilli2012}. However, \citet{Walther2019} found larger values for the slope and \citet{Gaikwad2020b} found slightly smaller slopes. Again our estimate is competitive without suffering from averaging in the redshift domain.

Lastly, we compare our estimate for the photoionization rate at redshift $z=2.5$ with the observations by \citet{Tytler2004}, \citet{Bolton2005}, \citet{Faucher2008}, \citet{Becker2013} and \citet{Telikova2019} in Fig. \ref{fig: phot_results}. Our result coincides best with the observation by \citet{Becker2013} and \citet{Bolton2005}. However, there is some scatter between the different measurements visible. The error in $\Gamma_{12}$ in our analysis is dominated by the uncertainty in the estimation of the neutral hydrogen density at mean baryonic density. Thus, we cannot improve the significance of the estimate as we did for the temperature at mean density. In particular, our estimate favors a large photoionization rate of $\Gamma_{12} \approx 1$.

\subsection{Discussion of increased Precision}
We increase the precision of the estimation of $T_0$ by our method. Note that we only used a $40\,\mathring{A}$ portion of each spectrum (such that there is only a negligible redshift error introduced) and gained estimates with an error of only approximately $\approx 1000\,\mathrm{K}$, which is competitive to current measurements drawn from much greater redshift bins (and thus a greater sample of absorption lines). We explain this improved precision in the estimation of $T_0$ using our method by the following three assertions:
\begin{itemize}
    \item In comparison to simple Voigt-profile fitting we use a complete fitting approach which uses the explicit formulation of the forward model. On the one hand, this fitting increases the coverage of fitted wavelength (we do not fit only absorption lines, but every feature in the spectrum) and enables us to correctly take into account the profile of the underlying neutral hydrogen density. On the other hand, this procedure is much more computationally expensive because finding suitable inversion parameters (regularization parameter, stopping criterion) can be difficult.
    \item Among fitting of the absorption profiles, for the reconstruction with the RPC method and with the IRGN algorithm we also assume some prior distribution for matter. In comparison to purely fitting tools we therefore include much more prior knowledge in order to infer the temperature-density distribution. Additionally the approach of treating the set $[\hat{n}_\mathrm{HI}, \gamma, T_0]$ instead of the set of parameters $[\Gamma, \gamma, T_0]$ allows us to disentangle the degeneracy between $\gamma$ and $T_0$, i.e. we were able to estimate $\gamma$ solely without $T_0$ with the RPC\_IRGN algorithm.
    \item The statistical analysis reduces the error. There is a reasonable scatter for the temperature estimates for different parameters $\gamma$, but the assumption of a linear relation within the IRGN\_$\chi^2$ method allows us to reduce the error for the predictive distribution by cross correlating with estimates for other temperatures.
\end{itemize}

\subsection{Inverted Equation of State}
\begin{figure}
    \centering
    \includegraphics[width = 0.5 \textwidth]{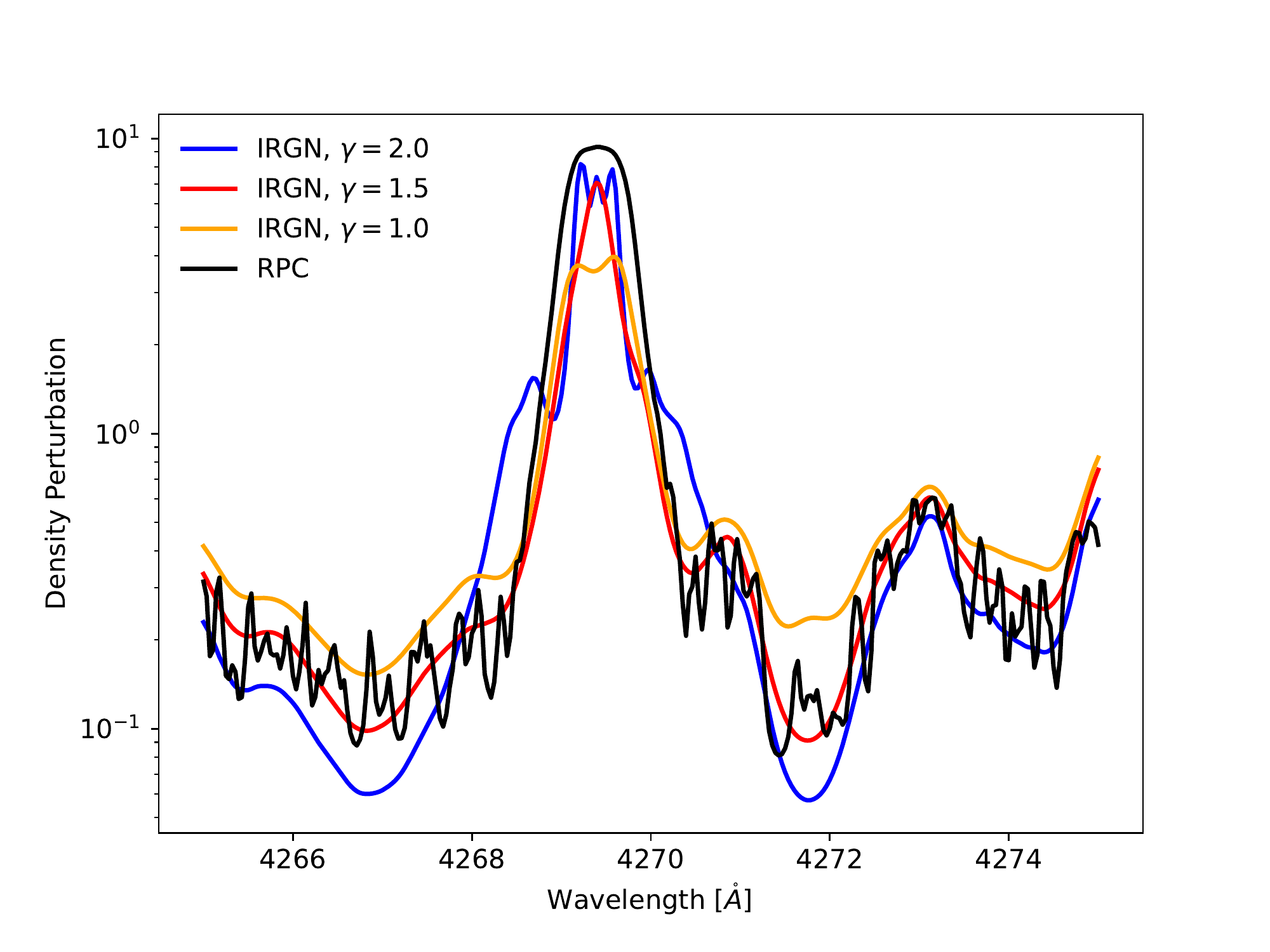}
    \caption{Inversion results for the QSO J101155+294141 with RPC (black line) and IRGN with varying $\gamma$. Only for $\gamma=1.5$ the RPC and the IRGN inversions match on all scales, i.e. the amplitude of the variations in density space match while being overestimated for $\gamma=2$ and underestimated for $\gamma=1$.}
    \label{fig: inverted_eos}
\end{figure}

Some studies reported that the match between observational data and simulations could be improved by assuming $\gamma \approx 1$ or even an inverted EOS ($\gamma < 1$), e.g. \citet{Bolton2008, Furlanetto2009}. While this is within the errorabrs of recent observations at redshift $z \approx 5$ \citep{Boera2019, Gaikwad2020}, measurements at smaller redshifts tend to favor a positive $\gamma$, see Fig. \ref{fig: gamma_results}. This fits the model of an asymptotic temperature-density relation \citep{McQuinn2016}. In fact, our findings do not support an inverted temperature density relation at redshift $z=2.5$. The RPC\_IRGN method shows a distinct minimum at values $\gamma > 1$ for the quasar spectra, compare for example the lower right panel in Fig. \ref{fig: fitting_demonstration}. In particular, for $\gamma = 0.9$ we measure logarithmic distances between the inversion results with the RPC method and with the IRGN method that are a couple of times larger than at the minimal value. The reason for that is that we underestimate the amplitude of variations in the neutral hydrogen density (while letting $\hat{n}_\mathrm{HI}$ unaffected) when assuming small values for $\gamma$ in the inversion procedure. An example of this effect on the inversion procedure is shown for QSO J101155+294141 in Fig. \ref{fig: inverted_eos}. Only for $\gamma \approx 1.5$ (red line) does the recovered density field span over the whole range of recovered overdensities with the RPC method (black line). At larger $\gamma$ (blue line) the range of possible values, i.e. the amplitude of density fluctuations, is overestimated, at smaller $\gamma$ (orange line) the range of possible values is underestimated. Overall our measurements do not support an inverted EOS.

\subsection{Outlook}
Our estimation method suffers from the major drawback that peculiar velocities are not included in the inversion algorithms along single lines of sight. This issue is common to most fitting algorithms used to recover the thermal history of the IGM. The unknown peculiar velocities introduce the largest error on $\gamma$ and thus induce the largest error in the estimation of $T_0$. However, if the peculiar velocities would be known a priori, they could be inserted in the IRGN inversion straight forwardly.

Although \citet{Pichon2001} studied a method to estimate the velocity field along a single line of sight in parallel, we were not able to reproduce these results. Gas streams towards large overdensities are possibly not detectable in the density along the line of sight if the overdensity is located transverse to the line of sight. In such cases the tangential projection of the stream leads to a redshift distortion which cannot be identified with an overdensity in the line of sight. However, these overdensities would be visible in neighboring lines of sight. Thus, the velocity field could be approximated by three dimensional Ly$\alpha$ forest tomography from a field of quasar spectra with small separation length, such as the COSMOS field \citep{Lee2018, Ata2021}. There are many recent approaches in recovering a three dimensional maps at low spectral resolution from the Ly$\alpha$ forest, among others \citet{Pichon2001, Kitaura2012, Lee2014, Cisewski2014, Stark2015, Lee2016, Lee2016a, Lee2018, Krolewski2018, Horowitz2019, Japelj2019, Porqueres2019, Porqueres2020} which would carry information regarding redshift space distortions.

\cite{Lee2014} discussed observational requirements for such tomographic reconstructions. In a nutshell, at mean separation length of $1\,h^{-1}\,\,\mathrm{Mpc}$ only moderate resolution ($R \approx 1000$) spectra are required. However, to fit the thermal profile of the absorption features in the Ly$\alpha$ forest by tomographic inversion we require high resolution spectra ($R \gtrsim 50000$). We observe that the velocity field is typically only varying on larger bins. We therefore propose to observe a field of close quasars with small transverse separation length at small resolution, estimate the peculiar velocity field from these observations and then subsequently observe one quasar at the center of the field with very high resolution. Then we propose to use the recovered redshift space distortions as approximation to the real distortions and perform the inversions with respect to these distortions. We expect that with this strategy one would be able to infer the temperature with some $100\,\text{K}$ precision using our estimation procedure.

\section{Conclusion}
We presented a novel estimation procedure to study the EOS of the IGM by Ly$\alpha$ forest tomography and applied this approach to a sample of 21 QSO spectra from the UVES\_SQUAD sample. In particular, we observed $T_0 = 13400^{+1700}_{-1300}\,\mathrm{K}$, $\gamma = 1.42 \pm 0.11$ and $\Gamma_{-12} = 1.1^{+0.16}_{-0.17}$ at redshift $z=2.5$. The corresponding (marginal) probability distributions are shown in Fig. \ref{fig: joint_estimate} and Fig. \ref{fig: marginal} Our estimates are more precise than existing estimates while drawn from significantly smaller redshift bins.

In contrast to classical absorption line fitting approaches we utilize a full inversion of the Ly$\alpha$ forest at high spectral resolution ($\approx 2.5\,\mathrm{km/s}$-bins) to constrain the EOS. In particular, we use the IRGN inversion algorithm (which strongly depends on the chosen model of the IGM) and the RPC method (which depends on less strong priors) to invert the spectra. The quality of the inversion results with the IRGN method depend strongly on whether the correct EOS has been used. We therefore compare in data space the recovered flux with the observed flux (IRGN\_$\chi^2$) and compare in density space the density recovered with the IRGN method and with the RPC method for different choices for the EOS (RPC\_IRGN). We take the set of parameters for the EOS for which the observed flux and the recovered density are described best.

It turns out that the RPC\_IRGN method is (to first order) only sensitive to $\gamma$, while the IRGN\_$\chi^2$ predicts a linear relation between the possible estimates $\gamma$ and $T_0$. Both methods solely are not capable of predicting the EOS due to a degeneracy between $\gamma$ and $T_0$. However, both methods together are complementary and provide unique estimates. The RPC\_IRGN method breaks the degeneracy between $T_0$ and $\gamma$ in the IRGN\_$\chi^2$ method. In principle both methods also could be combined with the results by classical fitting approaches or curvature methods to constrain the EOS.

Our estimates, in particular for the temperature, are more precise than most existing estimates. While sometimes this accuracy is achieved by extending the redshift range, our estimate does not introduce a significant error in redshift. We explain the achieved precision mainly by the fact that we combine several approaches to model the Ly$\alpha$ forest with varying assumptions. Hence, we combine prior knowledge regarding the model of the Ly$\alpha$ forest, the power spectrum, and the probability density distribution of matter overdensities. 

The full power of our method would be achieved if the peculiar velocities along the studied lines of sight could be known at least approximately. This might be possible from recent surveys of QSO spectra with small mean transverse separation length. Our method is expected to provide temperature measurements with a precision of several $100\,\text{K}$ for these data sets.

Nevertheless, suitable observations on which to apply our analysis are rare. They need to have a large spectral resolution and a high signal to noise ratio.

\section*{Acknowledgements}
We thank Simona Gallerani for helpful comments. DJEM is supported by the Alexander von Humboldt Foundation and the German Federal Ministry of Education and Research.

\section*{Data Availability}
The software to reproduce the analysis in this paper will be made publicly available under the url \url{https://github.com/hmuellergoe/reglyman} and upon request. The studied spectra are available in the UVES\_SQUAD survey \citep{Murphy2019}.




\bibliographystyle{mnras}
\bibliography{lymanalphaforest} 



\appendix

\section{Neutral Hydrogen Density} \label{app: neutral_hydrogen_density}
We describe the neutral hydrogen density in the quasi-linear regime of structure formation by the neutral hydrogen density fraction:
\begin{align}
    f_\mathrm{HI} = \frac{n_\mathrm{HI}}{n_\mathrm{b}} = \frac{n_\mathrm{HI}}{n_\mathrm{HI}+n_\mathrm{p}},
\end{align}
with the baryonic number density $n_\mathrm{b}$ and proton number density $n_\mathrm{p}$. Here, the helium mass density is ignored. Moreover, it is in local equilibrium:
\begin{align}
    \alpha n_\mathrm{p} n_\mathrm{e} = \Gamma n_\mathrm{HI},
\end{align}
with recombination rate $\alpha$. 
All together (when assuming that the Helium is fully ionized and that recombination rate and collisional ionization are much smaller than the photo-heating rate), it follows:
\begin{align}
    n_\mathrm{HI} = f_\mathrm{HI} n_\mathrm{b} = \frac{\alpha n_\mathrm{e} n_\mathrm{b}}{\alpha n_\mathrm{e} + \Gamma} \approx \frac{ \alpha n_\mathrm{e} n_\mathrm{b}}{\Gamma} = \frac{\mu_e \alpha n_\mathrm{b}^2}{\Gamma},
\end{align}
where $\mu_e = 2(2-Y)/(4-3Y)$ with helium mass fraction $Y \approx 0.24$ accounts for helium in the IGM. The recombination rate is modeled as a power law \citep{Black1981, Rauch1997} which had been subsequently been used for simulations of the Ly$\alpha$ forest \citep[e.g.][]{Choudhury2001, Gallerani2006}:
\begin{align}
    \alpha = \alpha_0 T^{-0.7},
\end{align}
where the temperature is represented in units of Kelvin and $\alpha_0 = 4.2 \cdot 10^{-10.2}\,\mathrm{cm}^3\mathrm{s}^{-1}$. With the power-law EOS Eq. \eqref{eq: eos} it follows:
\begin{align}
    \alpha = \alpha_0 T_0^{-0.7} \Delta^{0.7-0.7 \gamma}.
\end{align}
Thus:
\begin{align}
    n_\mathrm{HI} = \frac{\mu_e \alpha_0 n_0^2}{\Gamma T_0^{0.7}} \Delta^{2.7-0.7\gamma}, \label{eq: neutral_fraction_app}
\end{align}
where $n_0$ is the mean baryonic density defined by $n_0(z) = \frac{\Omega_\mathrm{b} \rho_\mathrm{c}}{\mu m_\mathrm{p}} (1+z)^3$. Here $\Omega_\mathrm{b}$ is the baryoinc density parameter, $\rho_\mathrm{c}$ the critical density, $\mu$ the mean molecular weight and $m_\mathrm{p}$ the proton mass. Eq. \eqref{eq: neutral_fraction_app} shows Eq. \eqref{eq: neutral_fraction}. We get:
\begin{align}
    \mu_e \alpha_0 n_0^2\approx 1.871 \cdot 10^{-14}\,\mathrm{m}^{-3}\mathrm{s}^{-1} \cdot 
\end{align}
at redshift $z=2.5$ which proves Eq. \eqref{eq: mean_dens}.

\section{Inversion Schemes} \label{app: inversion_schemes}
The Inversion schemes IRGN and RPC were intensively studied for their numerical performance, their accuracy and their resilience against systematic and statistical uncertainties in the publication \cite{Mueller2020}. We refer the interested reader to this publication. In a nutshell, IRGN is the most accurate inversion scheme and is robust against statistical noise. But it also requires the strongest assumptions which makes it vulnerable to misfitting the forward operator, e.g. by wrong estimates for $T_0$ and $\gamma$. The RPC algorithm on the other hand is slightly more inaccurate, but depends on weaker priors (i.e. not on $T_0$ and $\gamma$) and is much faster. A strong advantage that RPC has in common with the IRGN approach is its robustness against noise.

\begin{figure}
    \centering
    \includegraphics[width = 0.5\textwidth]{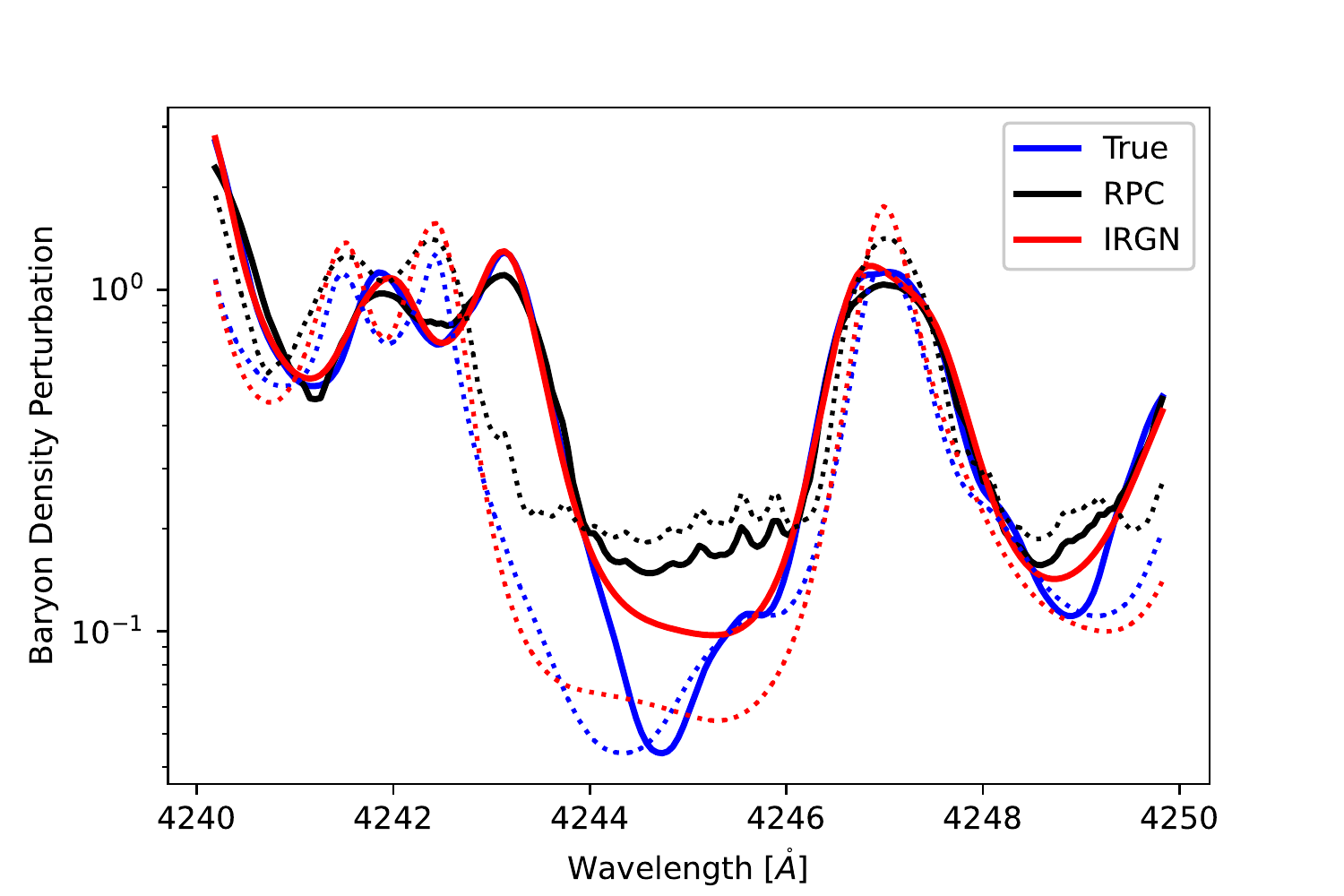}
    \caption{Inversion with the IRGN and with the RPC methods along a single line of sight on synthetic data. Dotted lines show the reconstruction from data which are distorted by peculiar velocities, the blue dotted line indicates the true density perturbation shifted to redshift space with peculiar velocities.}
    \label{fig: inv_schemes}
\end{figure}

We show in Fig. \ref{fig: inv_schemes} an exemplary inversion along a synthetic line of sight (spectral resolution $R=50000$, signal-to-noise ratio $SNR = 30$). Fig. \ref{fig: inv_schemes} indicates that while the inversion schemes succesfully recover the true density profile, they fail to recover very small densities, e.g. overdensities $\Delta \lesssim 0.2$. The same applies for very large densities due to line saturation. Thus, comparisons between reconstructions with RPC and with IRGN should be only done for moderate overdensities, e.g for $0.2 \lesssim \Delta \lesssim 2$. However, the majority of pixels in the spectra are in this regime.

\begin{figure*}
    \centering
    \subfigure[Density]{
    \includegraphics[width=0.4 \textwidth]{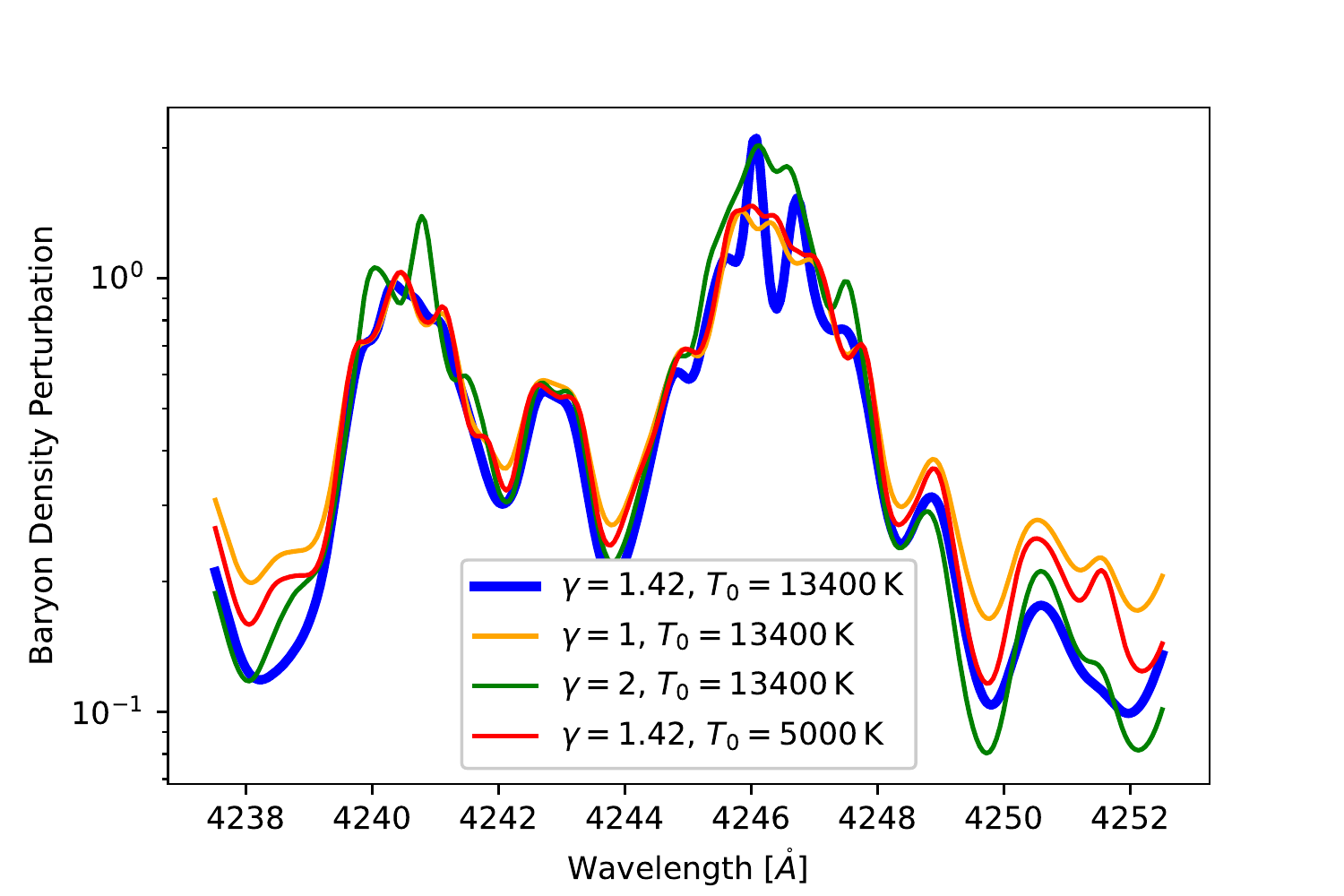}}
    \subfigure[Flux]{
    \includegraphics[width=0.4 \textwidth]{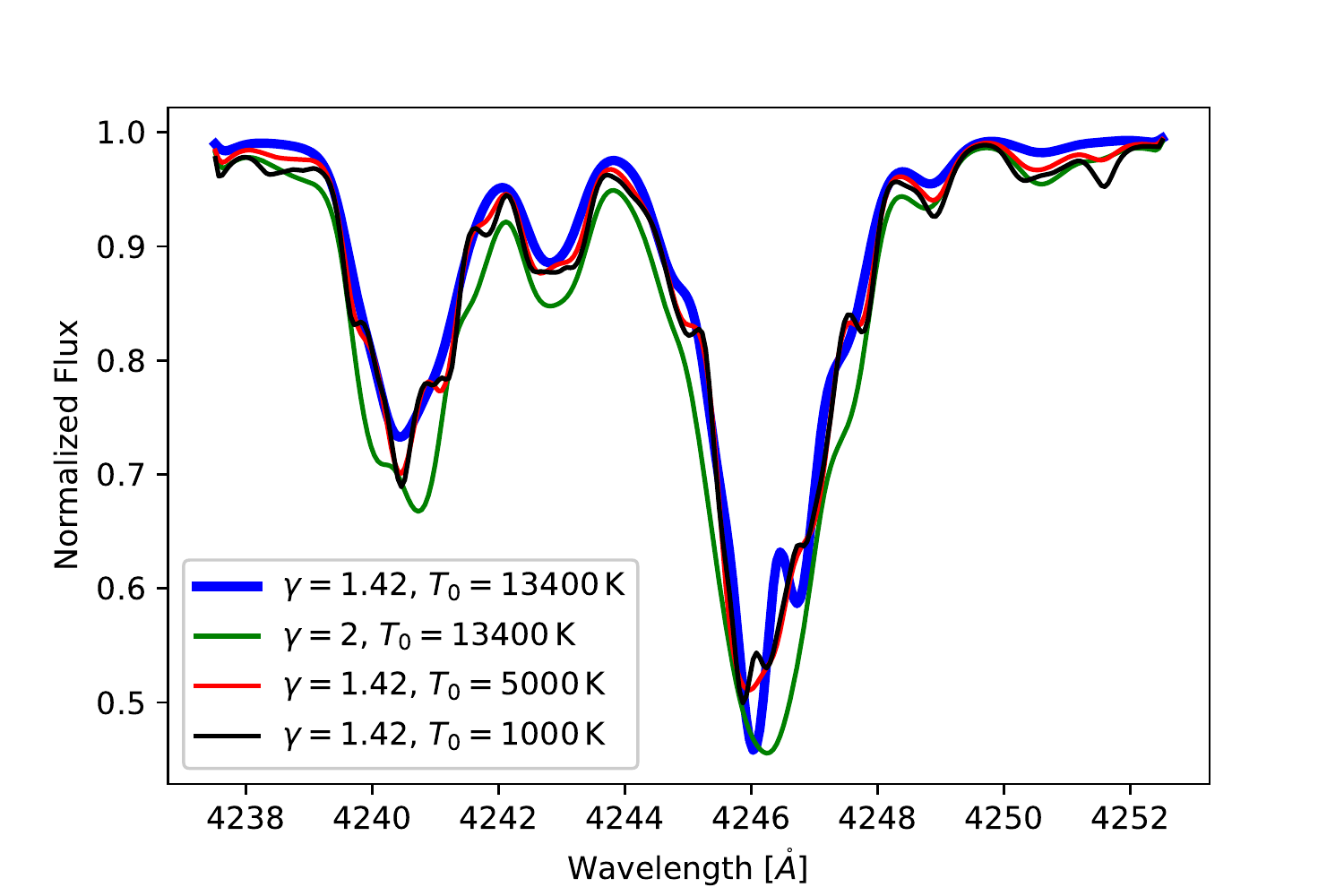}}
    \caption{Inversion results with the IRGN algorithm with different artificial parameter choices $\gamma$ and $T_0$. The correct values are $\gamma = 1.42$ and $T_0 = 13400\,\mathrm{K}$.}
    \label{fig: inv_schemes_gamma_temp}
\end{figure*}

We demonstrate in Fig \ref{fig: inv_schemes_gamma_temp} that the IRGN reconstruction depends strongly on the parameters $\gamma$ and $T_0$. This strong dependence will be the basis of the RPC\_IRGN approach. For demonstration purposes we show a different line of sight than in Fig. \ref{fig: inv_schemes} and reduce the signal to noise ratio to $SNR = 10$ to highlight how different assumptions of $\gamma$ and $T_0$ affect denoising in the recovered flux (and hence the reduced $\chi^2$). In panel (a) we show the reconstruction results of the baryonic density perturbation. The recovered density clearly depends on the chosen value for $\gamma$, but less on the choice of $T_0$. This leads to the fact that RPC\_IRGN will be to first order only sensitive to $\gamma$, compare Sec. \ref{sec: estimation} and in particular Fig. \ref{fig: rpc_eos}. However, the quality of the flux reconstruction depends also on $T_0$ as visible from the red and the black line in panel (b) of Fig. \ref{fig: inv_schemes_gamma_temp}. When the assumed temperature goes down the recovered flux starts to oscillate on small scales while fitting the small scale noise contribution. This leads to a reduced $\chi^2$ which is smaller than one. This is the basic idea of the IRGN\_$\chi^2$ algorithm \citep{Rollinde2001}.

Peculiar velocities have a great impact on the reconstruction. We demonstrate in Fig. \ref{fig: inv_schemes} with dotted lines the inversion results with the RPC method and with the IRGN method from spectral data that are distorted by peculiar velocities. Peculiar velocities shift the density in redshift frame, but they also affect width and strength of the absorption peaks in the spectrum. Fig. \ref{fig: inv_schemes} demonstrates some additional discrepancy between RPC and IRGN inversion due to peculiar velocities which will result in a bias in the estimation of $\gamma$, see Sec. \ref{sec: pec_vel}. Fig. \ref{fig: inv_schemes} reinforces the assertion that very small underdensities ($\Delta \lesssim 0.2$) and densities at the boundary of the computation box should be excluded from the RPC\_IRGN scheme. However, the reconstruction of moderate overdensities (e.g. see the density peaks at around $4242\,\mathring{A}$ and at around $4247\,\mathring{A}$) stays robust).

\section{Fitting Procedure} \label{app: fitting_procedure}
We discuss now the IRGN\_$\chi^2$ step in the procedure presented above (find the parameters for which $\chi^2 \approx 1$). 

When the temperature goes to zero the reduced $\chi^2$ typically converges to a constant lower limit $\chi^2 \approx 0.85$. This is expected. The Gaussian kernel in \eqref{eq: tau} turns into a $\delta$ distribution for very small temperatures. The forward operator turns into a one-to-one correspondence between the optical depth and the neutral hydrogen density as described by the fluctuating Gunn-Peterson approximation (Eq. \eqref{eq: gunn_peterson}):
\begin{align}
    \tau(z) \propto n_\mathrm{HI}(z).
\end{align}
Hence, if the prior in Eq. \eqref{eq: posteriori} would be absent, we would be able to exactly (over-)fit the noisy data, i.e. to achieve $\chi^2 = 0$. But as the second term in Eq. \eqref{eq: posteriori} is present, i.e. we assume a specific auto-correlation as prior, for very small temperatures the IRGN method computes a best estimator for the optical depth which satisfies the prior covariance $\bm{\mathsf{C}}_0$. Thus, we compute a smooth curve fit to the noisy observed data. Therefore, the reduced $\chi^2$ converges for small temperatures to a lower limit which is unequal to zero. The exact value of this lower limit depends on the prior covariance, $\bm{\mathsf{C}}_0$, the noise distribution, $\bm{\mathsf{C}}_d$ and the explicit choice of the forward model, i.e. the amount of Gaussian blurring due to thermal broadening.

\begin{figure}
    \centering
    \includegraphics[width=0.5 \textwidth]{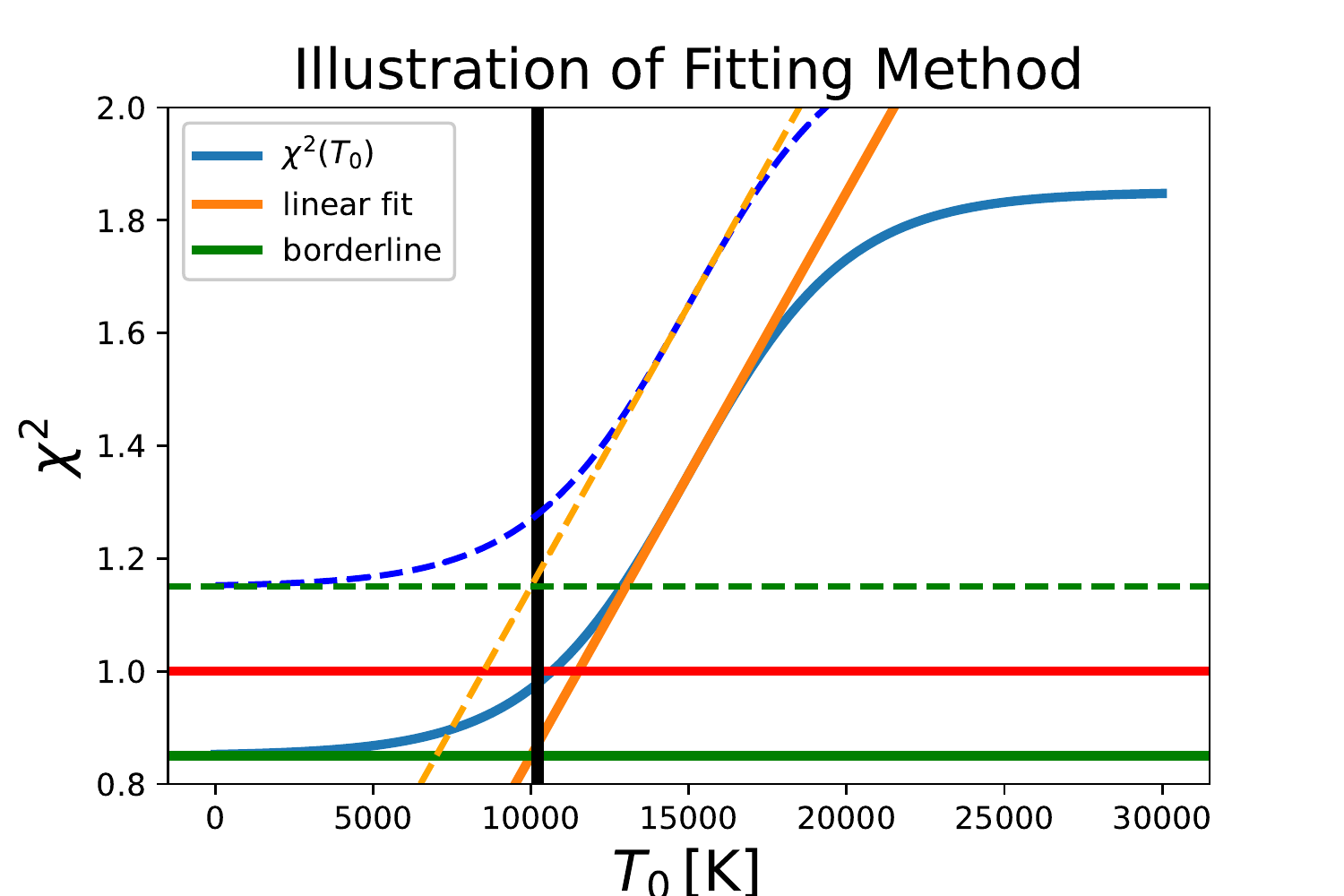}
    \caption{Schematic illustration of the fitting procedure used to find the temperature $T_0$ at which $\chi^2=1$ (red horizontal line). $\chi^2(T_0)$ (blue line) typically converges to a lower limit for small temperatures (green line). For slightly greater temperatures $\chi^2$ could be estimated well by a linear fit (orange line). The intersection between the linear fit and lower limit borderline roughly matches the exact value (black line). The dashed curves represent a case where $\chi^2$ is overestimated. Our method, however, still recovers the correct value of $T_0$.}
    \label{fig: fitting_procedure}
\end{figure}

For temperatures slightly bigger than the exact temperature, the reduced $\chi^2$ drops nearly linearly. The exact value for the temperature lies in between these two regimes: the regime of linear decay at slightly higher temperatures and the regime of a constant lower limit at slightly smaller temperatures (see the illustration in Fig. \ref{fig: fitting_procedure}). To find a proper estimate for the temperature we fit the borderline, the asymptotic lower limit (green line in Fig. \ref{fig: fitting_procedure}) and the linear regime (yellow line in Fig. \ref{fig: fitting_procedure}) separately. We take the intersection between these two lines as estimator for the temperature.

This procedure has the special advantage that it is unaffected by shifting the reduced $\chi^2$ for one of the reasons stated above (e.g. in the situation that a small fraction of the line of sight is badly estimated). This is also demonstrated in Fig. \ref{fig: fitting_procedure} by dashed lines. Due to the constant shift there is no intersection with the $\chi^2=1$ line anymore. But as the fitted borderline and also the linear fit function are shifted in the same manner, the estimate with our fitting procedure does not change. 

We study the precision of this fitting procedure on synthetic data in Sec. \ref{sec: synthetic_fitting}. 

\section{Lognormal Model} \label{app: just_sim}
We justify in this Section the use of the lognormal model to analytically describe the quasi-linear density distribution (see Sec. \ref{sec: rpc_inversion}) and to simulate the transmitted flux in the Ly$\alpha$ forest (see Sec. \ref{sec: synthetic_data}).

The lognormal model is well motivated \citep{Coles1991, Bi1992, Bi1997} and has been frequently used to describe the distribution of overdensities. In fact the model was used for modelling the cosmic density field \citep[e.g.][]{Choudhury2001, Viel2002, Gallerani2006, McDonald2006, Ribera2012, Hand2018, Karacayli2020} and for tomographic reconstruction \citep{Pichon2001, Kitaura2008, Caucci2008, Kitaura2012, Ata2015, Mueller2020} of the density field with the Ly$\alpha$ forest. In particular, the \small{NBODYKIT} toolbox \citep{Hand2018} which implements the lognormal model is currently used in a large number of publications as a reference model of the cosmic density field.

\begin{figure}
    \centering
    \includegraphics[width=0.5\textwidth]{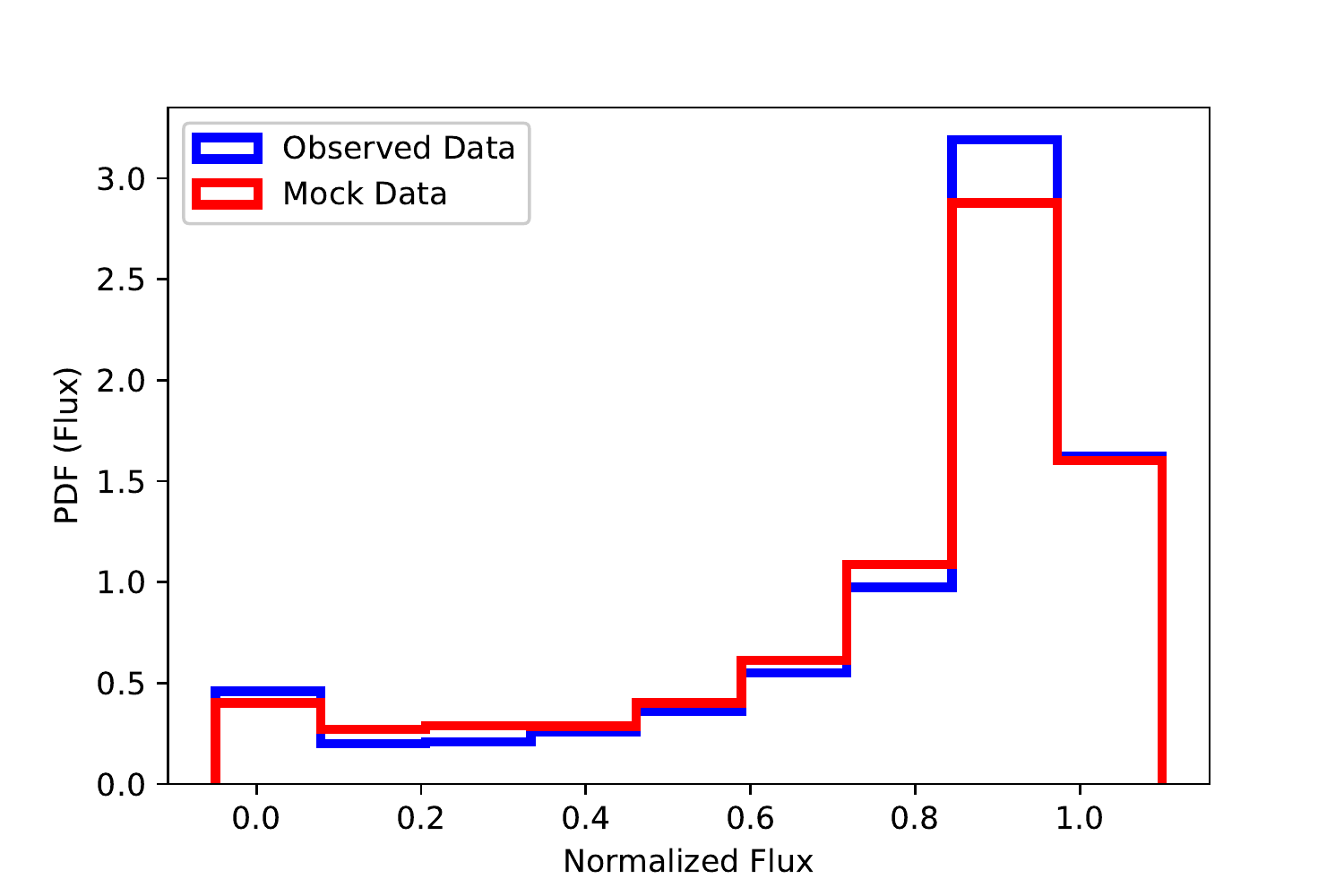}
    \caption{Histograms (normalized) of the flux density for the observational data sample (blue) and the corresponding mock data sample from Sec. \ref{sec: mock}(red).}
    \label{fig: mock_observed}
\end{figure}

We check that the simulation statistically reproduces the structure of real spectra. We compare the probability density distribution of the flux of observed spectra (see Sec. \ref{sec: data}) with our mock data set which mimics the spectral properties of the observed spectra (see Sec. \ref{sec: mock}) in Fig. \ref{fig: mock_observed}. The simulation probability density distribution matches the observational one sufficiently well.

\begin{figure}
    \centering
    \includegraphics[width=0.5\textwidth]{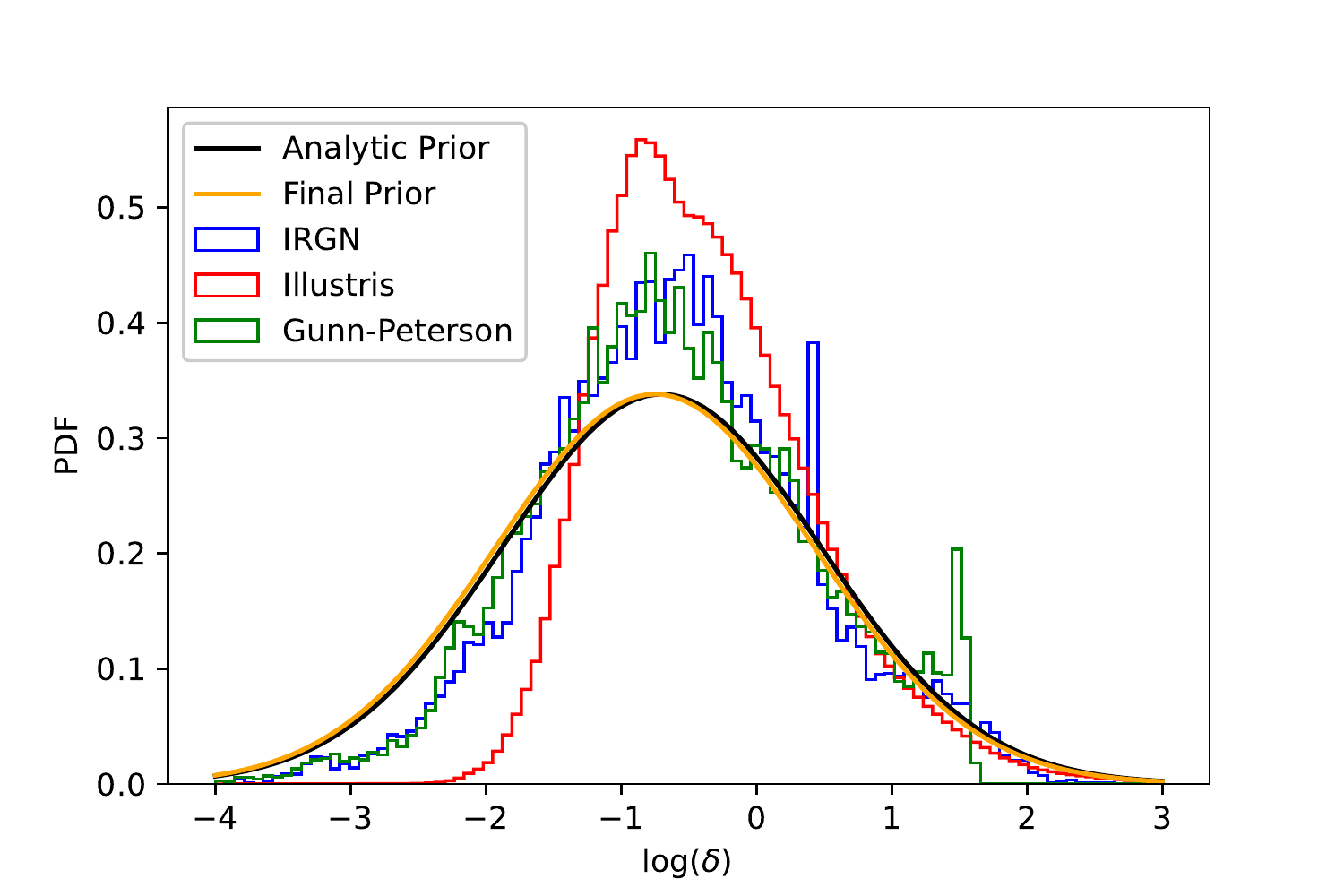}
    \caption{Probability density distribution of the baryonic density perturbation: measured from the first subbox of the third run of the illustris simulation (red), measured from the densities recovered with the IRGN inversion method (blue) and measured from the densities recovered with the fluctuating Gunn-Peterson approximation (green). Additionally, we show the analytic lognormal model (black) and the final prior that was used by us for observational data that are distorted by peculiar velocities (orange).}
    \label{fig: dens_hist}
\end{figure}

We compare the probability density distribution of the density perturbation in the sample of observed spectra in Fig. \ref{fig: dens_hist} with that of current hydrodynamic simulations, first subbox of third run of Illustris \citep{Nelson2015}, and the analytic prior lognormal model (black) which coincides with the probability density distribution computed in our simulation toolbox with \small{NBODYKIT}. We recover the pdf of the density perturbation with the IRGN inversion method and the best fit parameters $\gamma$ and $T_0$ (blue histogram in Fig. \ref{fig: dens_hist}). As the IRGN inversion results vary considerably with the assumptions on the thermal history of the IGM (see Appendix \ref{app: inversion_schemes}) we examine the probability density function computed by the fluctuating Gunn-Peterson approximation (green line), i.e. by Eq. \eqref{eq: gunn_peterson}. The Gunn-Peterson provides a less precise reconstruction, but does not depend on prior assumptions on the density pdf. Hence, the Gunn-Peterson approximation applied to observational data is a well probe to verify that observational data fit the prior density pdf. The results from the illustris simulation are tested at the resolution of the Jeans length. Moreover, we show the prior lognormal model that was used to fit data that are distorted by peculiar velocities (orange). We observed on synthetic data that the coincidence between IRGN and RPC inversions is improved for lognormal models with slightly decreased means. The lognormal model describes the overall probability distribution quite well, in particular the high density tail, but fails in the recovering the skewness in the distributions: the low density tail decreases to zero more quickly than the high density tail in real observations. However, the prior lognormal distribution matches well the mean and the interquartile range expected from simulations and from observations (note that there is some significant scatter between the three measurements on observations and in simulation).

\begin{figure}
    \centering
    \includegraphics[width=0.5\textwidth]{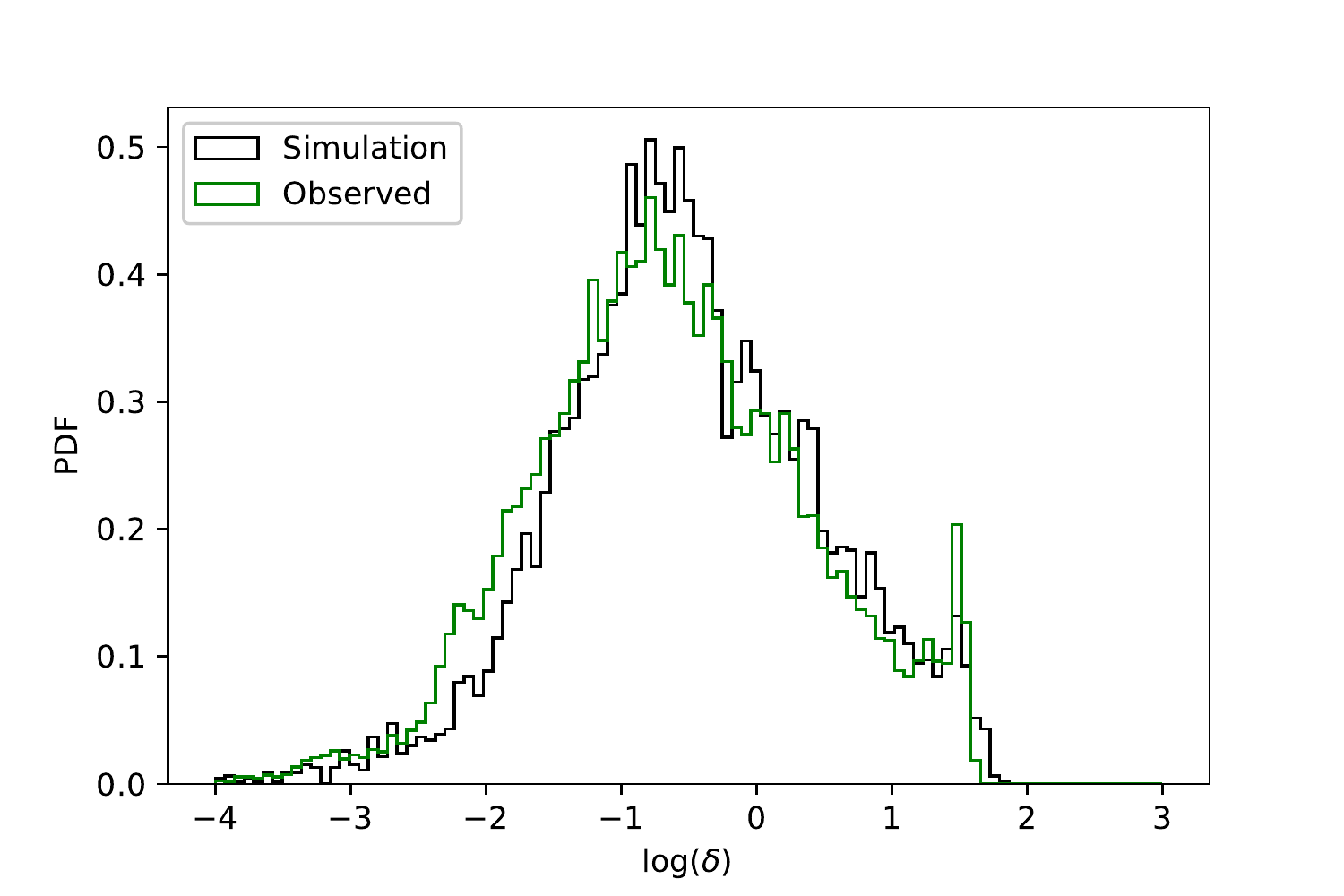}
    \caption{Probability density distribution of the baryonic density perturbation: measured with the fluctuating Gunn-Peterson approximation from observations (green) and measured with the fluctuating Gunn-Peterson approximation from simulations (black).}
    \label{fig: dens_hist_gunn}
\end{figure}

The mismatch of the low density tail of the distribution is (partly) explained by not including peculiar velocities in the analysis. Peculiar velocities enhance clustering of the matter in redshift space and thus prefer a more narrow probability density distribution. The probability distributions of observational data in Fig. \ref{fig: dens_hist} show this issue. The observed flux is more clustered in redshift space. When peculiar velocities are not included during inversion procedure (as they are unknown from observational data along single lines of sight) the conversion between redshift space and real space (and the related clustering) is not modelled correctly. Hence, the recovered density when not including peculiar velocities appears more clustered than it is truly. The recovered densities in Fig. \ref{fig: dens_hist} (blue and green histograms) are distorted in comparison to the analytic prior due to not including peculiar velocities. To measure this particular effect, we simulate the flux in the Ly$\alpha$ forest in 25 lines of sight with our lognormal based simulation (including peculiar velocities, using $\gamma=1.42$ and $T_0=13400\,\mathrm{K}$ as estimated in this paper) and recover the densities with the fluctuating Gunn-Peterson approximation. We compare our simulation with the observed density probability distribution in Fig. \ref{fig: dens_hist_gunn}. The coincidence between simulation and observations improves significantly in comparison to Fig. \ref{fig: dens_hist}. Overall Fig. \ref{fig: dens_hist_gunn} demonstrates that our lognormal based simulation of the cosmic overdensity field is able to describe observational data at high resolution. 

\begin{figure}
    \centering
    \includegraphics[width=0.5\textwidth]{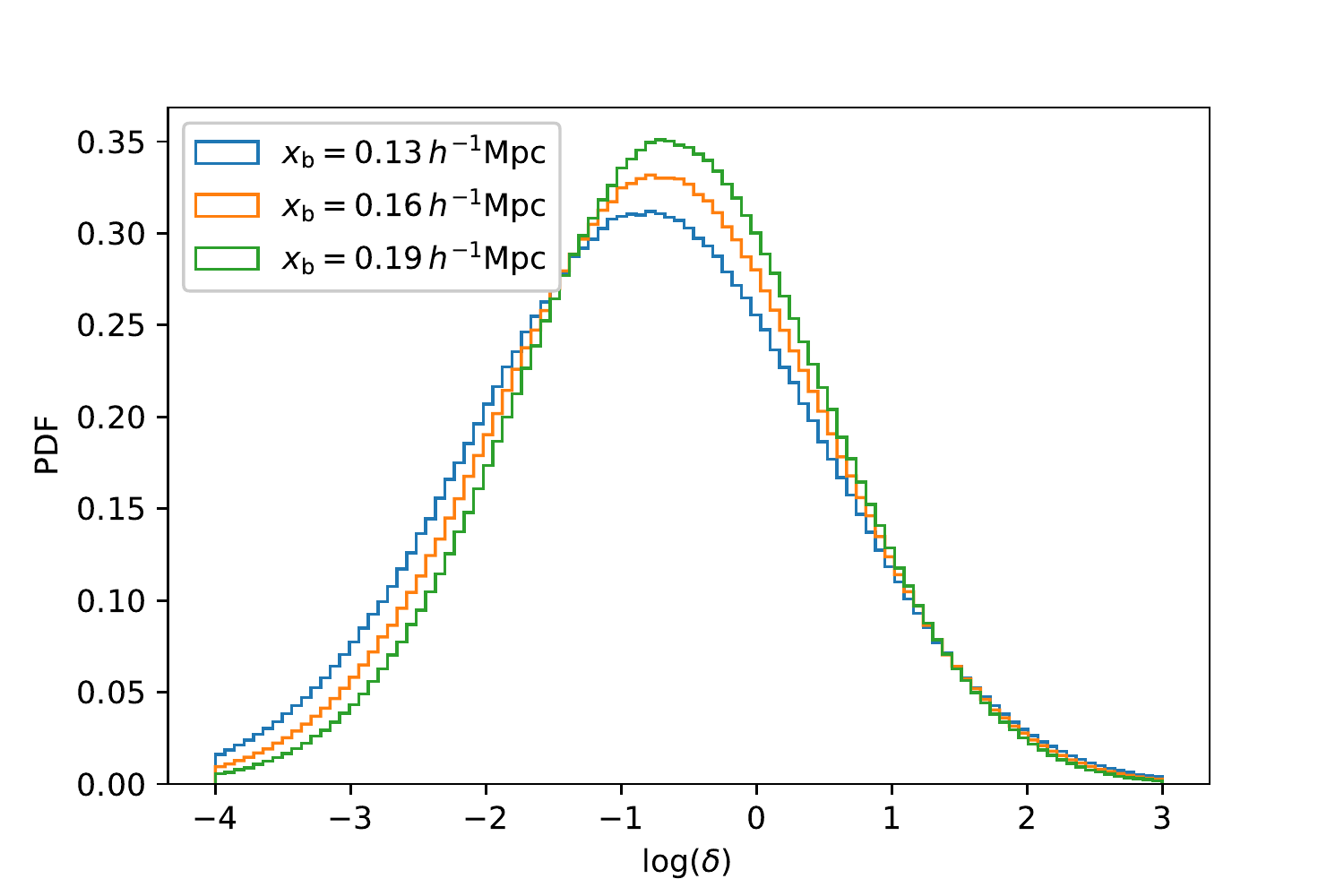}
    \caption{The matter probability density function from our simulation as function of Jeans length $x_\mathrm{b}$ and the analytic prior used during this paper.}
    \label{fig: jeans}
\end{figure}

Moreover, we compare the probability density distribution computed with varying smoothing scales with our simulation in Fig. \ref{fig: jeans}. The Jeans length indeed has an effect on the observed density probability density function. We find for all three smoothing scales the lognormal density distribution, but with varying mean and standard deviation. The value of $x_\mathrm{b} \approx 0.16h^{-1}\mathrm{Mpc}$ was found by \citet{Zaroubi2006} and is used throughout this publication. The analytic prior was chosen to match this \small{NBODYKIT} simulation with $x_\mathrm{b} \approx 0.16h^{-1}\mathrm{Mpc}$.

\section{Temperature Measurements} \label{app: meas_hist}
\begin{figure*}
\centering
Temperature Measurements \\
\subfigure[$\gamma = 1.3$]{
\includegraphics[width=0.3\textwidth]{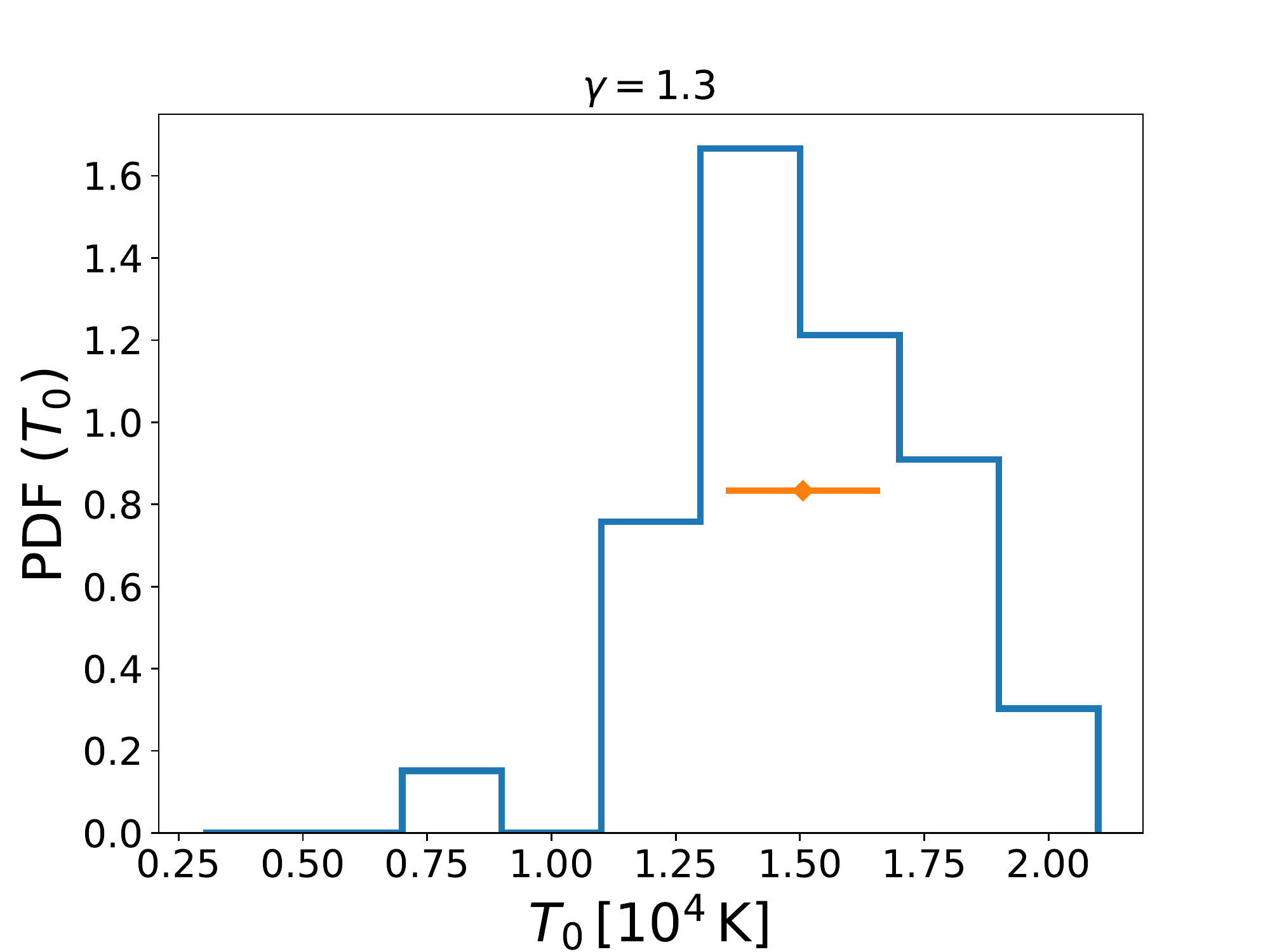}}
\subfigure[$\gamma = 1.35$]{
\includegraphics[width=0.3\textwidth]{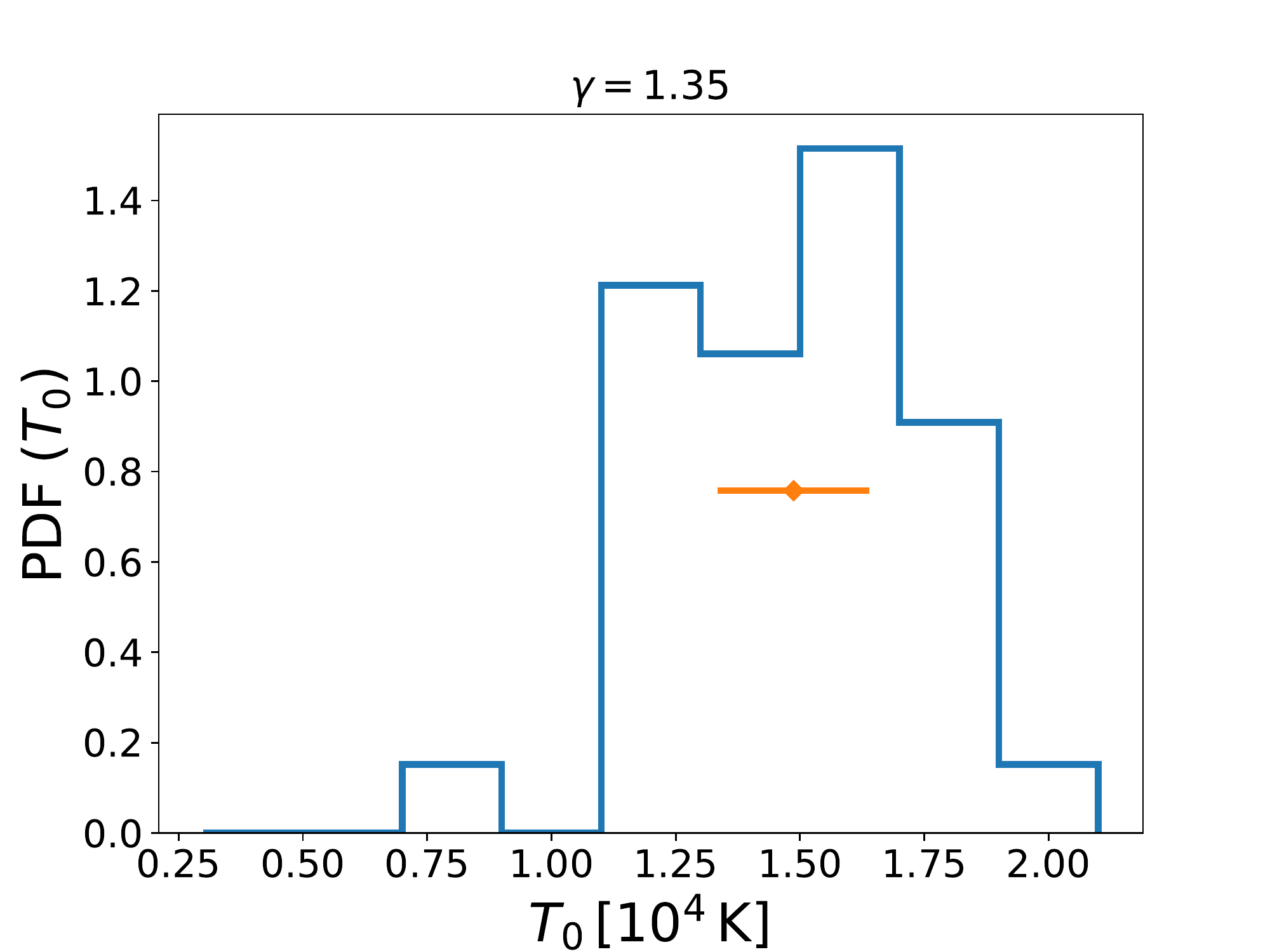}}
\subfigure[$\gamma = 1.4$]{
\includegraphics[width=0.3\textwidth]{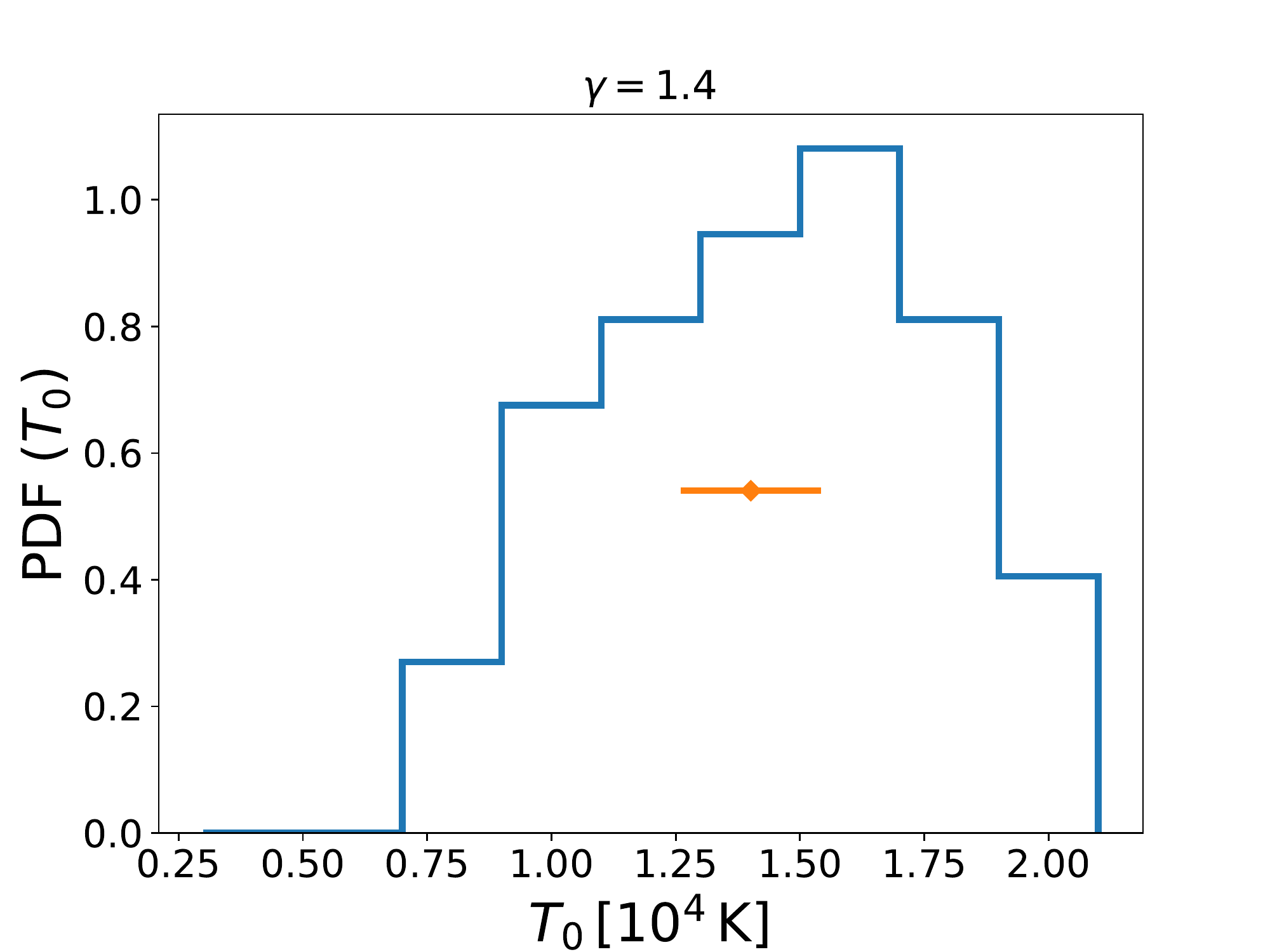}} \\
\subfigure[$\gamma = 1.45$]{
\includegraphics[width=0.3\textwidth]{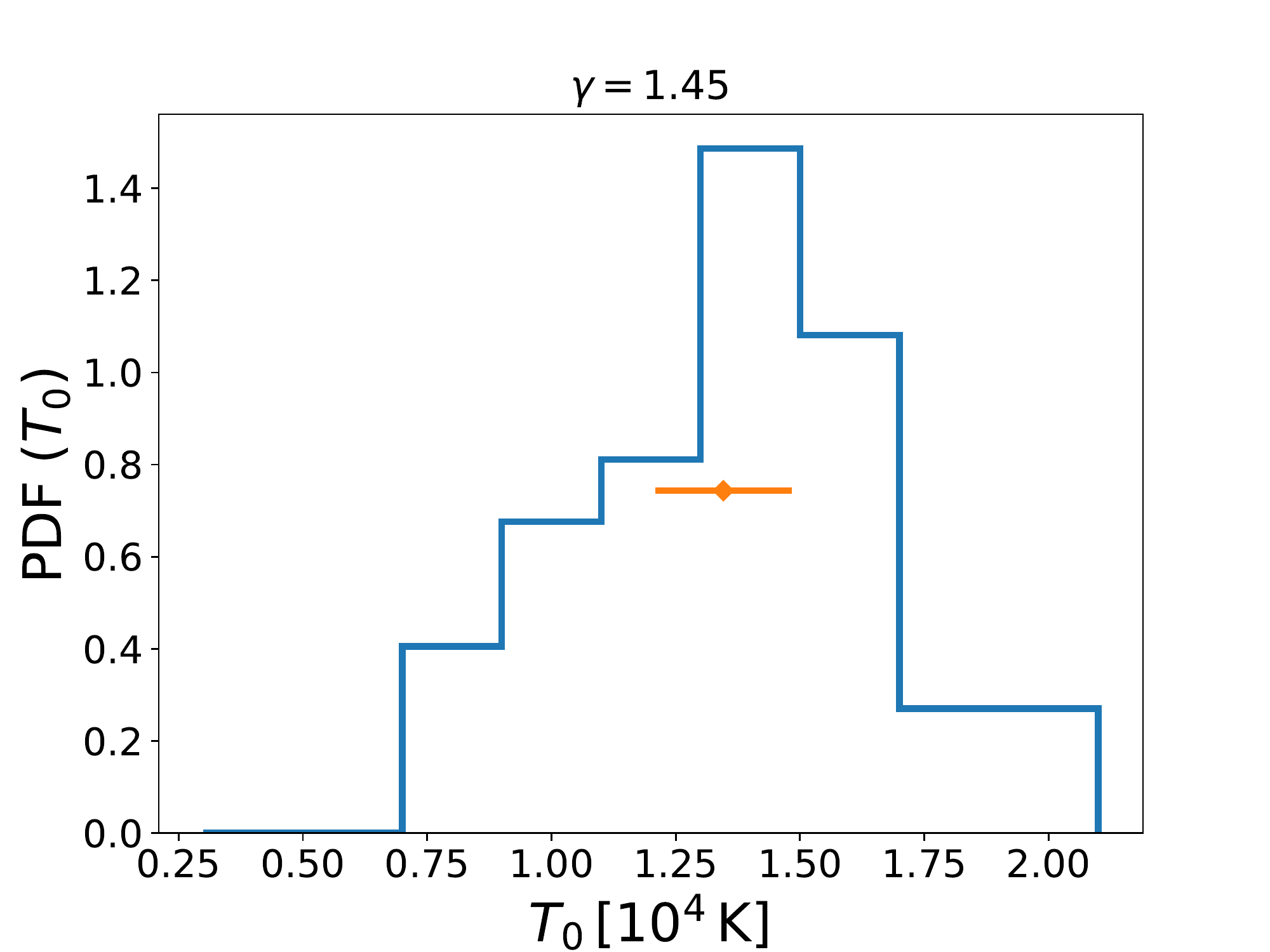}}
\subfigure[$\gamma = 1.5$]{
\includegraphics[width=0.3\textwidth]{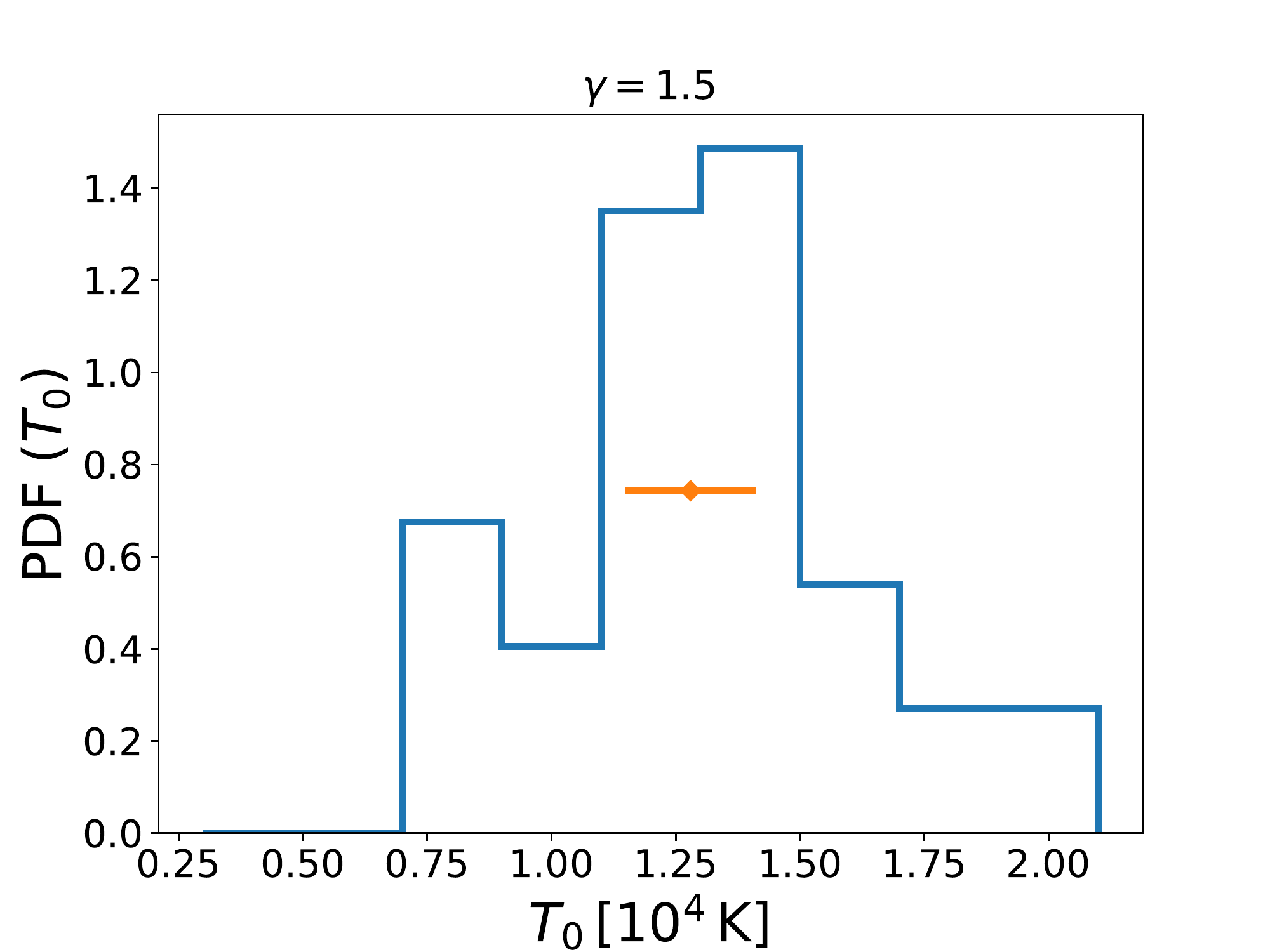}}
\subfigure[$\gamma = 1.55$]{
\includegraphics[width=0.3\textwidth]{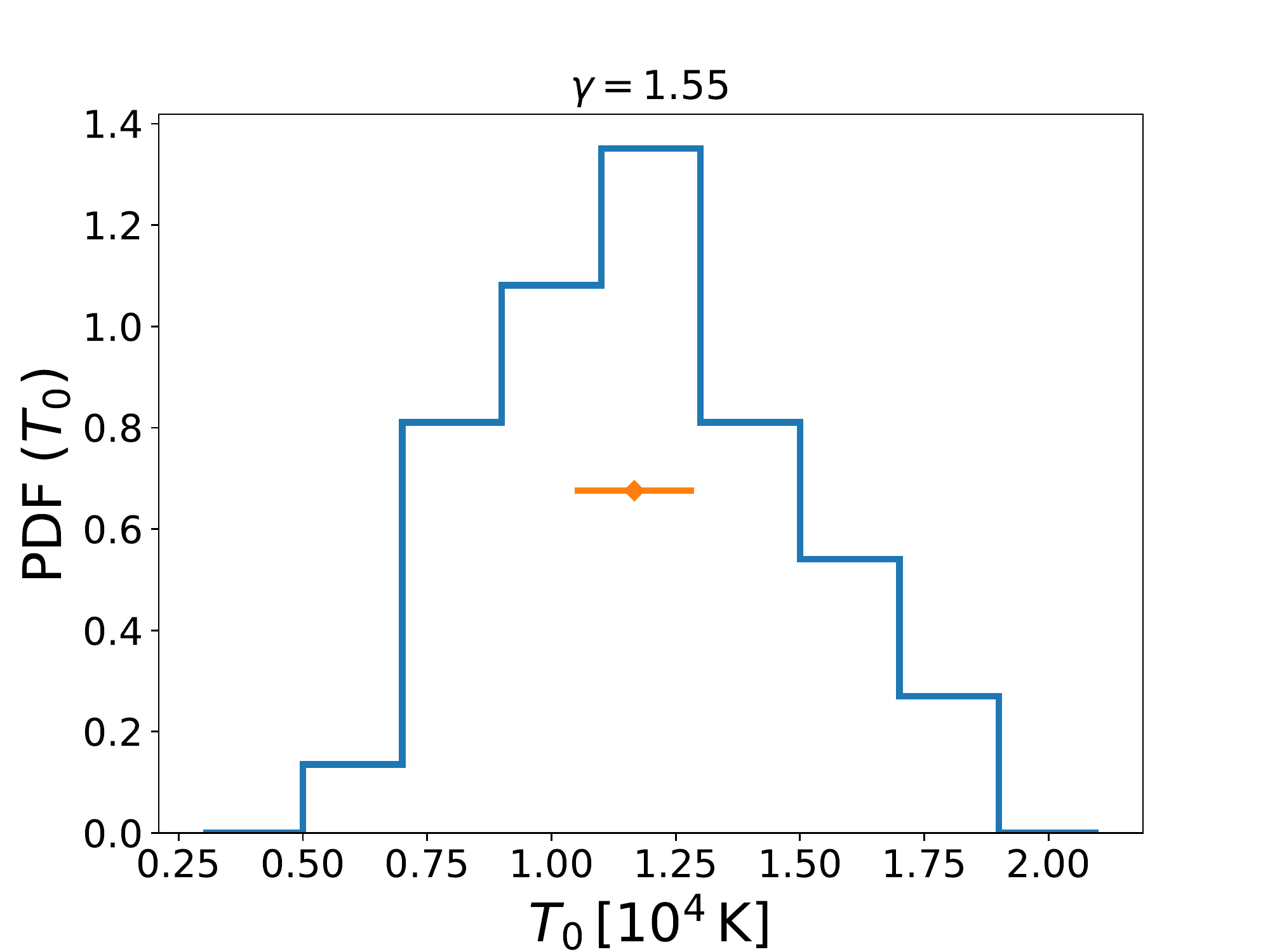}} \\
\subfigure[$\gamma = 1.6$]{
\includegraphics[width=0.3\textwidth]{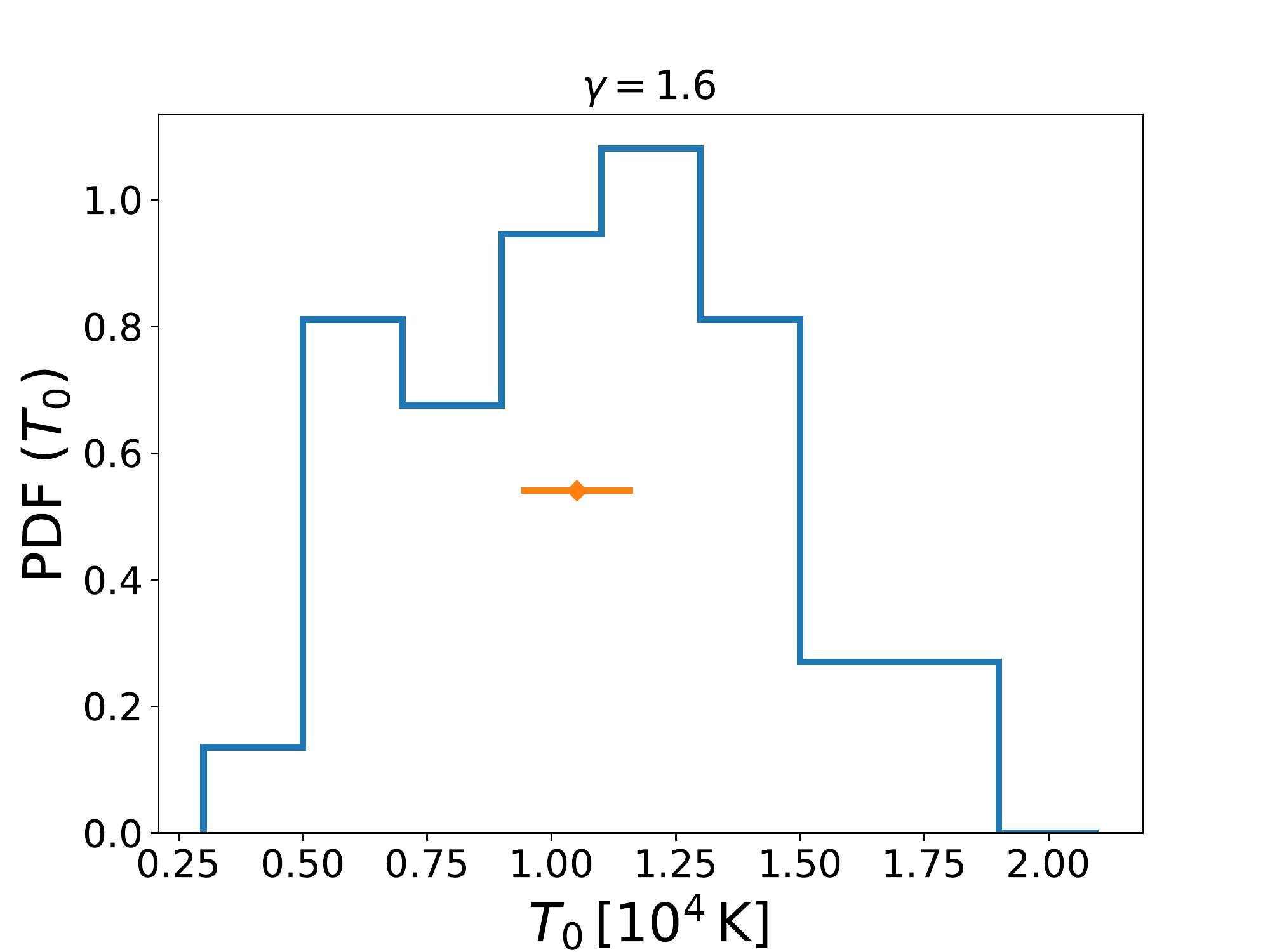}}
\caption{Histograms (blue) of temperature estimates for fixed values $\gamma$. We show with the yellow errorbar the weighted mean and the error of the weighted mean.}
\label{fig: meas_hist}
\end{figure*}

We show in Fig. \ref{fig: meas_hist} histograms of our temperature measurements with observational data for single values of $\gamma$. The probability distributions can be approximated Gaussian, which justifies the use of the (weighted) average as estimator for the temperature for some fixed values for $\gamma$. The yellow errorbars show the weighted (weighted by the number of unmasked pixels per set) of mean temperature and the statistical error of the weighted mean. 


\bsp	
\label{lastpage}
\end{document}